\documentclass{article}
\usepackage{graphicx}
\usepackage{xcolor}
\usepackage{amsfonts}
\usepackage{amssymb}
\def\ligne#1{\hbox to\hsize{#1}}
\def\leurre{\noindent\leftskip0pt\small\baselineskip 10pt}
\newtheorem{thm}{\textbf{Theorem}}
\newtheorem{fig}{\textbf{Figure}}
\newtheorem{tab}{\textbf{Table}}

\author{Maurice {\sc Margenstern}}
\title{A strongly universal cellular automaton on the heptagrid with seven states}
\begin{document}
\maketitle

\begin{abstract}
In this paper, we prove that there is a strongly universal cellular automaton on the 
heptagrid with seven states which is rotation invariant. This improves a previous paper 
of the author where the automaton required ten states.
\end{abstract}

\section{Introduction}

   The first paper about a universal cellular automaton in the pentagrid, the tessellation
$\{5,4\}$ of the hyperbolic plane, was \cite{fhmmTCS}. This cellular automaton was also rotation
invariant, at each step of the computation, the set of non quiescent states had infinitely many
cycles: we shall say that it is a truly planar cellular automaton. But, the neighbourhood
was \textit{\`a la von Neumann} and it had 22~states. 
This result was improved by a cellular automaton with 9~states in~\cite{mmysPPL}.
Recently, I improved this result with 5~states, see~\cite{mmpenta5st}. A bit later,
I proved that in the heptagrid, the tessellation $\{7,3\}$ of the hyperbolic plane,
there is a weakly universal cellular automaton with three states which is rotation 
invariant and which is truly planar, \cite{mmhepta3st}. Later, I improved the result
down to two states but the rules are no more rotation invariant.
In the present paper, I construct a cellular automation with seven states which is
rotation invariant and truly planar.

%In a paper under printing, see~\cite{mm11.3.2st}, 
%I proved that there is such a cellular automaton in the tessellation $\{11,3\}$ of the
%hyperbolic plane.

    In the result of~\cite{JAC2010}, the automaton evolves on a construction which 
follows a line, constructing it step by step, starting from the finite configuration.
On the line, that automaton simulates a Turing machine which, in its turn, simulates
a tag systems known to be universal when the deleting number is at least~2, 
see~\cite{minsky}. In the present paper we follow a different strategy. We explain it in 
Section~\ref{scenario}. In Section~\ref{scenar} we describe in detail the
structures which are involved in the new construction. In Section~\ref{rules}, we give
the rules which were checked by a computer program which also computed many figures
of the paper.
%together with the traces of execution of the computation performed by a
%computer program.
%%/ In Section~\ref{Program}, we give a few details on the algorithms
%which were implemented in the computer program I used to check the correctness of
%the automaton.
   
     All these sections constitute the proof of the following result:

\begin{thm}\label{letheo}
There is a strongly universal cellular automaton on the heptagrid
which is rotation invariant, truly planar and which has seven states.
\end{thm}

   The minimal introduction to hyperbolic 
geometry needed to understand the paper can be found in~\cite{smallbook}. We assume the 
reader is a bit familiar with that approach so that Sections~\ref{scenario}
and \ref{scenar} rely on it with no further reference.

\section{Hyperbolic railway and register machines}
\label{scenario}

    In the papers were I constructed weakly universal cellular automata in the pentagrid 
and in the heptagrid, I used the model devised by Ian Stewart in~\cite{stewart} to 
simulate the computation of a register machine. That model makes use of a circuit
constituted by switches connected by tracks to simulate the computation. The complexity
of the tracks used in those papers made it impossible to start from a finite 
configuration within a small number of states. However, the line constructed 
in~\cite{JAC2010} made use of a comparatively small number of states, a part of which 
being also used for the simulation of the particular Turing machine used in that
paper.

    As we start from a finite configuration, the idea is to implement the data stored
in a register as simply as possible. Here, the content of the register
which is supposed to be a natural number, is represented by that number written in
unary. If the content at time~$t$ is $N$, it is represented by $N$ cells in an 
appropriate state stored along a line of the hyperbolic plane. As seen in 
Figure~\ref{register}, such a line is easily characterised in the heptagrid. 
For the rest of the computation, our simulation follows the same lines as in the quoted 
papers: the instructions of the register machine and the access to the registered are
managed in the same way up to the adaption of the tracks we shall soon consider.

\subsection{The railway model}
\label{railway}

   As already mentioned, the railway model was introduced by Ian Stewart 
in~\cite{stewart}. It was initially devised to simulate a Turing machine, but I used it 
to simulate a register machine. For technical reasons, that latter model is easier to 
simulate as it is known that two registers are enough
to simulate any Turing machine on $\{0,1\}$, see~\cite{minsky}.

    The model consists of a railway circuit on which a single locomotive is running. The 
circuit contains crosses and switches and the state of all switches of the circuit 
considered at a given top of the clock constitutes the configuration of the circuit at 
that time. The circuit consists of tracks which are represented by assembling segments 
of straight lines, either horizontal or vertical ones and quarters of a circle. That 
representation lies in the Euclidean plane. In the hyperbolic one, things are necessarily
different. We turn to that point in Section~\ref{scenar}.
It contains crosses which allows tracks to intersect. It also contain switches in order 
to make suitable figures, in particular cycles. There are three kinds of switches 
illustrated by Figure~\ref{switches}. From left to right
in the figure, we can see the \textbf{fixed switch}, the \textbf{flip-flop} and the 
\textbf{memory switch}. We can see that
a switch realizes the intersection of three tracks, we shall denote them by~$a$, $b$ and~$c$.
Say that~$a$ is the single track which arrives at the switch and that $b$ and~$c$ are
those from which the locomotive may leave the switch, either through~$b$ or through~$c$.
The track taken by the locomotive to leave the switch is called the \textbf{selected track}.
When the locomotive arrives to the track through~$a$ and leaves it through the selected track,
we say that it is an \textbf{active} crossing. If the locomotive arrives through~$b$ or through~$c$,
leaving the switch through~$a$, we say that it is a \textbf{passive} crossing. In the figure,
each kind of switch is represented in two forms. Both of them will be used in our further
illustrations.
\vskip 10pt
\vtop{
\ligne{\hfill
\includegraphics[scale=0.8]{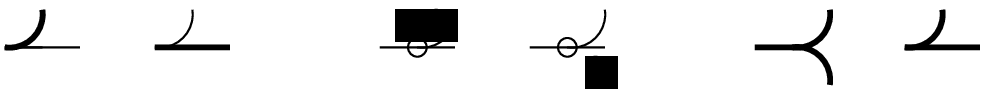}
\hfill}
%\vspace{-20pt}
\begin{fig}\label{switches}
\leurre
The switches used in the railway circuit of the model.
\end{fig}
}

   In the fixed switch, the selected track is determined once and for all. The switch can be
crossed both actively or passively. The flip-flop may be crossed actively only. When
it leaves the switch, the selected track changes: the previous non-selected track becomes
selected and the previously selected becomes non-selected. The memory switch can also be crossed
either actively or passively. The selected track is determined by the last passive crossing.
These switches are exactly those defined in~\cite{stewart}. We also take from 
paper~\cite{stewart} the configuration illustrated by Figure~\ref{basicelem} which contains 
one bit of information.

\vtop{
\ligne{\hfill
\includegraphics[scale=0.6]{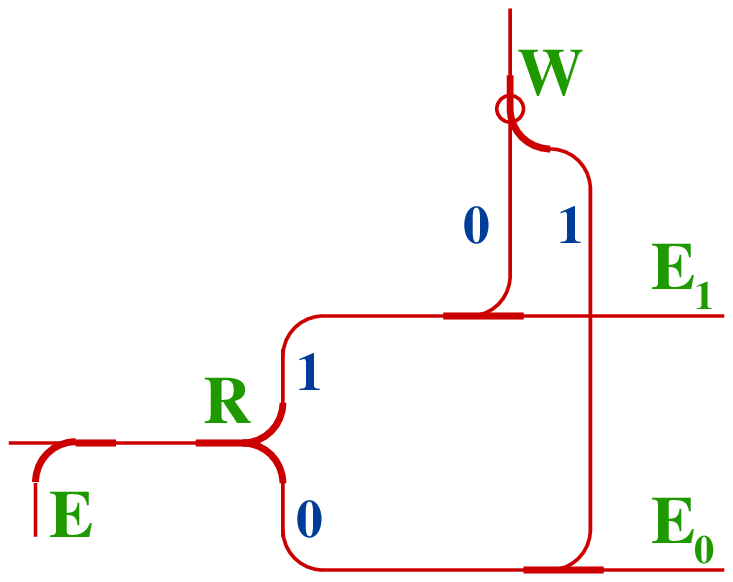}
\hfill}
%\vspace{-20pt}
\begin{fig}\label{basicelem}
\leurre
The basic element containing one bit of information.
\end{fig}
}

This configuration, called the \textbf{basic element}, is used as follows. If the 
locomotive arrives through~$R$, it goes either through track~0 or through track~1. If 
the locomotive goes through the track~$\alpha$ with 
$\alpha\in\{0,1\}$, it leaves the configuration through~$E_\alpha$. We say that the 
locomotive
\textbf{has read~$\alpha$}. If the locomotive arrives through~$W$, it goes through track~0 or~1.
Now, if it arrives through the track~$\alpha$, it passively crosses~$R$ through the track~$\beta$,
with $\beta\in\{0,1\}$ and \hbox{$\alpha+\beta=1$}. Eventually, the locomotive leaves the element 
through~$E$. Now, as the locomotive crossed~$W$,
the selected track at~$W$ is now~$\beta$. Consequently, we can say that the selected tracks 
at~$R$ and~$W$ are always the same when the locomotive is arriving at the configuration and
also when it leaves it. Accordingly, the locomotive may read~0 or~1 and also it may rewrite~$\alpha$
into~$\beta$.

\subsection{Register machine}

   Many instructions arrive to a register. In~\cite{smallbook} for instance, we 
described two dispatching structures: one of them to gather the instructions to
increment that register and the other to gather the instructions to decrement that 
register. The structure has a single output way to the register according or the
type of instructions arriving to it. On the way back, the structure sends out
the locomotive to the corresponding sending instructions.
Figure~\ref{gather} displays both structures: on the upper picture, the dispatcher of 
instructions for incrementing that register; on the lower picture it performs the same 
for the instructions which decrement the same register.

\vtop{
%\vspace{-15pt}
\ligne{\hfill
\includegraphics[scale=0.6]{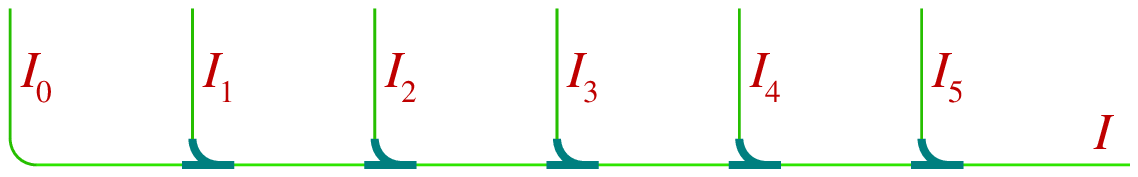}
\hfill}
\vspace{5pt}
\ligne{\hfill
\includegraphics[scale=0.6]{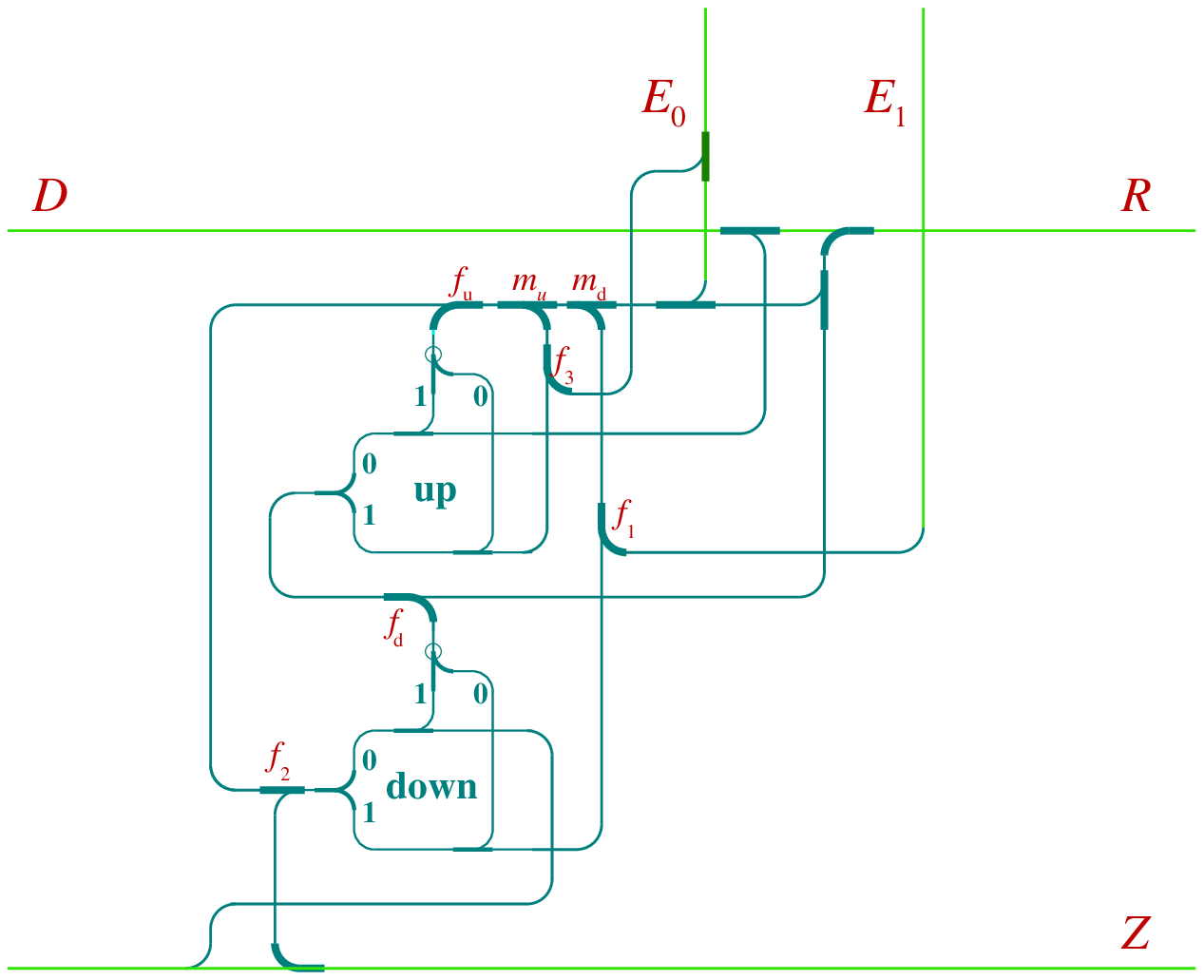}
\hfill}
\vspace{5pt}
\begin{fig}\label{gather}
\leurre
Gathering instructions on their way to a register and dispatching the return locomotive
to the appropriate way on its way back.
\end{fig}
}

   We can see on the figure the role played by the basic unit to register the
returning path by setting both basic units to~1 when the locomotive arrives from
an instruction to decrement the register. It is simpler for the instructions which 
increment the register: a memory switch is enough for that purpose. For an instruction
which decrements the register it is less simple as far as two cases may occur: the case
when it was possible to decrement the register and the opposite case when the register
was already zero at the arrival of the locomotive. It is the reason why to units are
used for each arrival from an instruction which decrements the register. Both units are
marked to~1 when the locomotive arrives from the instruction. One way is used when the 
registered was actually decremented the other way when it could not be. One unit is 
read by the locomotive when it could decrement the register, the other one is read
when it could not do it. When the unique couple of units set to 1 is detected, the
locomotive resets both units to~0 and it goes to the appropriate destination.

However, in the register, when the operation was applied the locomotive goes back to
the instructions through the same way. This is way that way goes through a special unit
which remembers whether the locomotive coming to the register has to increment it or to 
decrement it. The unit is~0 for an instruction which increments the register, it was set
to~1{} in the other case. And so, on the way back, the locomotive knows where it has to
go further. Note that the test to zero leads to the gathering units directly.

   Figure~\ref{memoinst} displays the circuit of that selection.

\vtop{
%\vspace{-15pt}
\ligne{\hfill
\includegraphics[scale=0.5]{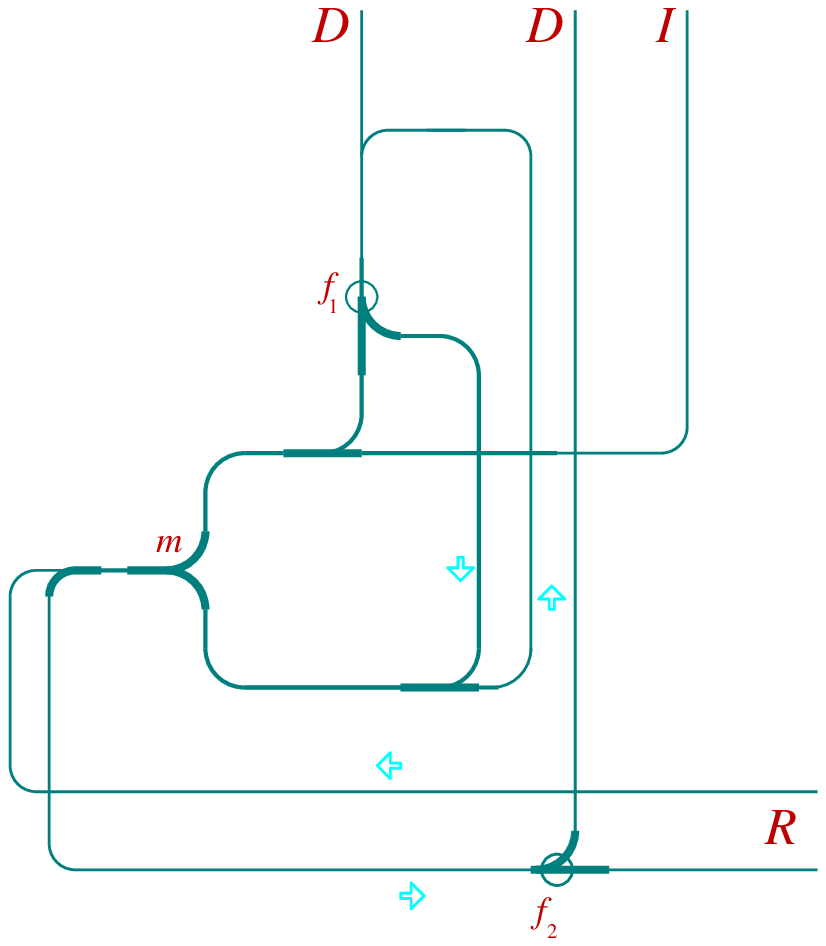}
\hfill}
\vspace{5pt}
\begin{fig}\label{memoinst}
\leurre
The unit which remembers whether the instruction was to decrement the register.
\end{fig}
}

\ifnum 1=0 {
As explained in~\cite{mmbook3}, using the basic unit illustrated by Figure~\ref{basicelem} 
and crossing with switches as in Figure~\ref{switches}, it is possible to build a circuit which
simulates the working of a register machine. Figure~\ref{toyexample} illustrates how such
a construction can be performed by implementing in this setting a small program of a register
machine involving three registers. The program performs the action denoted by the instruction
\hbox{$X := Y;$} of many programming languages. At any time, a configuration of the computation
consists in the position of all the switches of the circuit. However, note that at each time,
only finitely many of them are changed so that at each time, the computation can be described
with a finite word, despite the fact that the initial configuration is infinite. 

\vtop{
\ligne{\hfill
\includegraphics[scale=0.45]{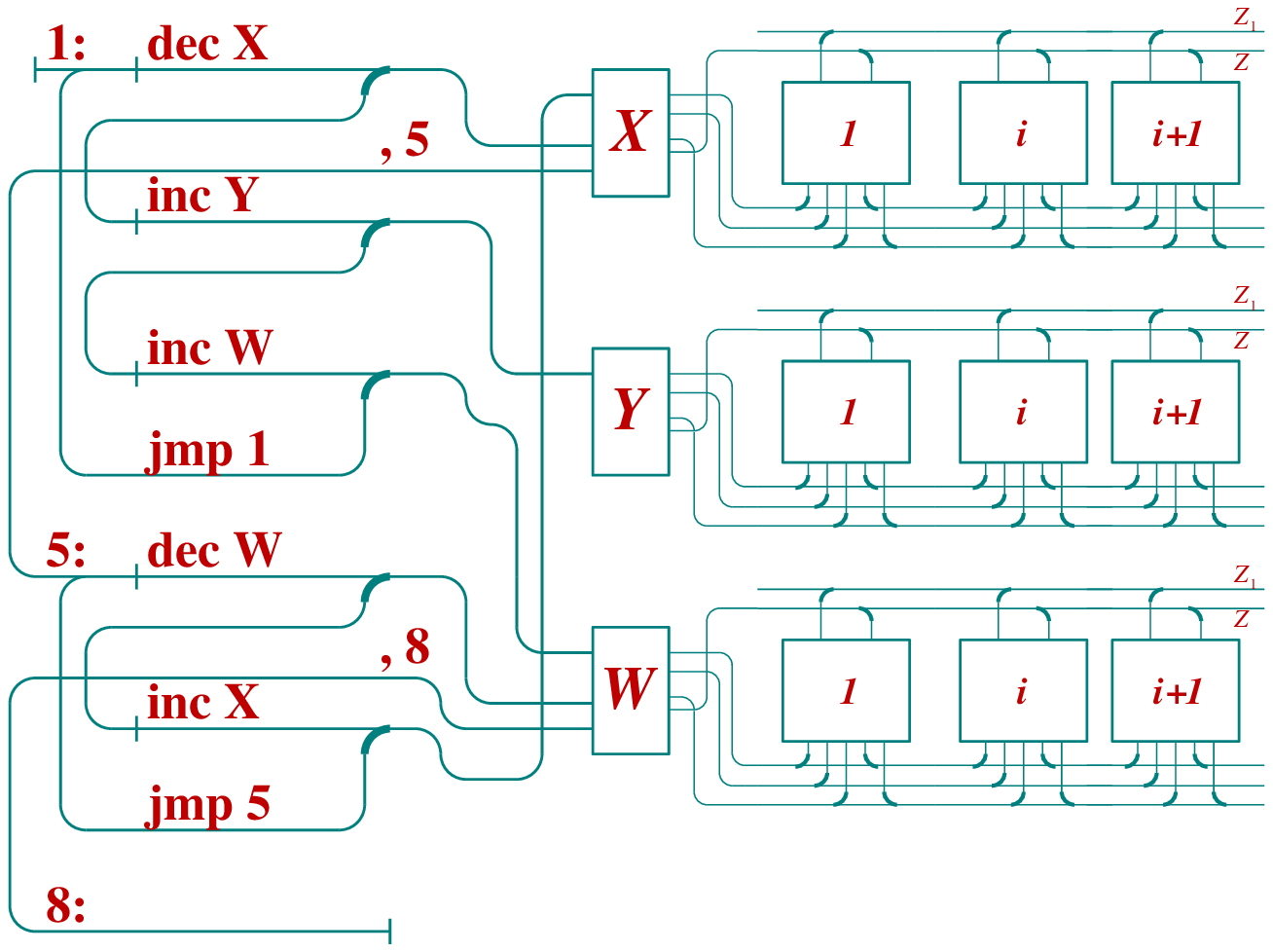}
\hfill}
%\vspace{-20pt}
\begin{fig}\label{toyexample}
\leurre
Implementation of a toy-register machine associated to a toy-program.
\end{fig}
}
} \fi

\subsection{Hyperbolic implementation of the model}
\label{sbimplement}

    The model of computation described in Sub-section~\ref{railway} is described in~\cite{stewart} as a circuit lying in the Euclidean plane. It cannot been transported into
the hyperbolic plane without detailed explanations. There are two steps in this translation. First, we need to have a global view of how to implement the circuitry which will simulate
the computation. This is what is discussed in Sub-section~\ref{global}. Once this question
is solved, we can look at the details which is the purpose of Sub-section~\ref{basics}.

\subsection{Global view}
\label{global}

    In order to fix things, we consider the Poincar\'e's disc model as the frame of our
investigation. We take this model for two reasons. The first one is that it preserves angles, 
which allow us to see something. The second reason is that the model gives us a global view on
the hyperbolic plane. However, we must be aware of two big restrictions. The first restriction
is that the model imposes a distortion on distances: when we are close to the centre of the 
disc, the distances are small and the closer we go to the border, the bigger are the distances.
Also, second restriction, there is no similarity in the hyperbolic geometry. There, distances
are absolute. This means that if we change our unit to measure distances by another one,
the new distances are not simply multiplied by a simple factor. 

    A good image is the following one. We have to see Poincar\'e's disc as a window over 
the hyperbolic plane, as flying over that plane in some spacecraft. Accordingly, the 
centre of Poincar\'e's disc is not the centre of the hyperbolic plane: that plane has no 
centre. The centre of the disc is the centre of our window in the spacecraft: it is the 
point of the plane on which our attention is focusing. %at which we are looking. 
But what is close to the border is not affected by changing the point at which we are 
looking. If we change of unit, this amounts to change our altitude over the hyperbolic 
plane in our image. We shall have a larger view close to the centre but what is close to 
the border still remains out of our view. 

   Figure~\ref{uhcaglobal} shows how to implement a two-registered machine.
    
\vtop{
\ligne{\hfill
\includegraphics[scale=0.5]{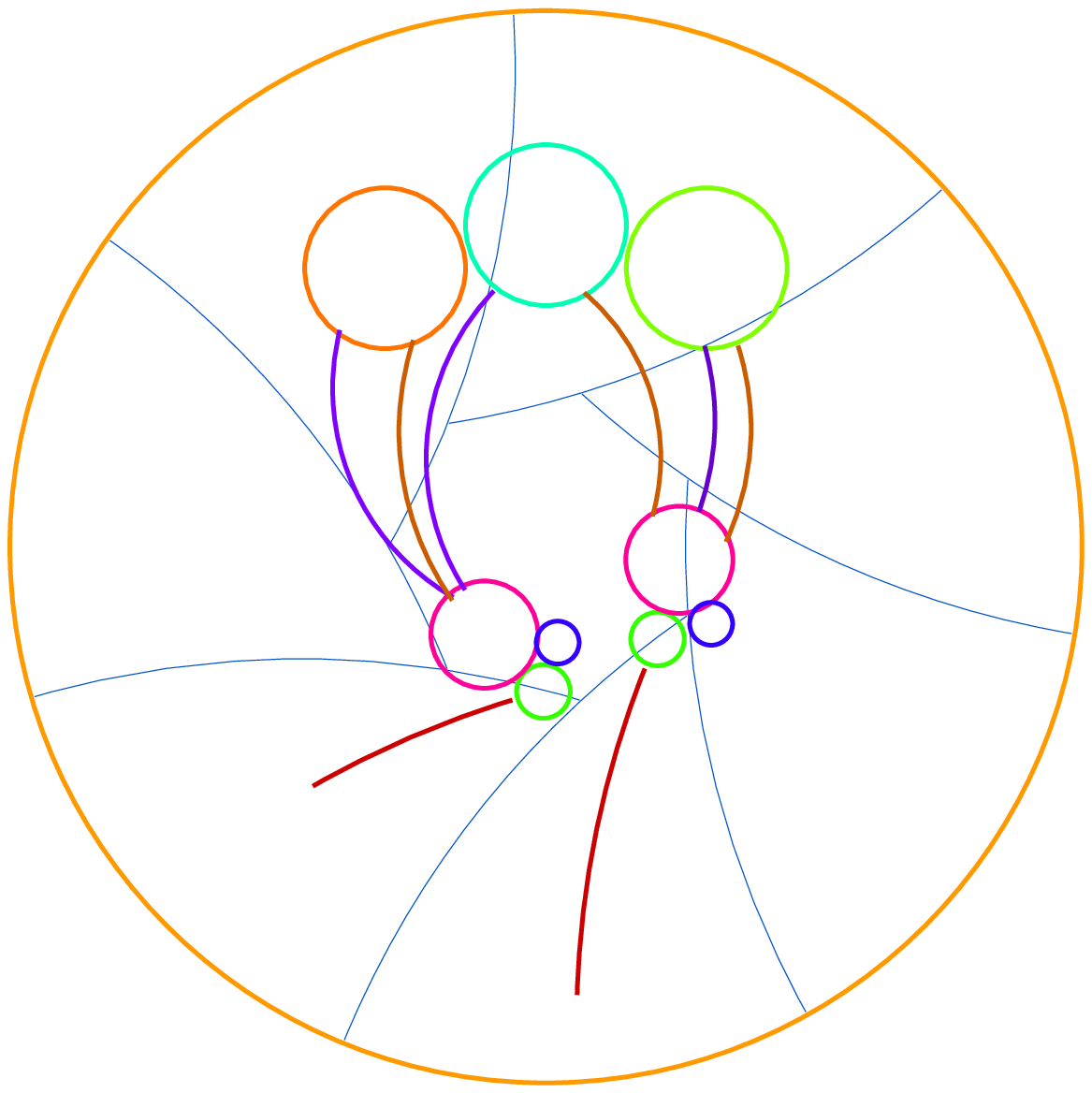}
\hfill}
\vspace{-20pt}
\begin{fig}\label{uhcaglobal}
\leurre
A global look at our implementation. For the sake of simplicity, the represented machines
has six instructions and two registers. We have to imagine other circles inside the three
circles from which are drawn the arcs representing the instructions going to the 
registers. Blue arcs represent incrementing instructions, brown ones represent those
that decrement registers.
\end{fig}
}

In Sub-section~\ref{basics} we explain the particular features of Figure~\ref{uhcaglobal}.
we know turn to that sub-section.

\ifnum 1=0 {
The vertical lines of Figure~\ref{toyexample} are replaced by rays issued from the centre of 
the disc and horizontal lines of that figure are replaced by arcs of circles centred at the 
centre of Poincar\'e's disc.

In a similar way, we can implement the basic element on which all the other structures are
built, see Figure~\ref{hca54elem}. 
} \fi

%   However, o
Our implementation is not directly performed in the hyperbolic plane, but in
a tiling of this plane, more precisely the \textbf{heptagrid}, the tessellation $\{7,3\}$
of the hyperbolic plane which is illustrated by the left-hand side part
of Figure~\ref{hepta}. 

%\vtop{
%\ligne{\hfill
%\includegraphics[scale=0.3]{hca54_element.ps}
%\hfill}
%\vspace{-5pt}
%\begin{fig}\label{hca54elem}
%\leurre
%Implementation of the basic element of Figure~{\rm\ref{basicelem}} in the 
%pentagrid.
%\end{fig}
%}

It is important to know that the tiling can be generated by a tree. The tree can be seen
on each sector of the right-hand side picture of Figure~\ref{hepta} by the following 
property. Green and yellow tiles have three sons: blue, green and yellow; while 
blue tiles have only two sons: blue and green.
On the left-hand side picture of Figure~\ref{hepta}, we can see the 
\textbf{lines of the heptagrid}: they are the lines
which joins mid-points of sides meeting at a vertex. On that figure, we especially drawn
the lines which delimit a sector.
Remember that, in Poincar\'e's
disc model, lines are represented by traces in the disc of diameters or circles which are
orthogonal to the border of the disc. A particular tree plays a key role. It generates the
tiles which are inscribed in a sector of the hyperbolic plane which is delimited by two 
lines as previously defined. There are seven such sectors
around the central tile of Figures~\ref{hepta}. 

\vskip 10pt
\vtop{
\ligne{\hfill
\includegraphics[scale=0.75]{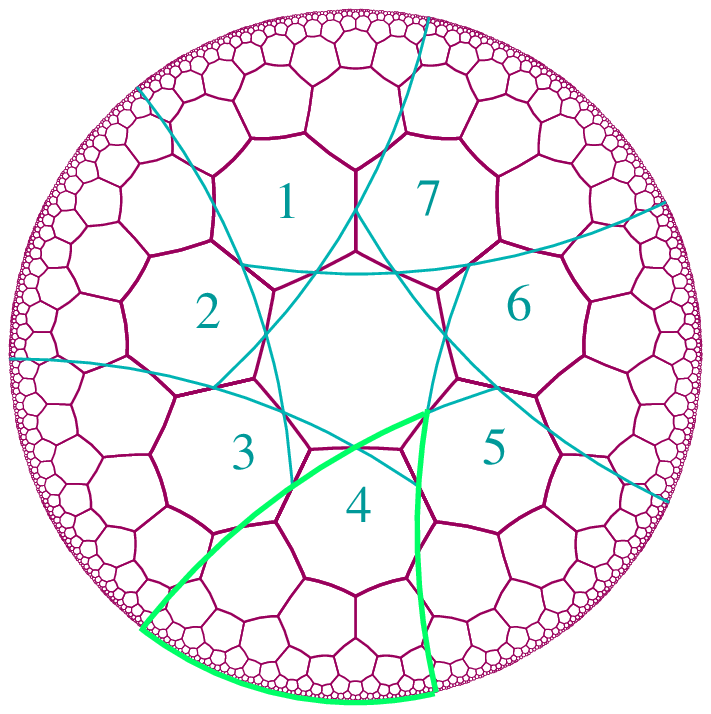}
\raise 2pt\hbox{\includegraphics[scale=0.46]{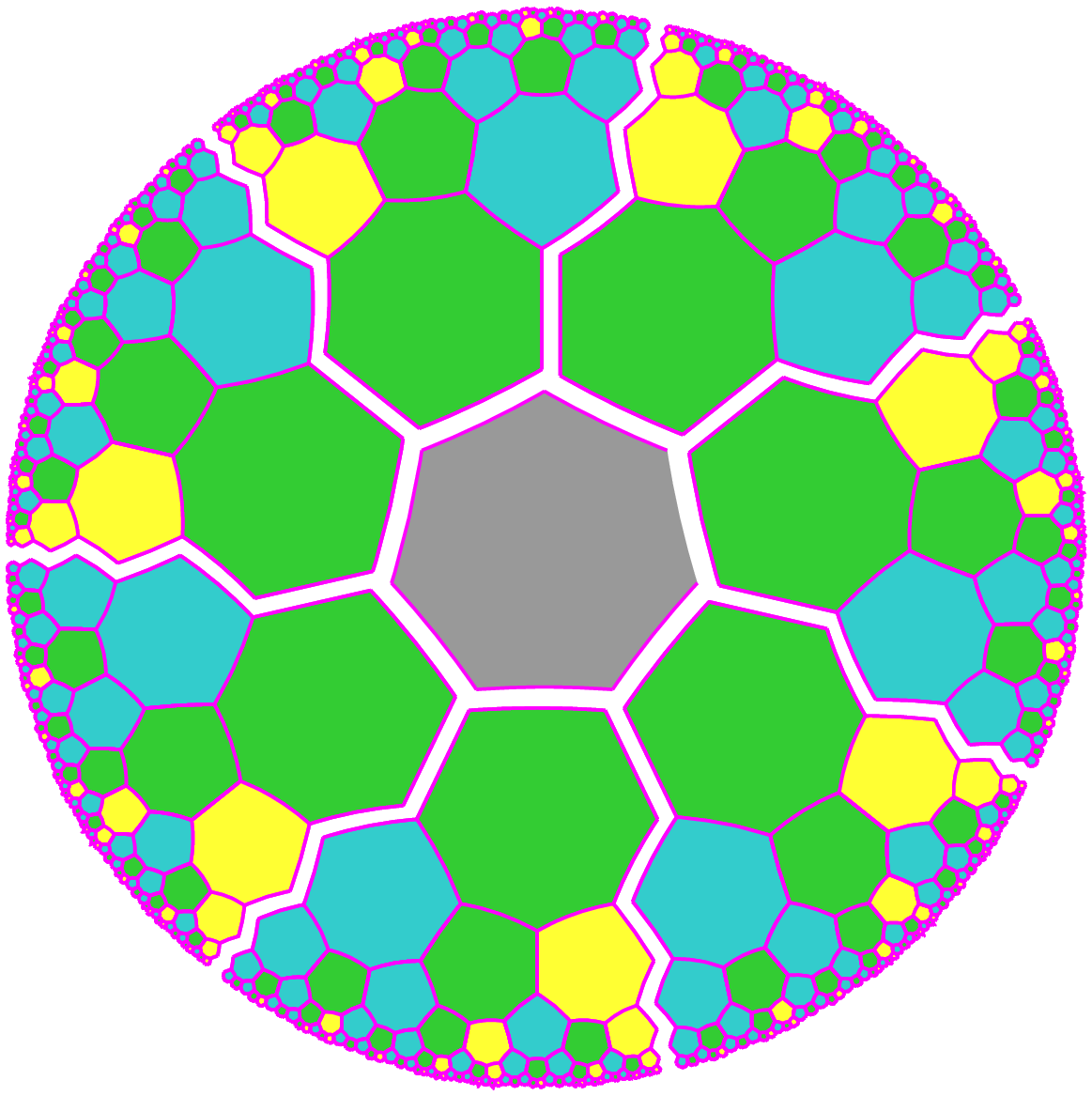}}
\hfill}
\vspace{-5pt}
\begin{fig}\label{hepta}
\leurre
To left: the heptagrid; to right: showing the sectors around a central tile.
\end{fig}
}

\subsection{Basic principles}
\label{basics}

    Presently, we turn to the implementation of the railway circuit in the heptagrid.
Picture~\ref{register} illustrates the way we implement a register.
The leftmost picture indicates a way which can be used to implement lines with tiles.
However, such a line of tiles in that way requires a two-celled locomotive in order to
define the direction of its motion. A possible solution would be to reinforce the line
as indicated on the central picture of the figure. However, we need another way for 
the return of the locomotive to the instructions once it had performed the operation
involved by the just executed instruction. The solution is given by the rightmost picture.
The blue line defines the content of the register. The tiles below the line, in yellow
on the picture, are used for the execution of the instruction. The tiles above the line,
in orange on the picture, are devoted to the return of the locomotive.

\vskip 10pt
\vtop{
\ligne{\hfill
\includegraphics[scale=0.6]{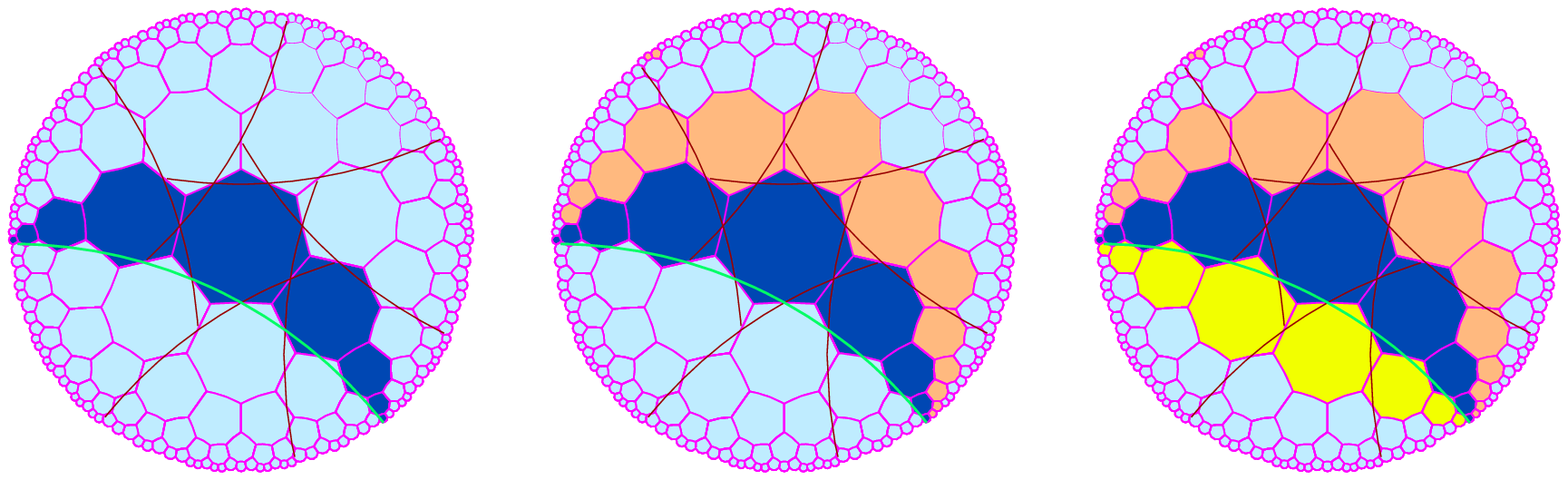}
\hfill}
\vspace{-5pt}
\begin{fig}\label{register}
\leurre
A line in the heptagrid: possible solution on the rightmost picture.
\end{fig}
}

    Our first focus will be the implementation of the tracks. Without tracks, there is no
way to convey information between the parts involved in the simulation of the register
machine. This is why any project for constructing a universal cellular automaton
must first solve the question of the circuit implementing the circulation of information.

    At this point, we have to remind that we are looking after a cellular automaton 
whose rules are rotation invariant: the new state of a rule is the same if we perform a 
circular 
permutation on the states of the neighbours. It appeared, during my search of a solution,
that using lines as described by the central picture of Figure~\ref{register} a confusion
could occur with rules in another situation. This is why I had to devise another solution
for the tracks which consists in taking arcs of 'circles'. 

In the tiling, a path between the tiles~$A$ and~$B$ is a finite sequence 
\hbox{$\{T_i\}_{i\in[0..n]}$} of tiles such that \hbox{$T_0 = A$}, \hbox{$T_n = B$}
and $T_i$ shares a side with~$T_{i+1}$ for all $i$ with \hbox{$0\leq i < n$}.
In such a situation, $n$ is called the {\bf length} of the path. In the 
tiling, the distance between two tiles $A$ and~$B$ is the length of a shortest path
between~$A$ and~$B$. We call {\bf circle} in the tiling, a set
of tiles whose distance to a fixed tile, called its {\bf centre} is constant. 
It will turn out that the rules for the motion of the locomotive along an arc of a circle
are rotation invariant, keeping this property with the rules we shall construct for the 
other structures. 

    Among the details we have now to look at, the first task in the 
implementation deals with the tracks. The main reason is that without tracks, our 
locomotive can go nowhere, so that there cannot be any simulation. The second reason is 
that it may happen that we can construct the switches of the circuit with say $k$~states 
but that we need a $k$+1$^{\rm th}$ one to build the tracks. Subsection~\ref{stracks} is 
devoted to that problem. Once this problem
is temporarily solved, we shall see the main features of the crossings and of each kind
of switches in Sub-sections~\ref{roundabout}, \ref{fix}, \ref{sbsfrkdbl} 
and~\ref{forcontrol}.

\subsection{The tracks}
\label{stracks}

   As already mentioned, the tracks are implemented by following arcs of circles.
Such tracks are illustrated by Figure~\ref{ftracks}. As can be seen on the figure, we 
define a track by a first arc at distance~$k$ from a tile which plays the role of the
centre of the circle supporting the arc and there is a second arc at distance~$k$+1.
As shown also by the figure, on the left-hand side picture there are blue tiles and
mauve ones. On the right-hand side picture, the mauve tiles are in contact with orange 
ones. Those two colours indicate the direction of the motion performed by the
locomotive. The locomotive is a single-tiled element. The blue tiles support a track 
on which the motion is counter-clockwise while the motion is clockwise on a track 
supported by orange tiles.

\vskip 10pt
\vtop{
\ligne{\hfill
\includegraphics[scale=0.7]{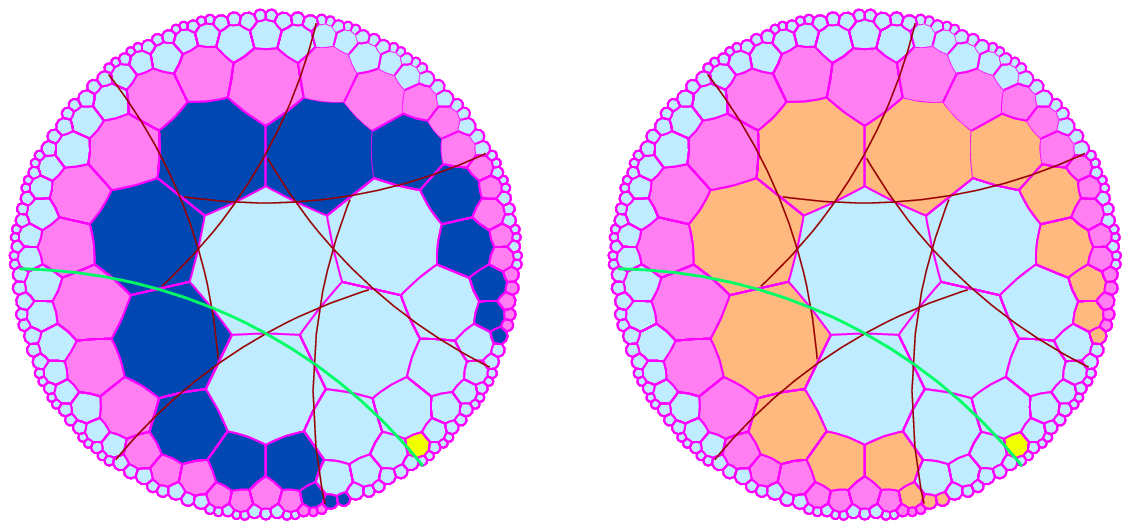}
\hfill}
\vspace{-5pt}
\begin{fig}\label{ftracks}
\leurre
Implementation of tracks in the heptagrid. On the left-hand side, for a counter-clockwise
motion along the arc. On the right-hand side, for a clockwise motion. The yellow tile
is the centre of the circles supporting the arcs. The blue and orange tiles are at 
distance~$4$ from the centre. The mauve tiles are at distance~$5$.
\end{fig}
}

\vskip 10pt
\vtop{
\ligne{\hfill
\includegraphics[scale=0.35]{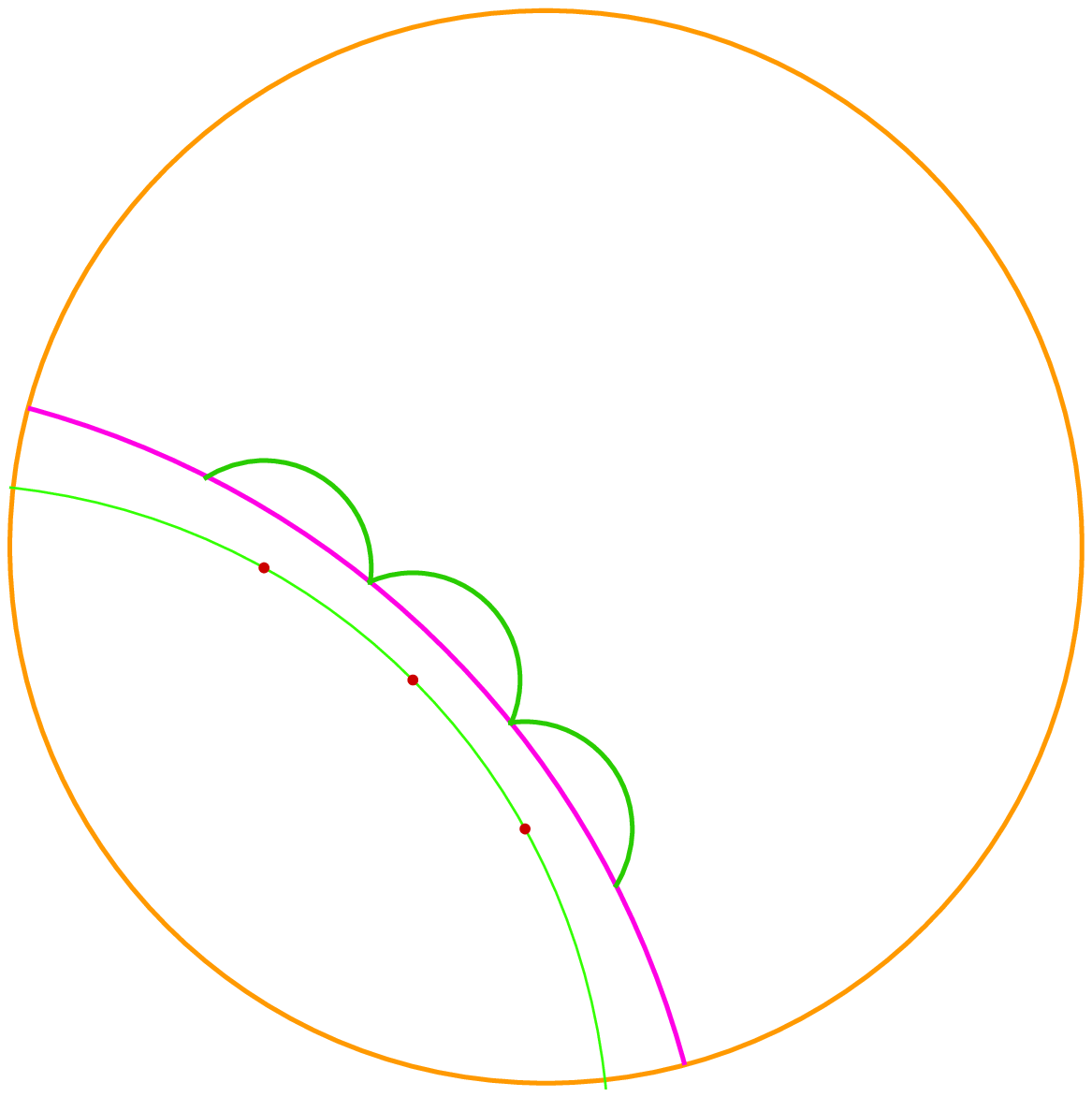}
\hskip 10pt
\includegraphics[scale=0.6]{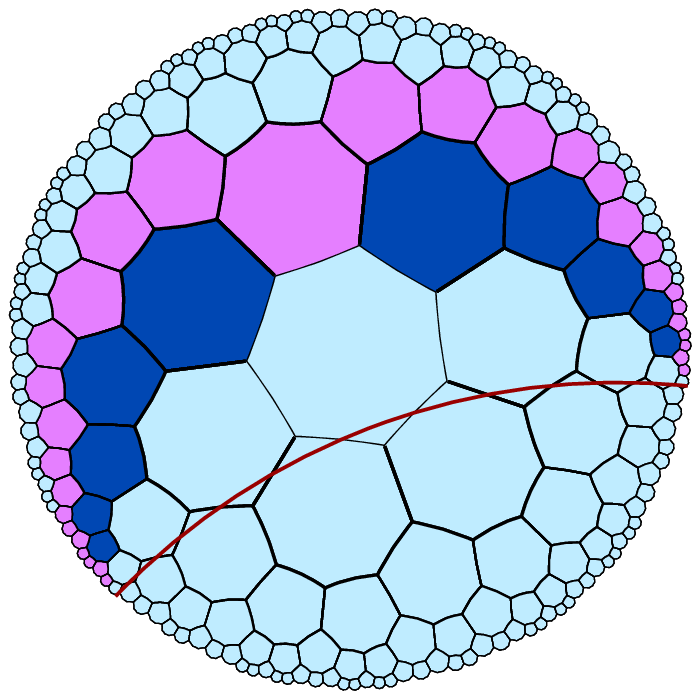}
\hfill}
\vspace{-5pt}
\begin{fig}\label{fzgzg}
\leurre
To left: the notion of a zig-zag line in green following a straight line, in purple.
To right: its implementation in the heptagrid.  
\end{fig}
}

   On the further figures of this section, we shall represent tracks by arcs of circles.
Blue arcs are counter-clockwise run, while red arcs are run in a clockwise motion.
This convention allows us to indicate the directions without using arrows.
As we shall have some discussion about delays, we need a which allow us to tune the length
of the path covered by the locomotive. Note that if the radius of an arc~$A_0$ is~$n$ 
and its angle is~$\alpha$, the length of the arc of the same angle and whose radius 
is~$n$+1 is more than twice the length of~$A_0$. The shortest path from a tile to another
one is not an arc of a circle, especially if the distance is big. An approximation of that
shortest path can be realized by a sequence of arcs which we call a {\bf zig-zag} line
illustrated by Figure~\ref{fzgzg}. The length of the shortest path following the line is
multiplied by~4 in the zig-zag line illustrated on the the right-hand side picture of
Figure~\ref{fzgzg}.

\section{The scenario of the simulation}
\label{scenar}

    With only three states at my disposal, I do not know how to directly 
implement switches where the needed sensors and controls are immediate neighbours of the 
cell where the tree tracks arriving at the switch meet. The simplest solution for that 
purpose is to devote a 
state for the tracks, different from the blank and then to give a series of states for 
each type of switch and for each step required for the crossing of the switch by the 
locomotive. This may require more states than the 22 states of the automaton 
in~\cite{fhmmTCS}.

   Here, we again take the idea of a special colour for the tracks, the mauve one.
Now, we take a single-celled locomotive, we shall say a {\bf simple} one. As we need
to use opposite directions, we have two different colours for the support of the track
as mentioned in the caption of Figure~\ref{ftracks}.
Note that the support is already needed for the motion of the locomotive in order to
observe the rotation invariance of the rules. Also, we indicated that there are 
instructions which decrement a register and others which increment it. We have a single 
track on the register in order to perform the instruction and a single one to allow
the locomotive to return to the instructions of the machine. The distinction is obtained
by defining two colours. The locomotive is {\bf green} when it has to increment a 
register and it is {\bf red} when it has to decrement it. The colours for the supports
of the tracks are blue and orange. Accordingly we shall speak of
a {\bf blue} and an {\bf orange path}.

Here too, the crossing is performed by a most expensive structure, the {\bf round-about}
described in Sub-section~\ref{roundabout}. Switches are decomposed into simpler
structures we shall indicate later. Although we follow the mainlines of~\cite{mmhepta3st},
we use here specific structures. All structures described in the present section
are under the {\bf idle configuration} which is their configuration at the starting time
of the computation.
 
\subsection{The round-about}
\label{roundabout}

    The round-about replaces the crossing, a railway structure, by a structure inspired by
road traffic. At a round-bout where two roads are crossing, if you want to keep the 
direction arriving at the round-about, you need to leave the round-about at the second 
road. Figure~\ref{froundabout} illustrates this features.

As mentioned in the legend of the figure, when the locomotive arrives through $B$, $A$,
it leaves the round-about through gate~2, 3 respectively. At $A$ and $B$, we have a 
special structure which we call the doubler which transforms the simple locomotive,
into a {\bf double} one which constitute of two contiguous cells of the same colour.
In the points 1, 2 and 3, the structure, which we call the {\bf selector}, sends
the locomotive further if it is a double one and sends it on the appropriate track if
it is a single one.

\vtop{
\ligne{\hfill
\includegraphics[scale=0.5]{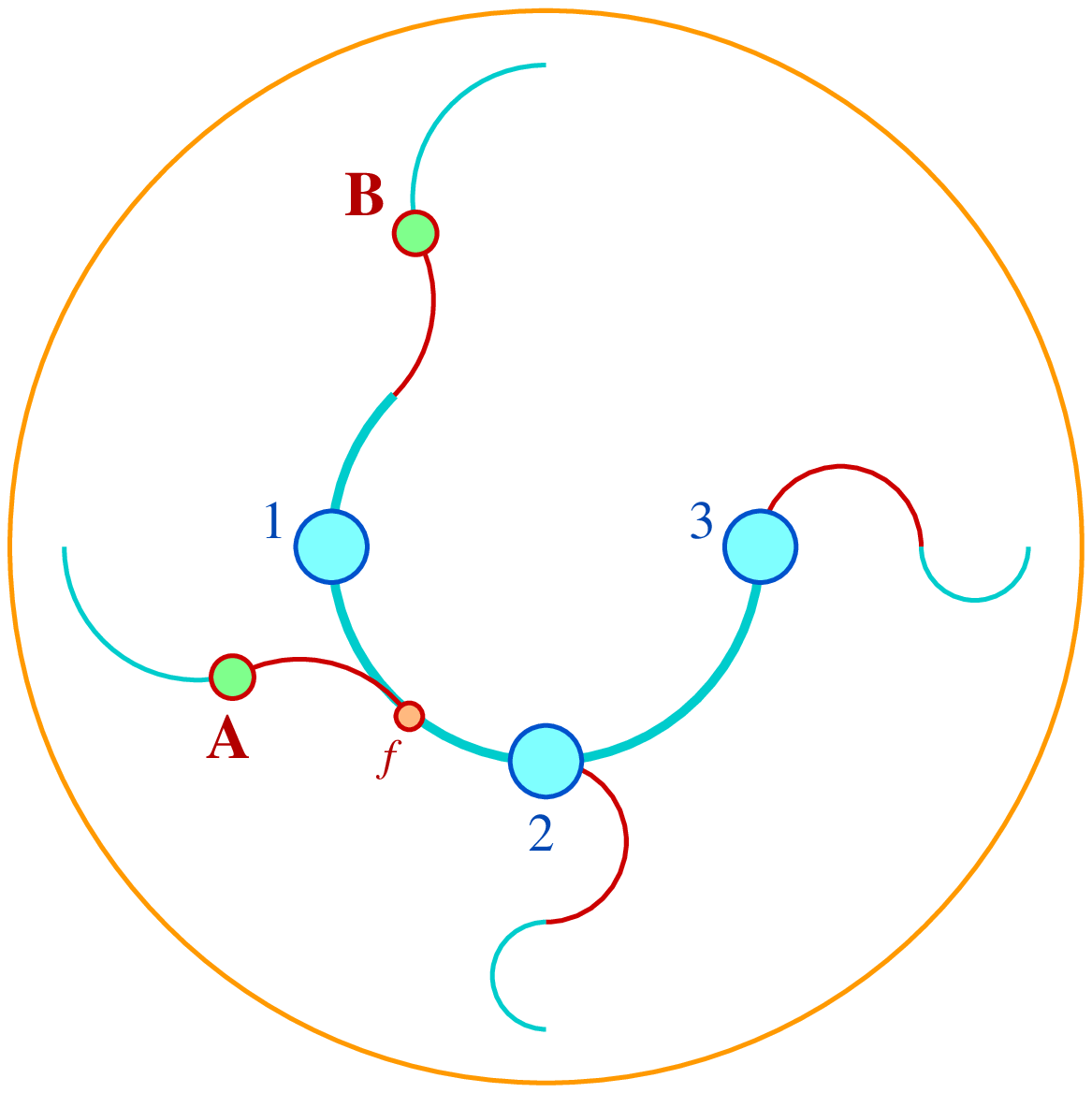}
\hfill}
\vspace{-5pt}
\begin{fig}\label{froundabout}
\leurre
Implementation scheme for the round-about. A locomotive arriving from $B$ leaves the
round-about through~$2$. If it arrives from~$A$, it leaves through~$3$. Note the fixed
switch $f$ in between the structures~$1$ and~$2$ allowing the track coming from~$A$ to
reach the round-about.
\end{fig}
}

The auxiliary structures we need are the fixed switch, see Sub-section~\ref{fix},
the fork, see Sub-section~\ref{fork}, the doubler, see Sub-section~\ref{double} and the 
selector, see Sub-section~\ref{select}.

\subsubsection{The fixed switch}
\label{fix}

    As the tracks are one-way and as an active fixed switch always sends the locomotive in
the same direction, there is no need of the other direction: there is no active fixed 
switch.
Now, passive fixed switch are still needed as just seen in the previous paragraph. 
Figure~\ref{stabfix} illustrates the passive fixed switch when there is no locomotive around. 
We can see that it consists of elements of the tracks which are simply assembled in the 
appropriate way in order to drive the locomotive to the bottom direction in the picture, 
whatever upper side the locomotive arrived at the switch. 
From our description of the working of the round-about, a passive fixed switch must be 
crossed by a double locomotive as well as a simple one. Also, it must be crossed by
such locomotives whatever their colour.

   Later, in Section~\ref{rules}, we shall check that the structure illustrated by
Figure~\ref{stabfix} allows these crossings.

\vtop{
\ligne{\hfill
\includegraphics[scale=0.8]{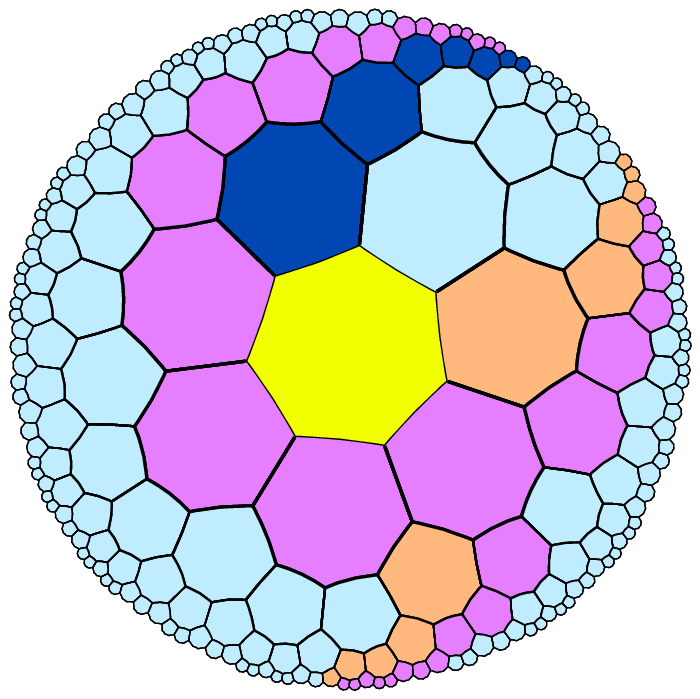}
\hfill}
\begin{fig}\label{stabfix}
The passive fixed switch in the heptagrid.
\end{fig}
}

\subsubsection{The fork} 
\label{fork}

   The fork is a structure which allows us to get two simple locomotives from a simple 
one. It also allows us to get two double locomotives from a double one. The structure
of the structure is illustrated by Figure~\ref{stabfrk}. As shown there, two arcs, 
one blue, the other red, abut a same tile to which a third one arrives. The arriving 
locomotive is duplicated on the orange and the blue paths

\vtop{
\ligne{\hfill
\includegraphics[scale=0.8]{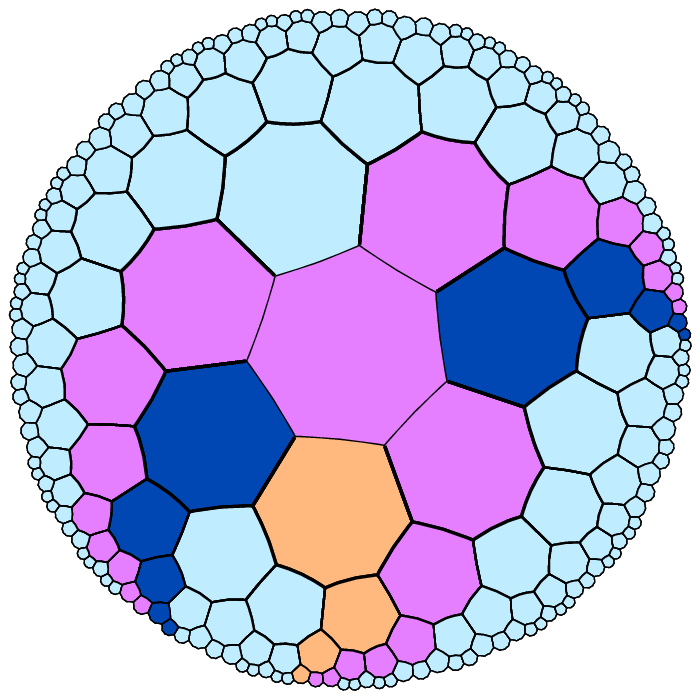}
\hfill}
\begin{fig}\label{stabfrk}
The fork in the heptagrid.
\end{fig}
}

    The fork is used later in several structures. We shall see it, namely, in the
doubler to which we turn now in Sub-subsection~\ref{double}. We can see on 
Figure~\ref{stabfrk} that the exiting paths, the red one and the blue one, have supporting
tiles which belong to the same circle.

\subsubsection{The doubler} 
\label{double}

The doubler  is illustrated by Figure~\ref{stab_dbl}. 

The left-hand side picture of the figure explains the structure of the doubler.
The picture shows us a circle whose radius is~3. The supported track lies on a circle
of radius~4.

\vskip 10pt
\vtop{
\ligne{\hfill
\includegraphics[scale=1]{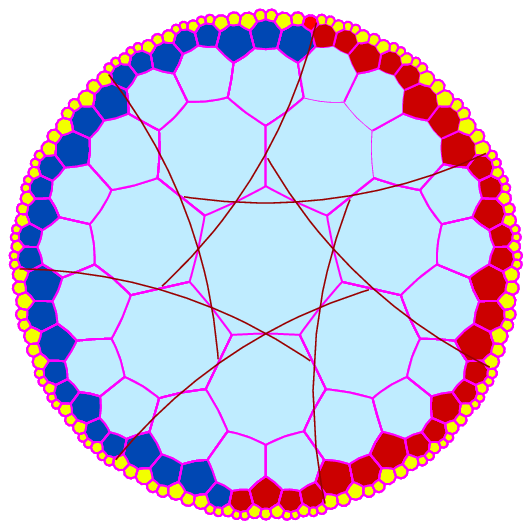}
\raise -5pt\hbox{\includegraphics[scale=0.74]{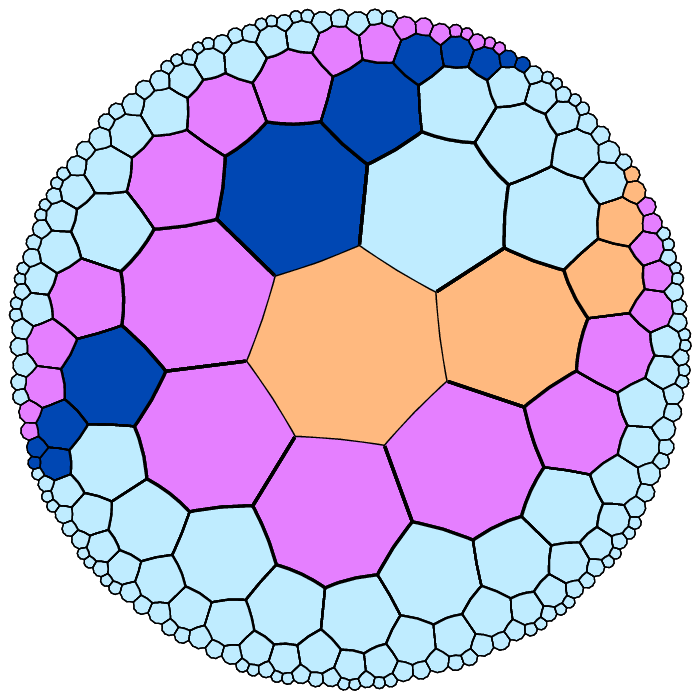}}
\hfill}
\begin{fig}\label{stab_dbl}
\leurre
To left: explanation of the implementation. To right: the doubler.
\end{fig}
}

On a circle of radius~4, there are
7$\times$21 = 147 tiles. Indeed, it was proved that within a sector, from one border 
line to the other, the number of tiles at the distance~$k$ is \hbox{$f_{2k-1}$} where 
$f_n$ is the Fibonacci sequence defined by the initial condition \hbox{$f_0 = f_1 = 1$}, 
see~\cite{mmbook1,mmJUCS}. Accordingly, an arc ending on the borders of a sector with 
radius~4 contains 21 tiles exactly. It can be checked in~\cite{mmbook1}. As the number 
of tiles is odd and as the duplication occurs at a tile~$T$ of the circle, we remain with
an even number of tiles so that each path on each side contains the same number of tiles, 
namely 73 of them. That same number of tiles means that both simple locomotives created 
at the same time~$t$ at both neighbours of~$T$ on the circle arrive at neighbouring 
tiles~$U$ and~$V$ so that, starting from time~$t$, they constitute a double locomotive.
The exit tracks allows that double locomotive to go further on its track.

That argument explains why the radius is important. The number of tiles on the circle
which supports the tracks should be odd and the smallest case avoiding complications is 
radius~4. This is why the left-hand side picture of Figure~\ref{stab_dbl} explains
the right-hand side picture.

\subsubsection{The selector}
\label{select}

In Figure~\ref{scheme_sel}, we have an illustration of the structures involved in
the selector. There are three structures: a fork at {\bf Fk}, a first controlling device
at $Sl$ and a second controlling device at $Dl$.

On {\bf Fk}, the fork which replicates the arriving
locomotive on both paths exiting from that point. If the locomotive is simple,
its copy on the blue path goes out through~$Sl$ which let it got. However, on the
orange path, the other copy does not cross $Dl$ which stops a simple locomotive. If the 
locomotive is double, its copy on the orange path goes out through~$Dl$ which let it go.
The other copy on the blue path does not cross $Sl$ which stops a double locomotive.

\vskip 10pt
\vtop{
\ligne{\hfill
\includegraphics[scale=0.5]{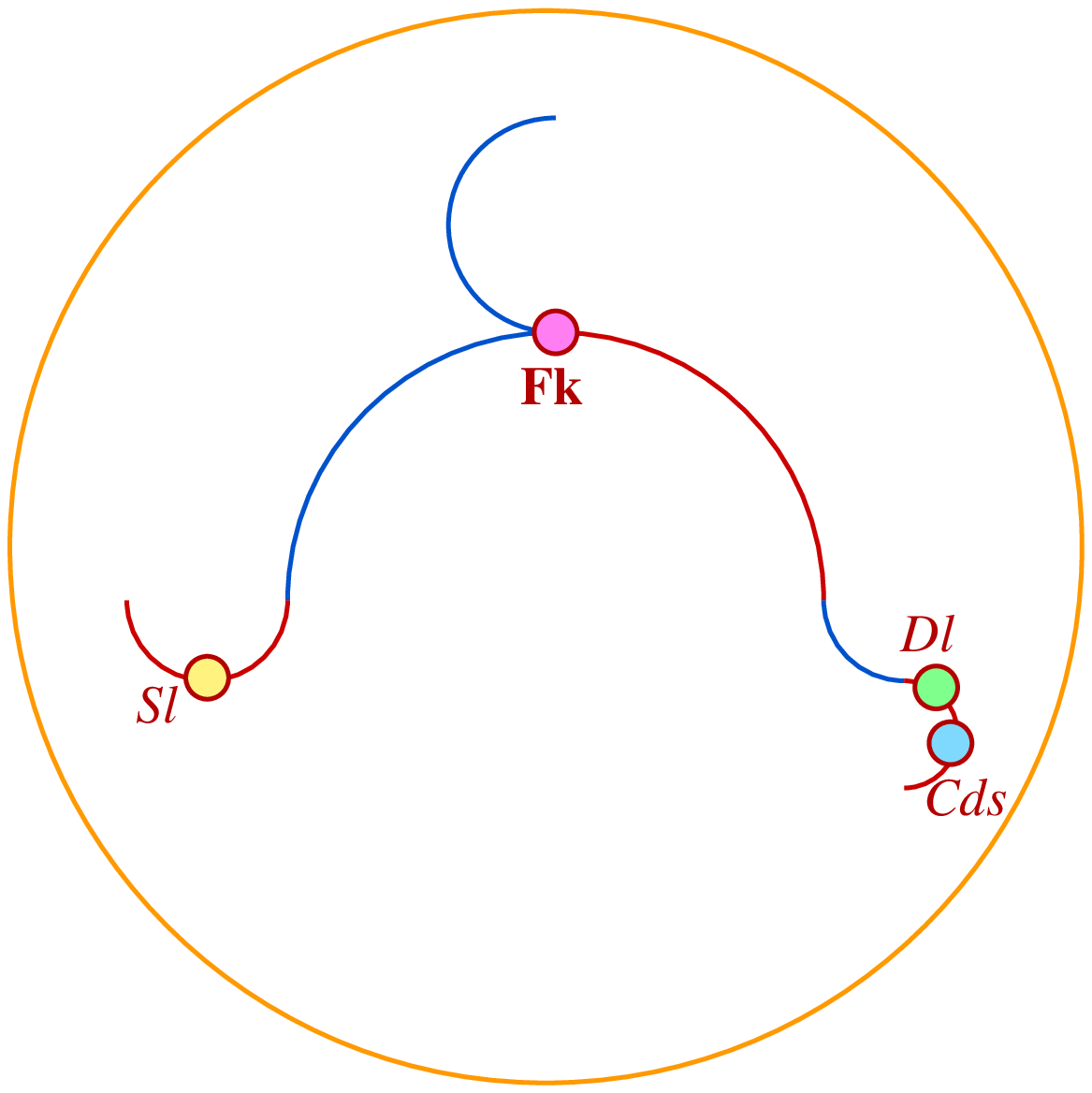}
\hfill}
\begin{fig}\label{scheme_sel}
\leurre
Scheme of the selector. Above, the fork. To left and to right the controlling devices.  
\end{fig}
}

On Figure~\ref{selctrl}, we indicate the implementation of the control structures of the
selector. On the left-hand side of the figure, we have the controlling device: it let
pass the appropriate locomotive. We note that both controllers have a similar structure
and that the converter is quite different.

According to what we said in the description of a round-about, the red path after
$Sl$ leads to the tracks which continue the track through which the locomotive arrived
at the round-about. On the other side of the scheme, the blue path after $Dl$ leads
to the tracks of the round-about which goes to the next selector. However, after $Dl$,
we can see another structure on the tracks denoted by $Cds$. It transforms the
double locomotive into a simple one, so that at the next selector, the simple locomotive
of the right colour is sent on the appropriate tracks. Note that the controller
structures must be placed on a red path.

\vskip 10pt
\vtop{
\ligne{\hfill
\includegraphics[scale=0.5]{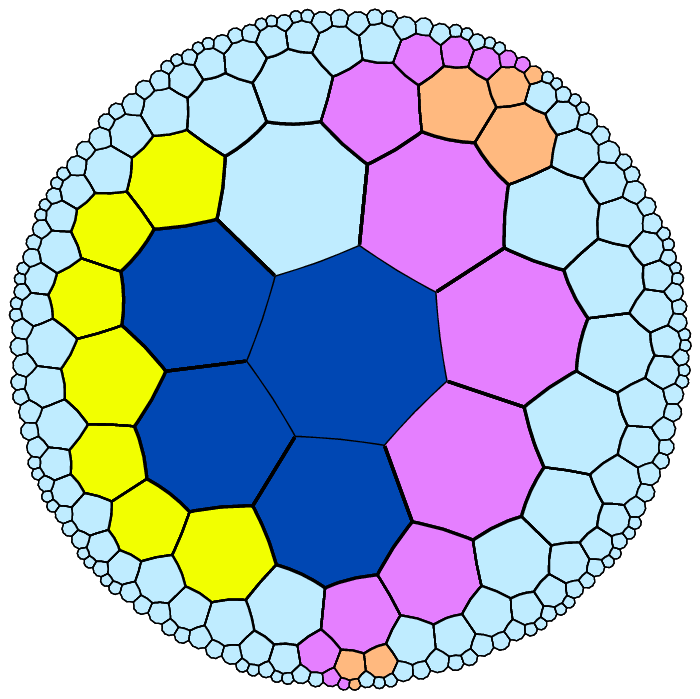}
\includegraphics[scale=0.5]{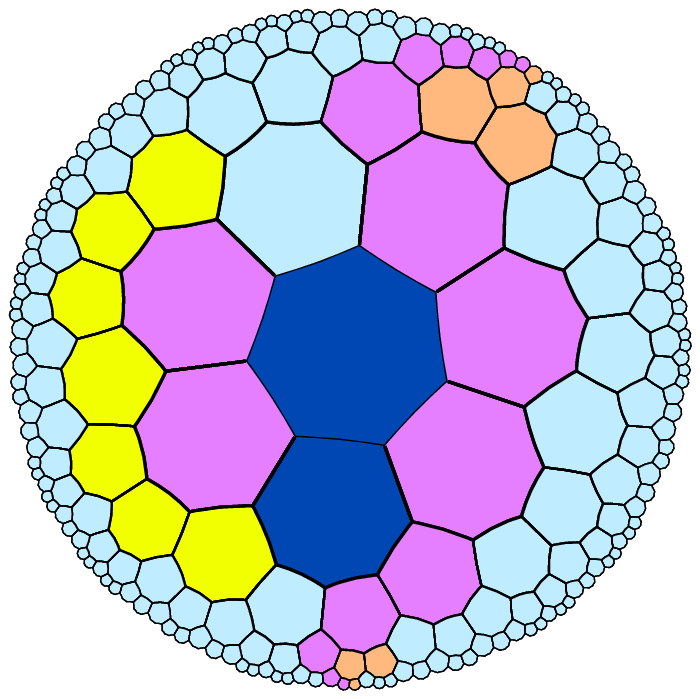}
\includegraphics[scale=0.5]{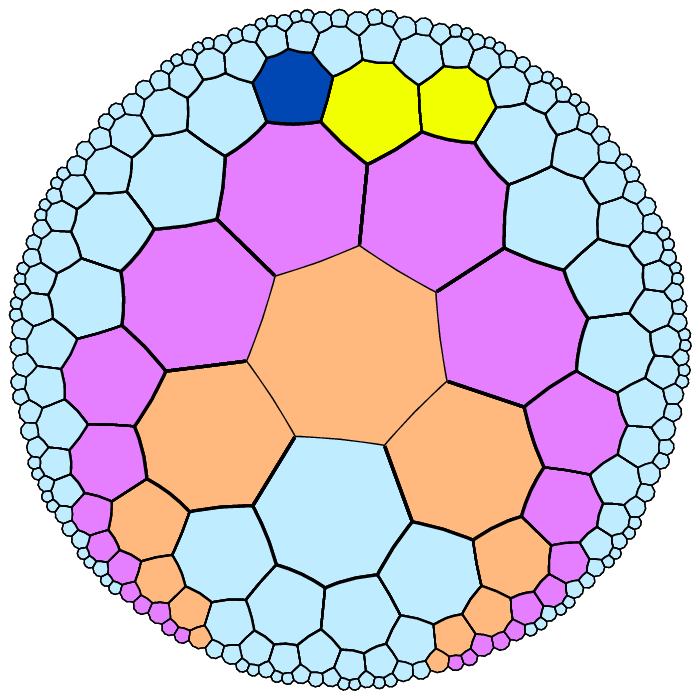}
\hfill}
\begin{fig}\label{selctrl}
\leurre
The control structures of the selector: to left, the control for letting pass a simple
locomotive, stopping a double one. In the middle: the control let a double locomotive
pass and stops a simple one. To right, the converter from a double locomotive to a 
simple one.
\end{fig}
}

\subsection{Flip-flop and memory switch: the use of controlling structures}
\label{forcontrol}

    In this Sub-section, we look at the decomposition of two active switches: the flip-flop 
and the active part of the memory switch. In both cases, we can split the working of the
switch by separating at two different stages, the bifurcation and the fact that the passage
in one direction is forbidden. It is not mandatory that both acts should happen 
simultaneously, as might be suggested by the definition of Sub-section~\ref{railway}. A fork
may dispatch two simple locomotives and, later, on one track, a controller let the locomotive
go on its way on the track and, on the other track, another controller stops the locomotive
and destroys it. The difference between the switches is the way in which the change of
selection is performed. Figure~\ref{activswitches} illustrates how to assemble forks
and controllers in order to obtain either a flip-flop, left-hand side of the figure, or
an active memory switch, right-hand side of the picture. The difference is important: the 
flip-flop immediately changes the selection once it was crossed by the locomotive. The active
memory switch changes the selection if and only if the passive switch ordered it to to do
so through a signal sent to the active switch. This is realized in 
Figure~\ref{activswitches}.

Indeed, in the active memory switch, when the simple locomotive arrives at the fork~$C$, 
it is duplicated into two simple locomotives, each one following its own path. One 
of these locomotives goes on its way to the controller, and the other is now a third 
locomotive which is sent to a second fork, labelled with~$A$ in the figure. When it is 
crossed by that locomotive, $A$~sends two locomotives~: one to the other controller and 
one to a third fork~$S$. 
The controllers are blue and red. The blue one let the locomotive
go its way. The red controller stops the locomotive and kills it. 
The locomotives sent by~$S$ go both to a controller, one to the blue controller, the other
to the mauve one. Now, when a locomotive sent by~$S$ arrives at a controller, it changes 
the blue one into a mauve one and the mauve one into a blue one, so that what should be 
performed by a flip-flop is indeed performed. It is enough to manage things in such a 
way that the locomotives arriving to the controllers from~$S$ arrive later than those 
sent by~$C$ and by~$A$.
\vskip 10pt
\vtop{
\ligne{\hfill
\includegraphics[scale=0.35]{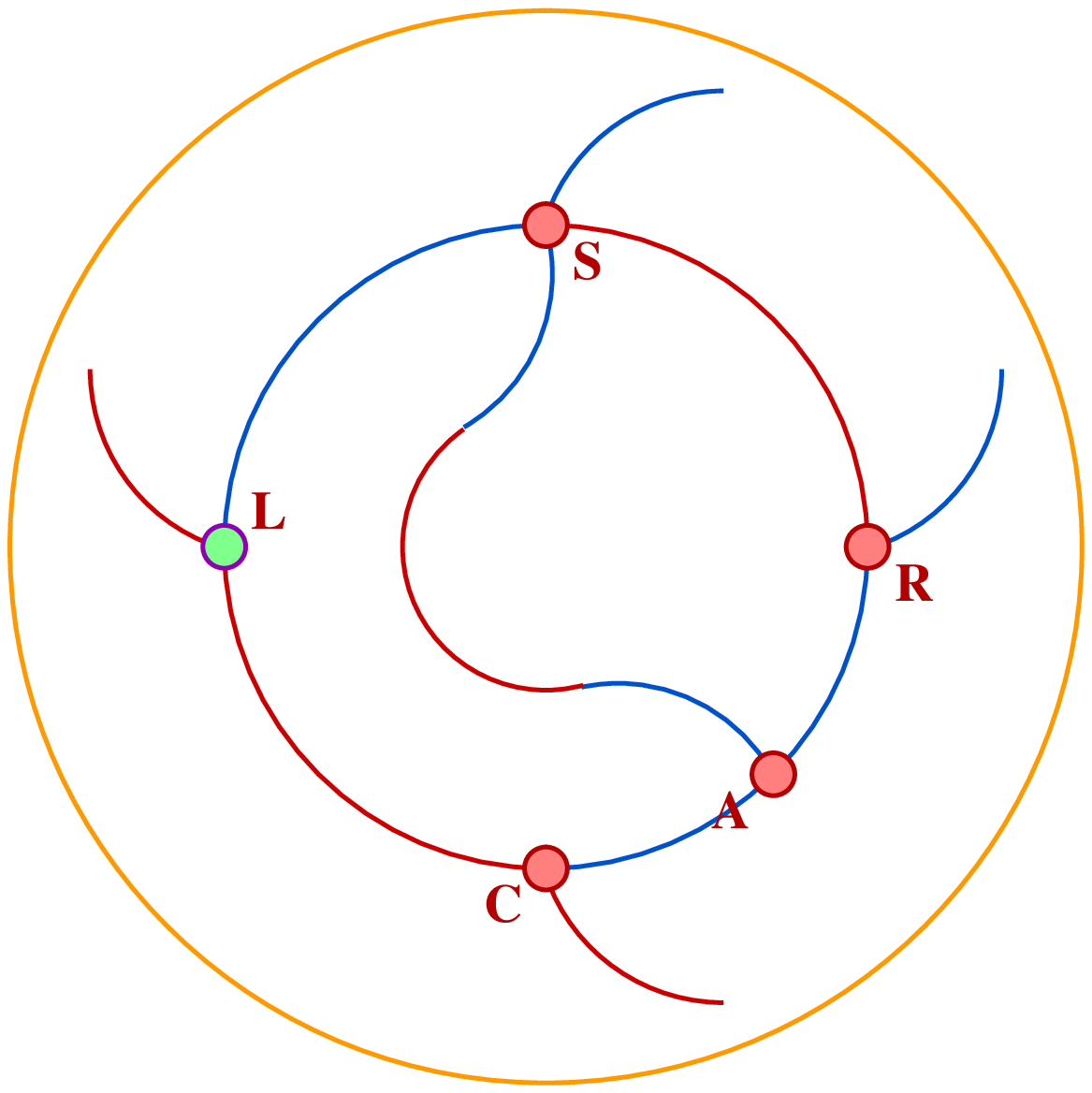}
\includegraphics[scale=0.35]{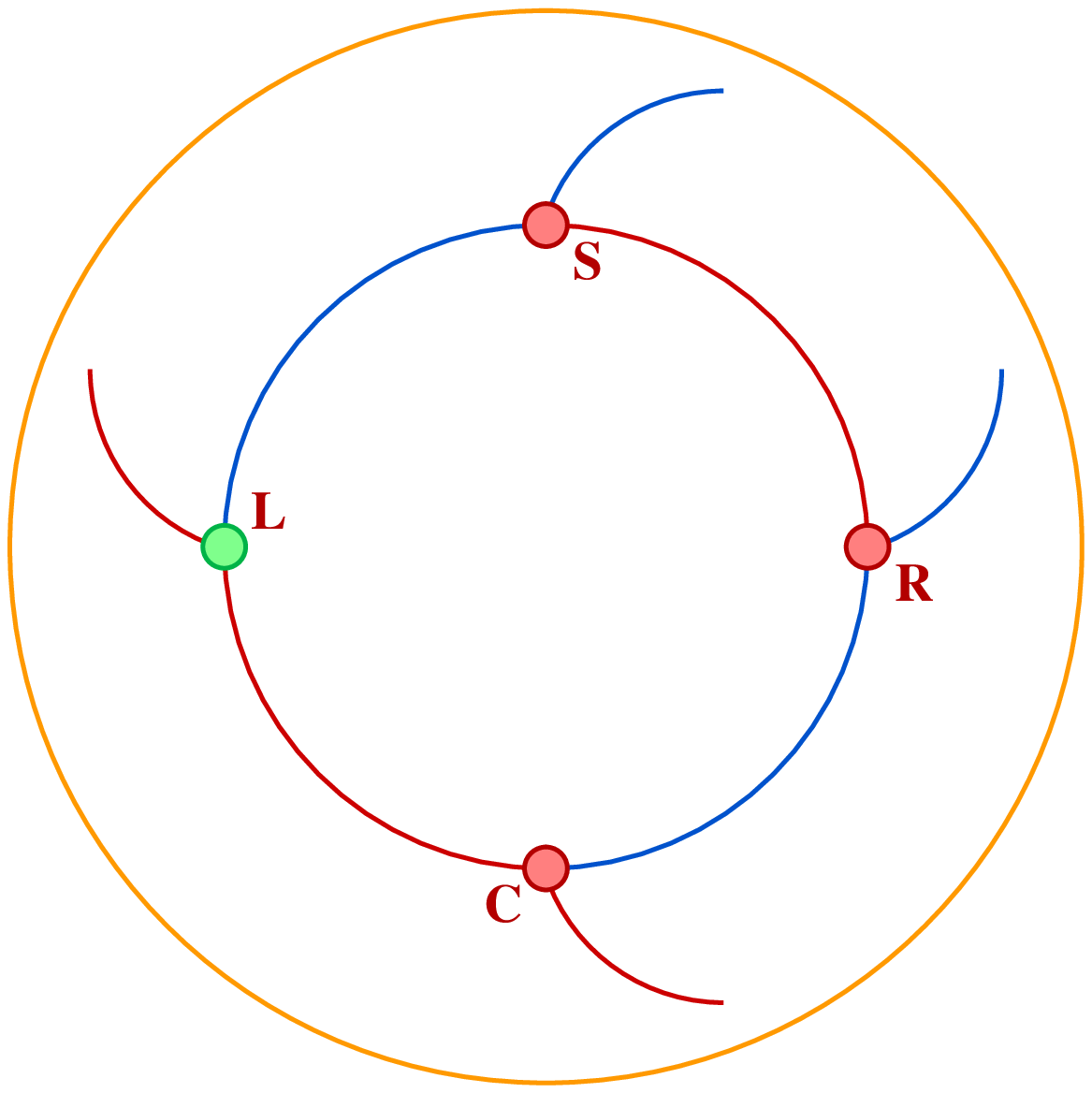}
\hfill}
\begin{fig}\label{activswitches}
\leurre
To left: the flip-flop. To right: the active memory switch.
\end{fig}
}

Figure~\ref{memoswitches} illustrates the working of the memory switch. As 
in~\cite{mmpaper2st}, the memory switch is split into two parts : the active one for
an active crossing and a passive one for the passive crossing of the switch. Consider an
active crossing. As shown on the left-hand side part of Figure~\ref{memoswitches},
the locomotive arriving to~$C$ is duplicated by a fork and each copy of the locomotive
is sent to a controller: one of them is $L$, the other is~$R$. One of them let the
locomotive go, whatever its colour, the other controller stops the locomotive by 
killing it. 

At present, consider a passive crossing, illustrated by the right-hand side
picture of Figure~\ref{memoswitches}. The locomotive arrives either to~$P$ or to~$Q$.
Assume that it arrives to~$Q$. A fork at~$Q$ sends a copy of the locomotive to~$F$
which is a fixed switch. The fork sends the other copy to~$R$, a controller. If
the corresponding track on the active switch can be crossed, the controller stops
that copy of the locomotive. If the corresponding track is barred on the active switch,
the controller let the copy of the locomotive go. The copy goes then to~$S$ and then
to~$U$. There, a fork sends a copy of the locomotive to from which 
the~$S$ of the active switch which is also a fork.
The other locomotive sent from~$U$ goes to~$T$ where another fork sends to copies
to~$L$ and~$R$ in order to change the function of those controllers which will thus be
in accordance with those of the active switch which will be changed accordingly when
the locomotive sent from~$U$ will reach them.
The fork at~$P$ and~$L$ play symmetrical roles when the 
locomotive arrives to the passive switch from the other side. Of course, the controllers
at~$L$ and~$R$ play opposite roles at any given time.

We can see on Figure~\ref{stabmemoctrl} that an auxiliary track abut the structure
without crossing it. It is the end of the paths taken by the locomotive-signal sent 
from~$U$ both in the passive and in the active switches. Those signals
change the controller. If the signal arrives at a blue controller, it becomes a mauve
one, if the signal arrives at a mauve controller, it becomes blue. The blue controller
let the locomotive go while the mauve one stops it. The controllers at
the active switch are the same structures as those indicate in Figure~\ref{stabmemoctrl}.
Simply, they are fixed on opposite functions: on the active switch, the blue controller
is on the track to which an arriving locomotive at~$C$ can go, see the left-hand side 
picture of Figure~\ref{memoswitches}, and the mauve controller is on the track whose 
access is forbidden. On the passive switch, the mauve, blue controller is set on the 
track which corresponds to the track of the active switch on which the blue, mauve 
controller lies respectively.
\vskip 10pt
\vskip 10pt
\vtop{
\ligne{\hfill
\includegraphics[scale=0.35]{disque_actmemo.ps}
\includegraphics[scale=0.35]{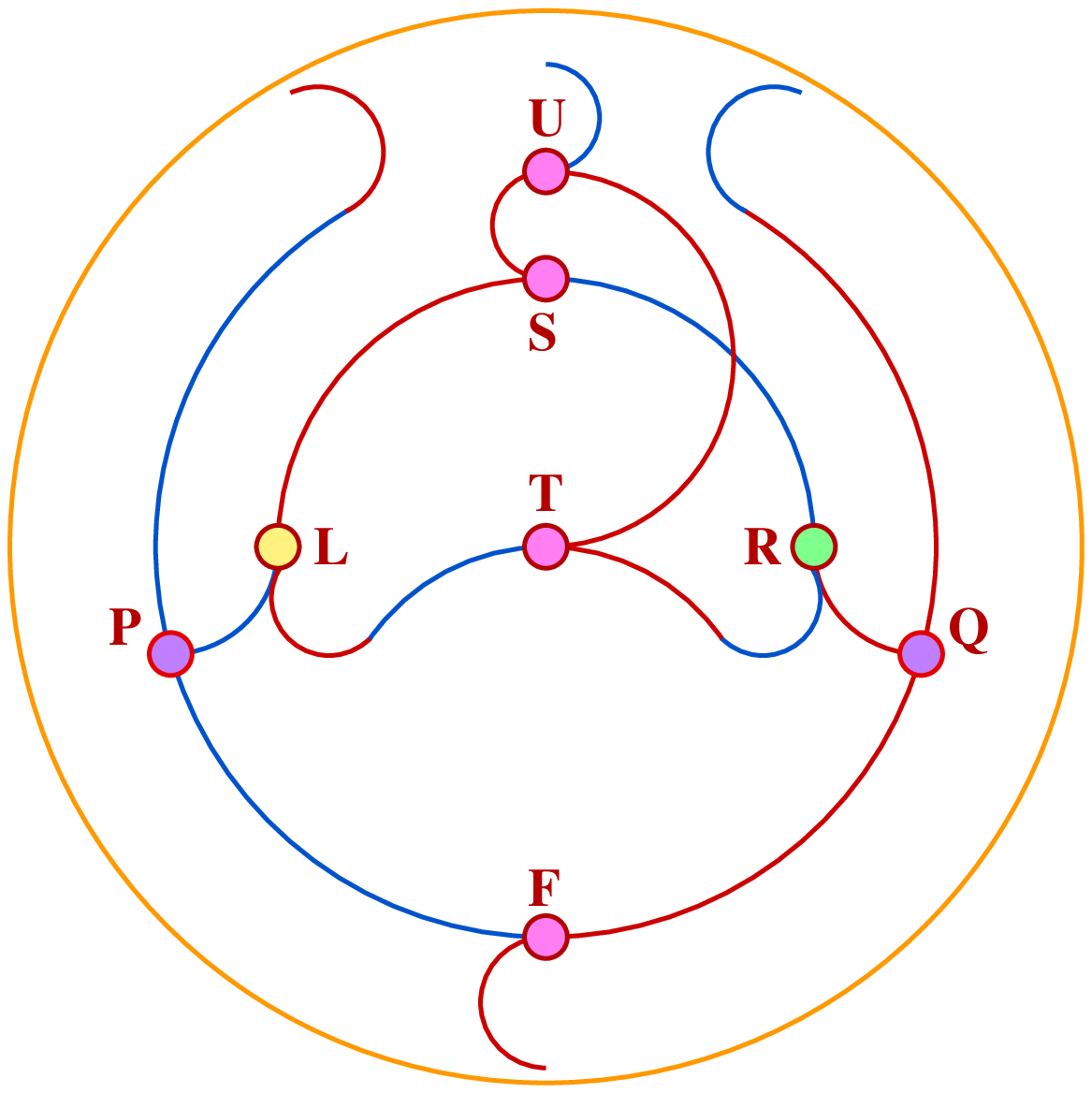}
\hfill}
\begin{fig}\label{memoswitches}
\leurre
Memory switch: to left, the active switch; to right the passive one.
\end{fig}
}

\vtop{
\ligne{\hfill
\includegraphics[scale=0.6]{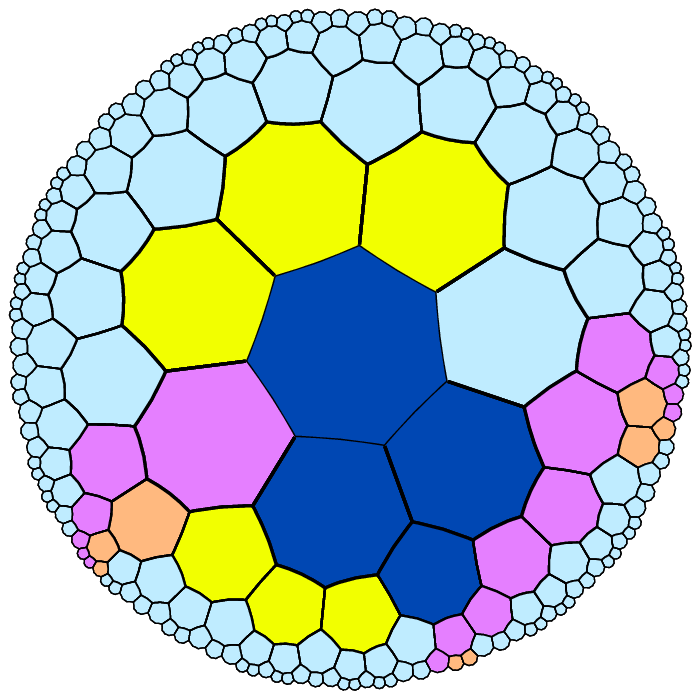}
\hskip 10pt
\includegraphics[scale=0.6]{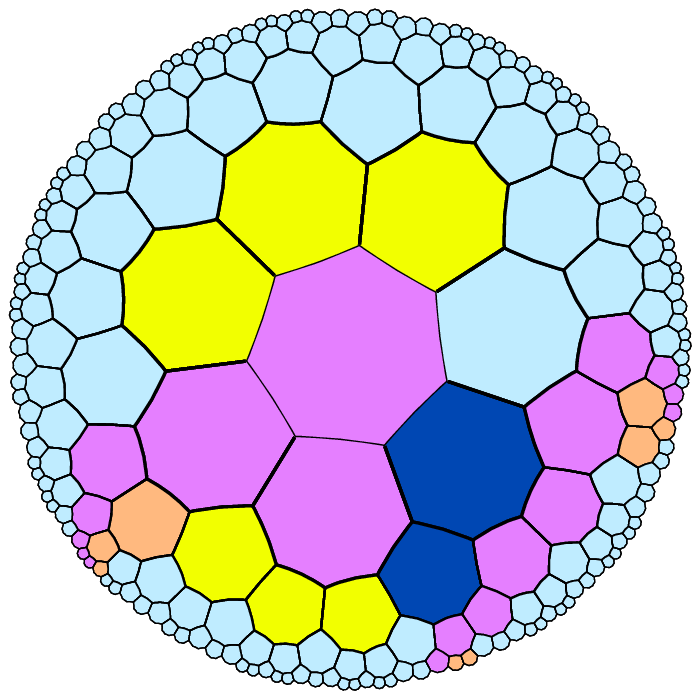}
\hfill}
\begin{fig}\label{stabmemoctrl}
\leurre
To left: the controller which let the locomotive go as a signal to the active switch. 
To right, the controller which stops the locomotive when there is no change to perform.
\end{fig}
}

Note that the controlling structures of Figure~\ref{stabmemoctrl} have a common point
with those of Figure~\ref{selctrl}. The structures of Figure~\ref{stabmemoctrl}
are programmable versions of those of Figure~\ref{selctrl}.

\subsection{Register and instructions}

Thanks to the tracks and the connected structures, switches and control devices,
the locomotive can go from the instructions to the register and then, from there
back to the instructions.

    The structures we considered up to now allow us to revisit the schemes we indicated
in Section~\ref{scenario}. We start with the one-bit memory which is the content of 
Figure~\ref{basicelem}. In Figure~\ref{fhypunit}, we remind that latter scheme and we
present how we implement it with the structures defined up to now.
\vskip 5pt
\vtop{
\ligne{\hfill
\raise 20pt\hbox{\includegraphics[scale=0.7]{elem_gb.ps}}
\hskip-20pt
\includegraphics[scale=0.475]{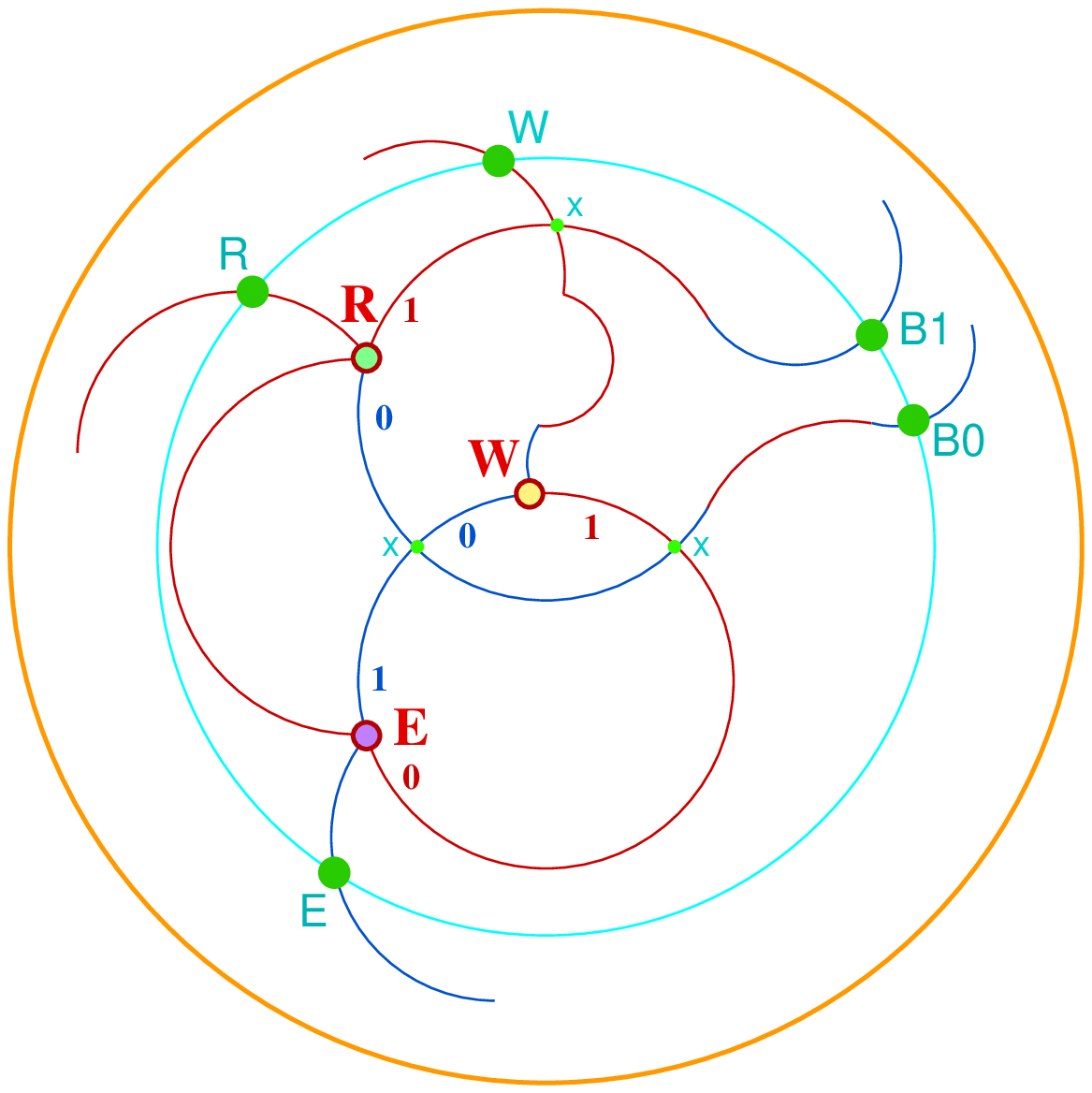}
\hfill}
\begin{fig}\label{fhypunit}
\leurre
To left, the scheme of Figure~{\rm{\ref{basicelem}}}, to right, its implementation
in our setting. In the hyperbolic picture, the colours at~$W$, $R$ and~$E$, namely
yellow, light green and light purple respectively represent a flip-flop switch, an
active memory switch and a passive memory switch. The green points with an {\tt x}
letter indicate crossings.
\end{fig}
}
\vskip 5pt
In both figures, the content of the memory is defined by the position of the
switches at~$W$, at~$R$ and at~$E$. There are five gates in both structures: $W$, $R$,
$E$, $0$ and $1$ in the right-hand side picture of the figure. The main difference with
the left-hand side picture is that  in that latter case, some portions of the tracks are
two-ways: the tracks between $R$ and the fixed switch in between $R$ and $E_1$ on one hand
and in between $R$ and $E_2$ on the other hand can be crossed in both directions by the
locomotive. In the setting of the present paper, tracks are one-way. The direction is 
indicated by the colour of the support. In our schemes, red tracks are clockwise run
while blue ones are counter-clockwise run. In both pictures, there are two possible 
entries: $R$ and $W$. Entering through~$R$ does not change the structure. In the left-hand
side scheme, a memory switch stands at~$R$. In the right-hand side picture, at~$R$,
we have an active memory switch. If the position of the switch at~$R$ is~0, the exit is
through gate~0, $E_0$ in the right-, left-hand side picture respectively. If the position
is~1, the exit is through gate~1, $E_1$ respectively. If the locomotive enters 
the structure through gate~$W$, as a flip-flop switch is placed at that point, the 
position
of the switch is changed by the passage of the locomotive. Then, the locomotive
exits through gate~$E$ after it has passively crossed the switch at~$E$, $R$ in
the right-, left-hand side picture respectively. This also changes the selected track in
that switch. Accordingly, if the locomotive enters the structure through~$W$, the 
positions of the switches are changed to the opposite position. It is also the case
in the right-hand side picture for the switch at~$R$ as its selected track is defined
by the track through which the locomotive passed at~$E$. On the right-hand side picture
of the figure, we can see a track going from~$E$ to~$R$ which represent the structures
we have seen for the memory switch in Sub section~\ref{forcontrol} allowing the passive
switch to change the position at the active switch when it is required.

In later Figures, we represent the one-bit memory unit as a circle with five gates,
$W$, $R$, $E$, $0$ and~$1$, two gates for entrance, $R$ and~$W$, and three exits,
$E$, $0$ and~$1$.

\vtop{
\ligne{\hfill
\includegraphics[scale=0.5]{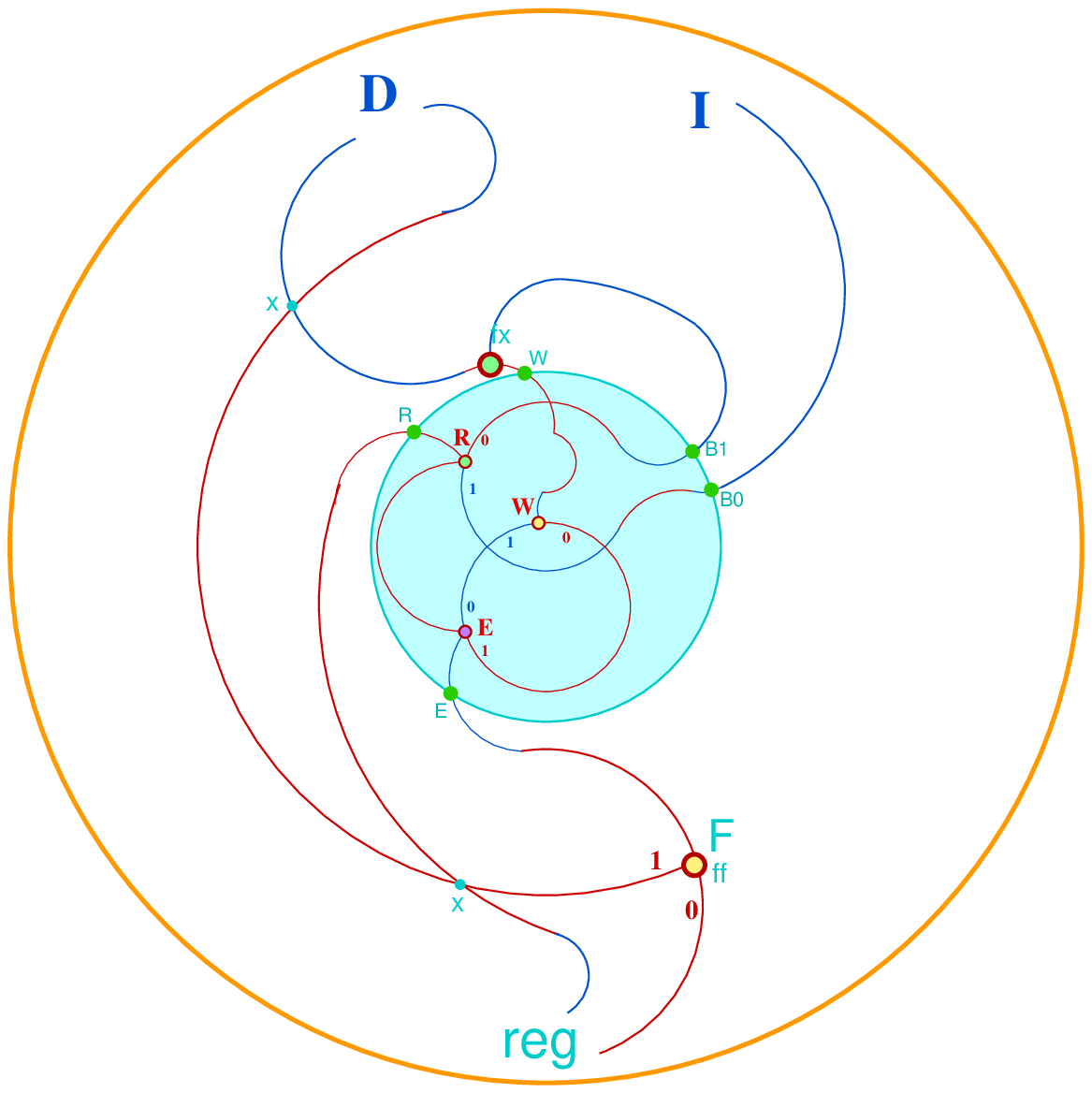}
\hfill}
\begin{fig}\label{fdiscrunit}
\leurre
A one-bit memory to remember whether the last instruction was to decrement
or to increment the register which was just operated. 
The conventions for colours are those of Figure~{\rm\ref{fhypunit}}.
\end{fig}
}

Figure~\ref{fdiscrunit} illustrates a structure belonging to a register which allows the
locomotive to return to another memory structure, that for the incrementing instructions
and that for the decrementing ones. The structure is based on a one-bit memory: 0 if
the instruction incremented the register, 1 if it decremented it. The idle configuration
is~0. When the locomotive goes to increment a register, it does not visit the structure
which remains~0. When the locomotive goes to decrement a register, it enters the structure
through its $W$-gate so that when it leaves it through the $E$-gate, it goes to the 
register. When it goes back from the register, it enters the structure through its
$R$-gate: if it reads~0, it is sent to the memory structure for incrementing instructions
which will send back the locomotive to the right instruction. If the locomotive reads~1
it leaves the structure through its 1-gate and it is sent to the $W$-gate of the 
structure, resetting it to~0. Now, the locomotive is not sent again to the register.
On its way to decrement the register after crossing the structure, the locomotive
crossed a flip-flop {\tt ff} whose 0-branch sent it to the register. And so, when it
again meets the flip-flop which is now indicating its 1-branch, the locomotive is sent
to the memory structure of the decremeting instructions and the flip-flop again indicates
its 0-branch which is its idle configuration.
\vskip 10pt
\vtop{
\ligne{\hfill
\includegraphics[scale=0.7]{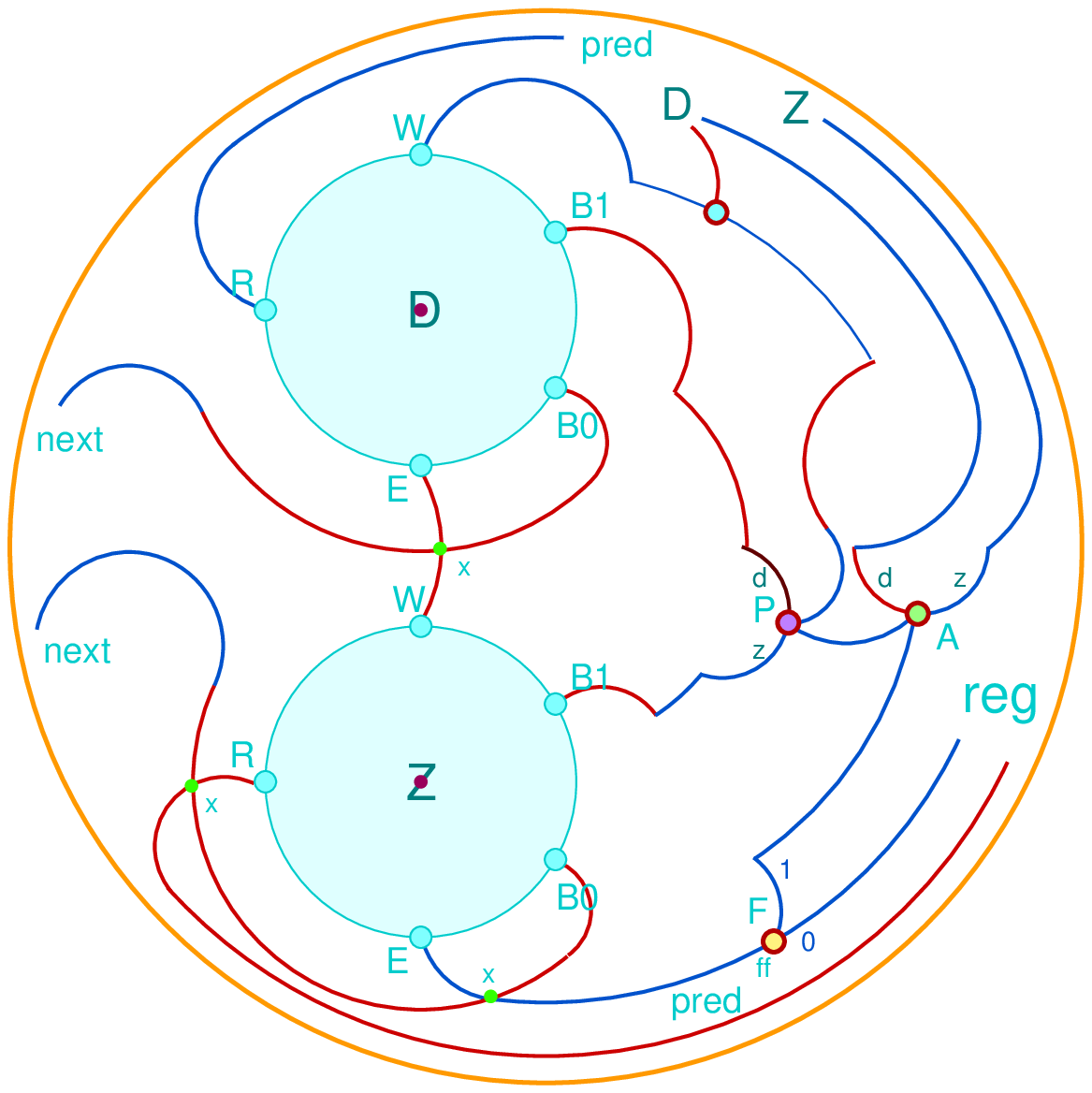}
\hfill}
\begin{fig}\label{fudisphyp}
\leurre
The unit in the memory structure of a register for the return from a decrementing
operation to the right place in the instructions.
The conventions for colours are those of Figure~{\rm\ref{fhypunit}} and of
Figure~{\rm\ref{fdiscrunit}}.
\end{fig}
}

\def\ttD{\hbox{\tt D}}
\def\ttZ{\hbox{\tt Z}}
Figure~\ref{fudisphyp} illustrates a unit of the memory structure which ensures a 
returning locomotive which decremented a register to go back to the appropriate place 
among the instructions. Such a memory structure contains as many units as there are
instructions in the program which decrement that register. As we can see, the unit 
contains two one-bit memories denoted by~\ttD{} and~\ttZ{} in the program, so that
we shall denote their $W$-, $R$, $E$, 0- and 1-gates by $W_{\ttD}$, $W_{\ttZ}$, 
$R_{\ttD}$, $R_{\ttZ}$, $E_{\ttD}$, $E_{\ttZ}$, $B0_{\ttD}$, $B0_{\ttZ}$, and 
$B1_{\ttD}$, $B1_{\ttZ}$ respectively. The tracks which connect $E_{\ttD}$ to $W_{\ttZ}$ 
shows us
that both~\ttD{} and~\ttZ{} are simultaneously either~0 or~1. The reason why we need two 
connected one-bit memory is that when the locomotive reaches the register in order to
decrement it, it may happen that the operation cannot be performed because the register
is empty. In that case at which we look a bit further, the locomotive is sent to the 
memory structure on track~$D$ if the register was actually decremented  and on another
track, say~$Z$, if the register was empty so that the locomotive could not decrement it.
The distinction is important as the next instruction depends on which the case was.
When it goes to decrement a register, the locomotive first goes to the memory structure.
It enters the \ttD{} memory through $W_{\ttD}$. Note that the locomotive enters into one
unit only so that all other units remain~0. As it enters the unit through~$W_{\ttD}$ we
know that both~\ttD{} and~\ttZ{} become~1. Next, from~$E_{\ttZ}$, the locomotive goes
to~$F$ at which a flip-flop switch stands which indicates its 0-branch. That branch is
a track leading to the register.

   When the locomotive succeeded to decrement the register, it goes back to the memory
along the $D$-track. The track enters a unit through $R_{\ttD}$. If the locomotive
reads~0 it is sent through $B0_{\ttD}$ to the $R_{\ttD}$ of the next unit. So that it
goes from one unit to the next one until it reads~1{} in the {\ttD} memory of a unit.
A similar motion happens for the locomotive if it arrives to the memory through the
$Z$-track: it enters the unit through $R_{\ttZ}$ and, it it reads~0, it is sent
through $B0_{\ttZ}$ to the $R_{\ttZ}$ of the next unit.

    Consider the case when the locomotive arrives to a unit containing~1. First, let us 
look the case when the locomotive arrives to $R_{\ttD}$. As it read~1, it goes through
$B1_{\ttD}$ which sends it to~$P$ at which a passive memory switch stands. It arrives to
the switch through a track named {\tt d}. Accordingly, another locomotive is sent
through~$P$ which reaches~$A$ at which an active memory switch stands. The locomotive
make the active switch indicate its branch also named {\tt d}. From~$P$ the former
locomotive is sent to $W_{\ttD}$ so that it will set \ttD{} and \ttZ{} to~0.
Now, from~$E_{\ttZ}$ the locomotive passes a new time through~$F$ so, that it takes
the 1-branch of the flip-flop switch which stands there and makes the flip-flop 
to indicate its 0-branch again. On the 1-branch, the locomotive goes from~$F$ to~$A$
where the active memory switch indicates the {\tt d}-track, leading the locomotive
to the right place in the program.

   Now, consider the case when the locomotive reads~1{} in the unit after it arrived
through $R_{\ttZ}$. Then it goes out through $B1_{\ttZ}$ which sends it to~$P$
at which it arrives through the track named {\tt z}. The memory switch at~$P$ sends 
another locomotive to~$A$ as seen previously but this time the flip-flop switch at~$A$
is made to indicate its {\tt z} branch. Meanwhile, the former locomotive goes from~$P$
to~$W_{\ttD}$ so that it resets both memories \ttD{} and \ttZ{} to~0. That locomotive
also goes through $E_{\ttZ}$ which sends it to~$F$. The flip-flop switch sitting there
indicates its 1-branch so that the locomotive goes to~$A$ and the switch again indicates
its 0-branch. At $A$ the locomotive is sent through the track named {\tt z} to the right
place in the instruction part of our implementation.

   On the same line, we have to make a bit more precise the memory structure used
for defining the right place in the program for the return of a locomotive after it
has incremented a register. There is such a structure for each register. Each memory
structure consists of as many units as there are instructions which increment the
considered register. Figure~\ref{fdiscrincr} illustrates such a unit. It makes use of a 
one bit memory. In the idle configuration, each unit contains~0. When the locomotive
arrives from a given instruction which increments the register, it enters the appropriate
unit of the memory structure of that register through the {\tt W}-gate. Accordingly, 
from now on, that unit contains~1. As it exits through the {\tt E}-gate, the locomotive 
arrives at~$A$ to a flip-flop switch which sends it on its 0-branch leading to the 
register, and the flip-flop switch indicates its 1-branch.
When the locomotive returns from the register,
it arrives to a unit through a track which leads to the {\tt R}-gate. If the locomotive
reads 1 there, it is sent through {\tt B0} to the {\tt R}-gate of the next unit. 
Accordingly the locomotive arrives to the unique unit containing~1. As it reads~1, it
leaves the one-bit memory through {\tt B1} which sends the locomotive to the {\tt W}-gate
through a fixed switch.
Accordingly, the locomotive changes the one-bit memory to~0 and it leaves that memory 
through the {\tt E}-gate. Accordingly, it is sent again to~$A$. The flip-flop switch 
which sits there, sends the locomotive to its 1-branch leading to the right instruction 
of the program and the flip-flop switch again indicates its 0-branch, the direction to 
the registers.

\vtop{
\ligne{\hfill
\includegraphics[scale=0.5]{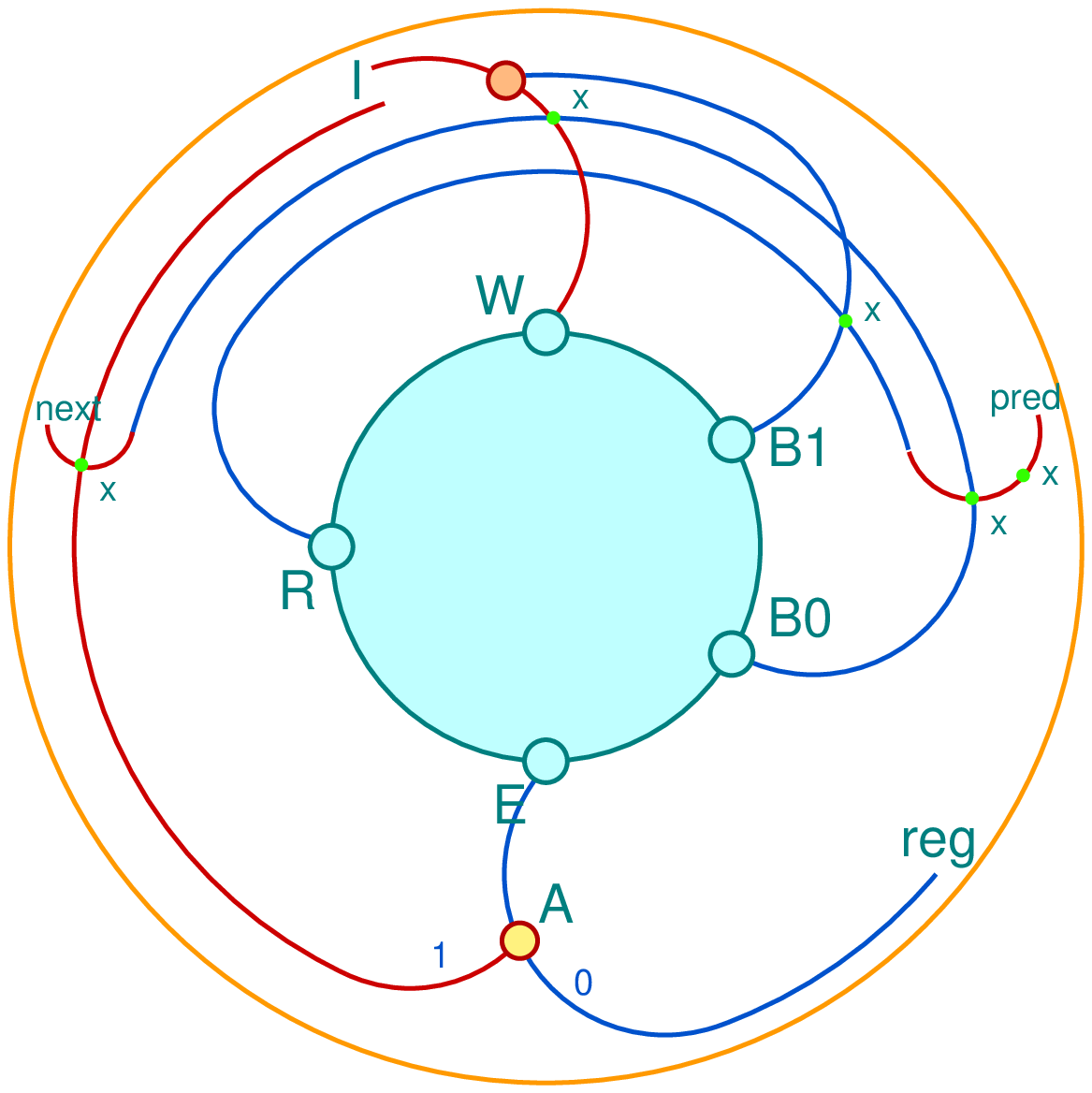}
\hfill}
\begin{fig}\label{fdiscrincr}
\leurre
The use of a one-bit memory in a unit for selecting the right place for an incrementing
instruction on its returning way. The conventions for the switches and the crossings are 
those of Figure~{\rm\ref{fhypunit}}. 
\end{fig}
}

It is time to look closer at what happens for a returning locomotive to the instructions.
When it comes back from a register, the locomotive is red. If it arrives to an 
incrementing instruction or to a jump instruction, it must become green. If the jump 
instruction arrives to a decrementing instruction, the locomotive must become red.
Accordingly, we need a structure which allows to convert a green locomotive to a red one 
and conversely. It is illustrated 
by Figure~\ref{stabconv}, on the left-hand side picture. The right-hand side picture
reminds us the converter of a double locomotive to a simple one.

Next, Figure~\ref{stabreg} shows us different configurations of a register. The rightmost
picture of the figure illustrates a standard configuration, when some positive number
is stored in the register. The leftmost picture illustrates an empty register, which means
that its content is~0. The middle picture shows a register which contains~1 exactly.
On the rightmost picture, we can see the structure of the register : the number stored
in the structure is the number of blue central cells. As already mentioned, the yellow
cells allows the locomotive to arrive at the end of the stored number in order to
append a new blue cell or in order to erase the last one. The process of appending or
of erasing has also to deal with the yellow and orange cells which sit around the blue
part. A green locomotive performs its action as a red cell which becomes later the
returning locomotive. A red locomotive performs its action as a green cell which
becomes a red locomotive again when the action is performed.

\vtop{
\ligne{\hfill
\includegraphics[scale=0.6]{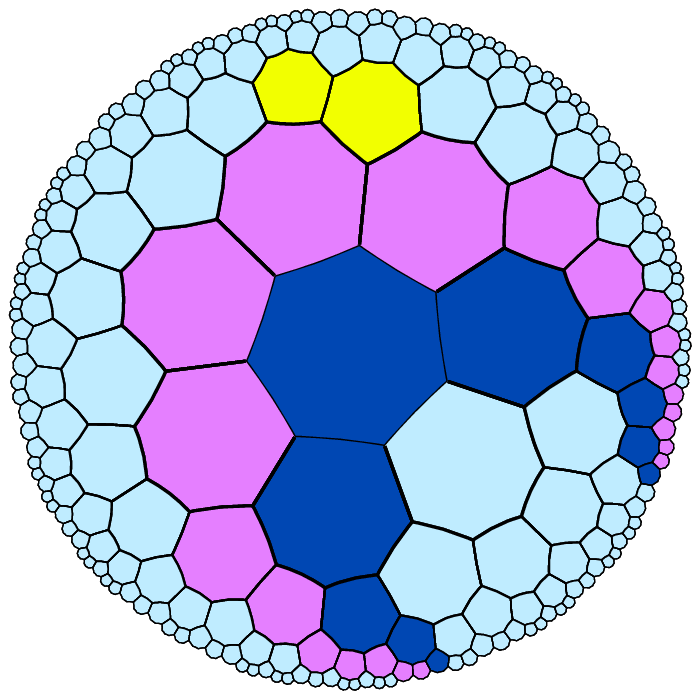}
\hskip 10pt
\includegraphics[scale=0.6]{stab_cvds.ps}
\hfill}
\begin{fig}\label{stabconv}
\leurre
Converters for the locomotive.
To left: the converter from green to red and from red to green colours. To right, 
the converter from double to simple forms.
\end{fig}
}

We conclude the subsection devoted to registers by showing two schemes on 
Figure~\ref{regtests}. To left, we have the scheme of zero. As clear from 
Table~\ref{rregmr}, when the red locomotive checks that the register is empty, it sends
two locomotive on the circuit. One locomotive takes the path to the structure which stand
to the right of the figure, a structure which distinguish between instructions which 
incremented and those which decremented. The second locomotive goes to the instruction
to be executed after a zero test. In the case when the locomotive realises that the
register was empty, the first locomotive must be stopped. The scheme of that action 
is illustrated by the right-hand side picture of Figure~\ref{regtests}.

\vtop{
\ligne{\hfill
\includegraphics[scale=0.5]{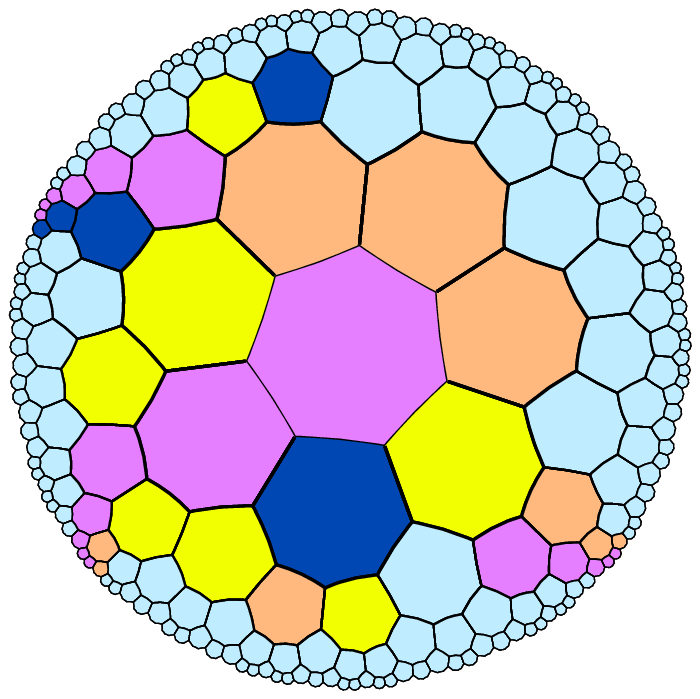}
\hfill
\includegraphics[scale=0.5]{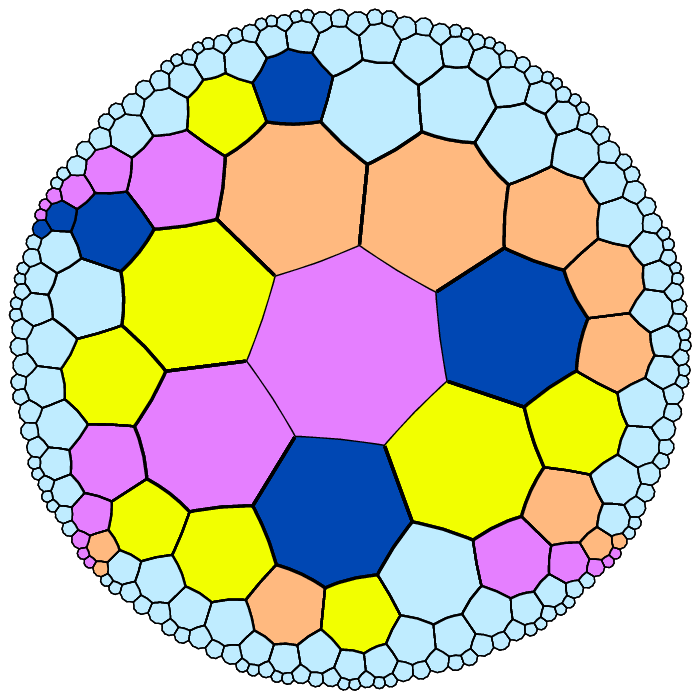}
\hfill
\includegraphics[scale=0.5]{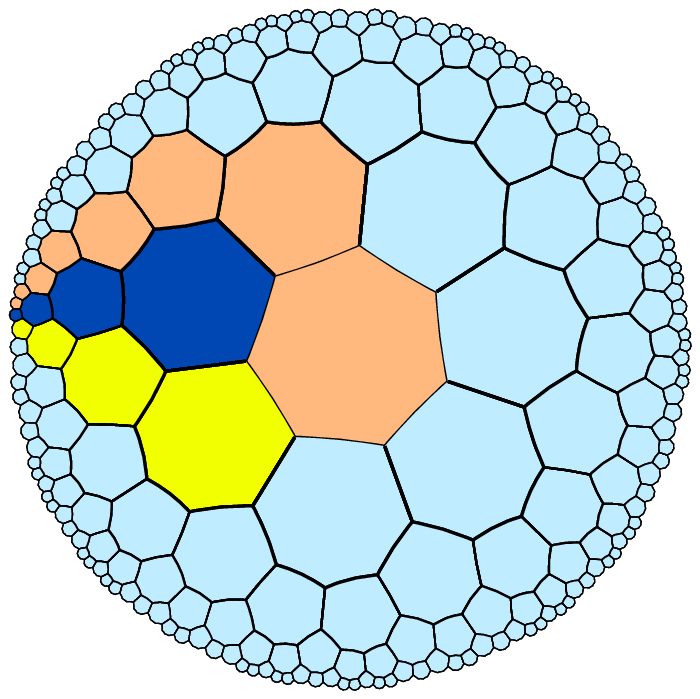}
\hfill}
\begin{fig}\label{stabreg}
\leurre
The register: to left, when it is zero; in the middle, when it contains $1$ exactly;
to right, when it has some number.
\end{fig}
}

On the way to the further treatment of the locomotive returning from zero, call it
the zero-locomotive, that locomotive sends at~$I$ a first copy to a controller which 
will stop the locomotive returning through the standard way, say the nonzero-locomotive. 
On its way to the controller, the first copy sends a second one to the
controller in order to let the next locomotive go. The paths are computed in such a way
that the non-zero locomotive arrives to the controller in between the
arrival of the copies sent by the zero-locomotive. Indeed, increasing the radius of a 
circle by one multiply the circumference by at least two, so that it is easy to manage
the circuit in order to obtain the expected action.

\vtop{
\ligne{\hfill
\includegraphics[scale=0.45]{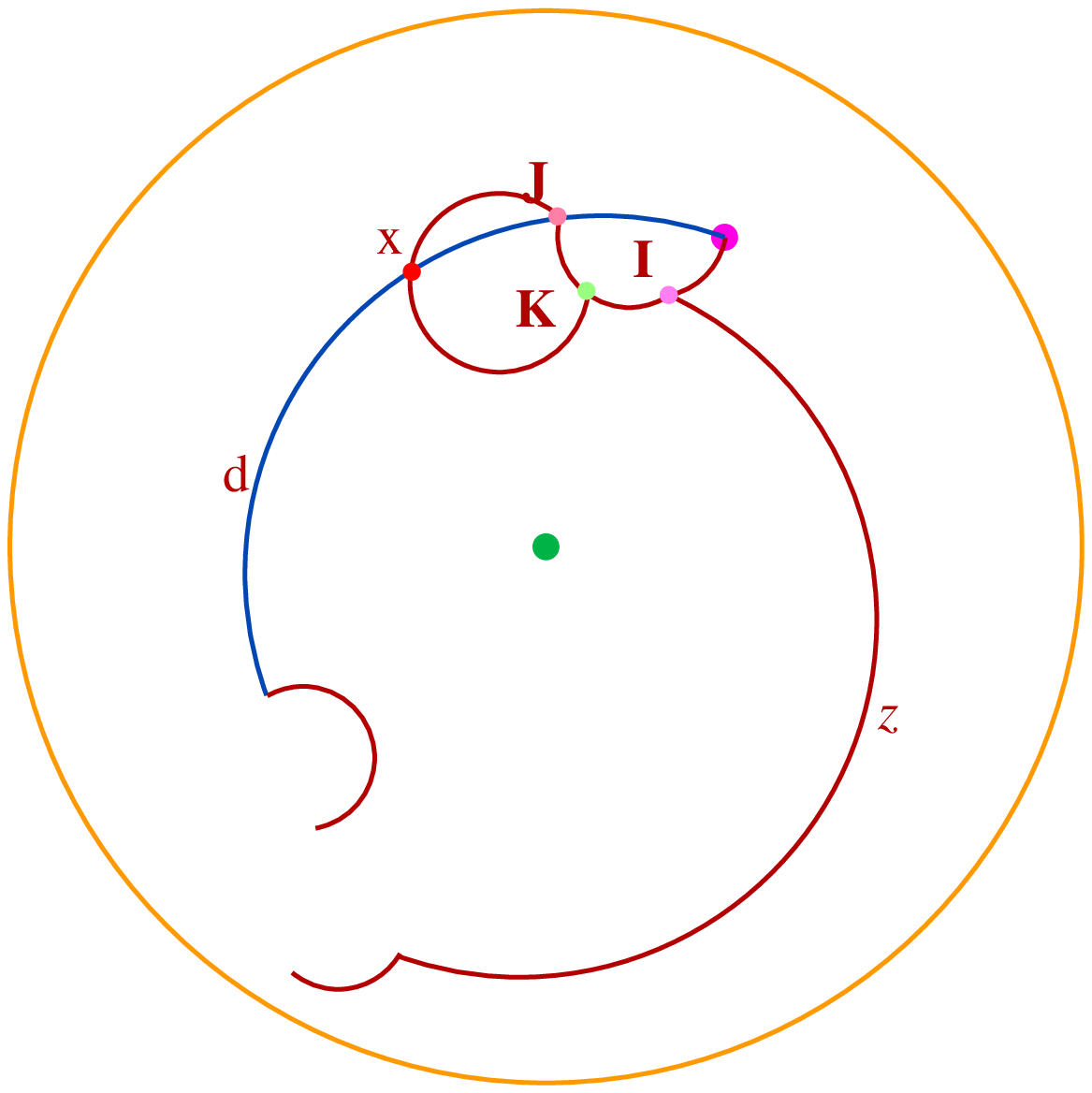}
\hskip-10pt
\includegraphics[scale=0.45]{disque_discrimret_vrai.ps}
\hfill}
\vspace{-10pt}
\begin{fig}\label{regtests}
\leurre
To left: the test of zero for a register. 
To right, the discrimination between 
instructions which increment and those which decrement when the locomotive returns
to the instructions. That structure uses the one-bit memory.
\end{fig}
}

On the right-hand side picture, we can see the implementation of the structure which
allows an instruction returning from a register to go back to the appropriate next
instruction. It makes use of a one-bit memory, see the middle picture of the figure,
which was set to~1 if a red locomotive crosses it on its way to the
register. A green locomotive does not cross that unit which, in that case, remains set
to~0. The returning locomotive reads the memory. If it is~0, it know it goes to the
selector of incrementing instructions in order to find out the appropriate next
instruction. If the memory is set to~1, the locomotive crosses it again which 
automatically sets the unit to~0, next it goes to the selector of the decremeting 
instructions. 
The locomotive is set again to the unit through a flip-flop switch
which was already crossed on the way to the register, so that the second crossing
sets back the switch to the right position. Note that a zero-locomotive goes directly
to the selector of decrementing instructions in order to go to the correct next 
instruction.

\section{Rules and figures}
\label{rules}

    The figures of Sections~\ref{scenario} and~\ref{scenar} help us to establish the 
rules. The rules and the figures were established with the help of a computer program 
which checked the rotation invariance of the rules and which wrote the PostScript files 
of the pictures from the computation of the application of the rules to the 
configurations of the various type of parts of the circuit. The computer program also 
established the traces of execution which
allow the reader to check the application of the rules.

   In comments of the figures, the central tile is denoted by~0(0).  The cells which are
neighbours of~0(0) by sharing an edge with that cell are numbered 1($i$), with 
\hbox{$i\in\{1..7\}$}, increasingly while counter-clockwise turning around 0(0), see the
left-hand side part of Figure~\ref{hepta}. 
In each sector delimited on that picture, the cell 1($i$) has three sons numbered 
2($i$), 3($i$) and 4($i$), blue, green and yellow respectively in the right-hand side part
of Figure~\ref{hepta} where the cell 1$(i)$ is the closest green tile to the central tile.
The cell 2($i$) has two sons: the cells 5($i$) and 6($i$), blue and green respectively. 
The cell 3($i$) has three sons numbered 7($i$), 8($i$) and 9($i$), blue, green and yellow
respectively on the picture while the cell 4($i$) has also three sons numbered
10($i$), 11($i$) and 12($i$), also blue, green and yellow respectively on the figure.
For further numbers, the reader is invited to look at~\cite{mmbook1}.

    Later, we provide the reader with tables of rules and figures in following the various
structures of the implementation, starting with the tracks, see 
Sub-section~\ref{sbstracks}.
Before turning to the study, let us mention notations and important properties for a 
better understanding of the rules and their application.

\def\ftt {\footnotesize\tt}
\def\disp #1 #2 #3 {\hbox{%
$\underline{\hbox{\ftt#1}}${\ftt#2}$\underline{\hbox{\ftt#3}}${\ }\hskip -7pt}
}

   First, we establish the format we use to denote the rules. A rule has the form
%\hbox{$\underline{\hbox{\ftt w}$_{\hbox{\ftt 1}}$}${\ftt w}$_\hbox{\ftt 1}$..{\ftt w}%
%$_\hbox{\ftt 7}$
%$\underline{\hbox{\ftt w$_{\ftt n}$}}$},
\disp {w$_o$} {w$_1$..w$_7$} {w$_n$} {}   
where \hbox{\ftt w$_o$} is the current state of the cell~$c$,
\hbox{\ftt w$_i$} is the current state of the neighbour~$i$ of~$c$ 
and \hbox{\ftt w$_n$} is the next state of~$c$ once the rule
has been applied. The rules must be rotation invariant which means that
the rule \disp {w$_o$} {w$_1$..w$_7$} {w$_n$} {} and the rule
\disp {w$_o$} {w$_{\pi(1)}$..w$_{\pi(7)}$} {w$_n$} {} 
are the same for any circular 
permutation $\pi$ on the set \hbox{$[1..7]$}. The cellular automaton have seven states,
{\ftt W$,$ B$,$ R$,$ Y$,$ G$,$ O {\normalsize\rm and} M} in that order. The states 
{\ftt G} and {\ftt R} are essentially used by the locomotives.
Consider the rules \disp {w$_o$} {w$_{\pi(1)}$..w$_{\pi(7)}$} {w$_n$} {}  with
$\pi$ running over the seven circular permutations on \hbox{$[1..7]$}. We can order these
rules according to the order we defined on the states as a rule can be read as a word. 
In the tables, we shall take the rule using the circular permutation which provides the 
smallest word.

\def\Bb#1{{\color{blue}{#1}}}
\def\Rr#1{{\color{red}{#1}}}
    There are two types of rules. Those
which keep the structure invariant when the locomotive is far from them, 
and those which control the motion of the locomotive.
We call \textbf{conservative} the rules of the former type and {\bf motion rules}
those of the latter one. Motion rules are applied to the cells of the tracks: the one
on which the locomotive stand and also the neighbours which will be occupied by the
locomotive at the next time. Among the conservative rules, there are rules in which 
the state of the cell does not change but the states of at least one neighbour is
affected by the motion of the locomotive. Those rules are called {\bf witness rules}.
A cell to which a witness rule applies has at least one neighbour on which the
locomotive may stand. The largest set of contiguous neighbours of a cell on which the
locomotive may stand on each of them is called a {\bf window} on the motion of the
locomotive. As an example, such cells occur in the supports of the tracks, see
Figure~\ref{ftracks}. As an example, on a blue path, many cells are applied the
conservative rule \disp {B} {WBMMMMB} {B} {}. The contiguous four {\footnotesize\tt M}
constitute a window, each cell of which is step after step occupied by the locomotive,
giving rise to four conservative rules. We shall mention the witness rules among the
conservative ones, but we shall not provide all the rules, indicating the window only
as follow, taking the previous example:
\disp {B} {WB:MMMM:B} {B} {}. That notation replaces the rules
\disp {B} {WB\Bb LMMMB} {B} {},
\disp {B} {WBM\Bb LMMB} {B} {},
\disp {B} {WBMM\Bb LMB} {B} {} and 
\disp {B} {WBMMM\Bb LB} {B} {} as well as the rules when a double locomotive is
passing through the window:
\disp {B} {WB\Bb{LL}MMB} {B} {},
\disp {B} {WBM\Bb{LL}MB} {B} {} and
\disp {B} {WBMM\Bb{LL}B} {B} {}. 
In the previous rules, we used symbol~{\ftt L} to replace~{\ftt G} and~{\ftt R}. Note 
that in the case of several occurrences of~{\ftt L} in a rule, all of them have to be 
replaced either by~{\ftt G} or by~{\ftt R}. That notation will be used in the sequel
in order to spare space. However, a single-symboled window will not use that notation.
In the tables for the rules, in order to avoid fastidious repetitions, we shall use
the notation displayed in \disp {B} {WB:{MMMM}:B} {B} {}. In those
cases, the successive symbols of the window can be replaced by~{\ftt G} or by {\ftt R},
as we can do when {\ftt L} has to be replaced by those symbols in the seven previous 
rules.

We can do the same remarks about similar rules when green or red, simple or double 
locomotives run on the blue path and on the orange path. We leave that to the reader as 
an exercise.

\def\affH #1 #2 #3 #4 {\hbox{\footnotesize\tt\hbox to 13pt{\hfill#1}
\hskip 5pt$\underline{\hbox{\tt#2}}$#3$\underline{\hbox{\tt#4}}$}
}

\def\aff #1 #2 #3 #4 {\ligne{\footnotesize\tt\hbox to 13pt{\hfill#1}
\hskip 5pt$\underline{\hbox{\tt#2}}$#3$\underline{\hbox{\tt#4}}$\hfill}
\vskip -2pt
}
\newcount\numr\numr=1
\def\affn #1 #2 #3 {\ligne{\footnotesize\tt\hbox to 13pt{\hfill\the\numr}
\hskip 5pt$\underline{\hbox{\tt#1}}$#2$\underline{\hbox{\tt#3}}$\hfill}
\vskip -2pt
\global\advance\numr by 1
}
\def\affRr #1 #2 #3 #4 {\ligne{\footnotesize\tt\hbox to 13pt{\hfill\Rr{#1}}
\hskip 5pt$\underline{\hbox{\tt#2}}$#3$\underline{\hbox{\tt#4}}$\hfill}
\vskip -2pt
\global\advance\numr by 1
}
\subsection{Rules and figures for the tracks}\label{sbstracks}

Table~\ref{rblank} provides us conservative rules which are applied to the white cells
which remain white in the computation and for cells which are completely surrounded
by white cells.

\vtop{
\begin{tab}\label{rblank}
\leurre
Conservative rules for white cells.
\end{tab}
\vspace{-10pt}
\ligne{\hfill
\vtop{\leftskip 0pt\parindent 0pt\hsize=61pt
\affn {W} {WWWWWWW} {W}
\affn {B} {WWWWWWW} {B}
\affn {R} {WWWWWWW} {R}
\affn {Y} {WWWWWWW} {Y}
\affn {G} {WWWWWWW} {G}
}
\hfill
\vtop{\leftskip 0pt\parindent 0pt\hsize=61pt
\affn {O} {WWWWWWW} {O}
\affn {M} {WWWWWWW} {M}
\affn {W} {WWWWWWB} {W}
\affn {W} {WWWWWWR} {W}
\affn {W} {WWWWWWY} {W}                %V 309
}
\hfill
\vtop{\leftskip 0pt\parindent 0pt\hsize=61pt
\affn {W} {WWWWWWG} {W}                %V 318
\affn {W} {WWWWWWO} {W}
\affn {W} {WWWWWWM} {W}                
}
\hfill}
% total partiel : 13
}
\vskip 10pt
Table~\ref{rtracks} gives the rules for ordinary tracks. By that expression, we mean
mean tracks which are supported by a circle. Later, we look at rules which allow
the locomotive to pass from an arc of a circle to that of another circle. As already 
mentioned, the simulation works on two kinds of tracks. We call {\bf $B$-, $O$-path}
a track whose support is an arc of a circle consisting of blue, orange cells respectively.

Figures~\ref{fBpath} and~\ref{fOpath} illustrate the application of the rules of
Table~\ref{rtracks} for $B$- and $O$-paths respectively.
In the table, two rules are numbered by red digits. It means that the same rules will
later be repeated with the same numbers. That convention will be used in the sequel.

\vtop{
\begin{tab}\label{rtracks}
\leurre
Rules for ordinary tracks
\end{tab}
\vspace{-10pt}
\ligne{\hfill Conservative rules\hfill\hfill witness rules\hfill}
%   pour mémoire : total partiel 13
\ligne{\hfill
\vtop{\leftskip 0pt\parindent 0pt\hsize=61pt
\affn {M} {WWWWMBM} {M}                %V 27
\affn {M} {WWWMBBM} {M}                %V 32
\affn {M} {WWWWMOM} {M}                %V 37
\affn {M} {WWWMOOM} {M}                %V 42     +4
}
\vtop{\leftskip 0pt\parindent 0pt\hsize=61pt
\affn {W} {WWWBBBB} {W} 
\affRr {19} {W} {WWWWBBB} {W}                %V 131
\affn {W} {WWWOOOO} {W} 
\affRr {21} {W} {WWWWOOO} {W}                %V 123    +4
}
\hfill\hfill
\vtop{\leftskip 0pt\parindent 0pt\hsize=70pt
\affn {B} {WB:MMMM:B} {M}              %         +8
\affn {B} {WWB:MMM:B} {M}              %         +6
\affn {O} {WO:MMMM:O} {M}              %         +8
\affn {O} {WWO:MMM:O} {M}              %         +6
}
\hfill}
%                                        partielle :  +36
\vskip 9pt
\ligne{\hfill Motion rules\hfill}
\ligne{\hfill $B$-path \hfill\hfill $O$-path\hfill}
\ligne{\hfill
\vtop{\leftskip 0pt\parindent 0pt\hsize=61pt
\affn {M} {WWWWMBL} {L}
\affn {L} {WWWWMBM} {M}                %V 14     
\affn {M} {WWWWLBM} {M} 
\affn {L} {WWWWMBL} {L}
\affn {L} {WWWWLBM} {M}
%                                                     +10
}
\hfill
\vtop{\leftskip 0pt\parindent 0pt\hsize=61pt
\affn {M} {WWWMBBL} {L} 
\affn {L} {WWWMBBM} {M}                %V 15
\affn {M} {WWWLBBM} {M} 
\affn {L} {WWWMBBL} {L} 
\affn {L} {WWWLBBM} {M} 
%                                                     +10
}
\hfill
\vtop{\leftskip 0pt\parindent 0pt\hsize=61pt
\affn {M} {WWWWLOM} {L} 
\affn {L} {WWWWMOM} {M}                %V 16
\affn {M} {WWWWMOL} {M} 
\affn {L} {WWWWLOM} {L}
\affn {L} {WWWWMOL} {M} 
%                                                     +10
}
\hfill
\vtop{\leftskip 0pt\parindent 0pt\hsize=61pt
\affn {M} {WWWLOOM} {L} 
\affn {L} {WWWMOOM} {M}                %V 17
\affn {M} {WWWMOOL} {M} 
\affn {L} {WWWLOOM} {L} 
\affn {L} {WWWMOOL} {M} 
%                                                     +10
}
\hfill}
%                                                     +10
%                                        partielle    +36+40
% total partiel : 13+76 = 89 
}

\vskip 10pt
\vtop{
\ligne{\hfill
\includegraphics[scale=0.5]{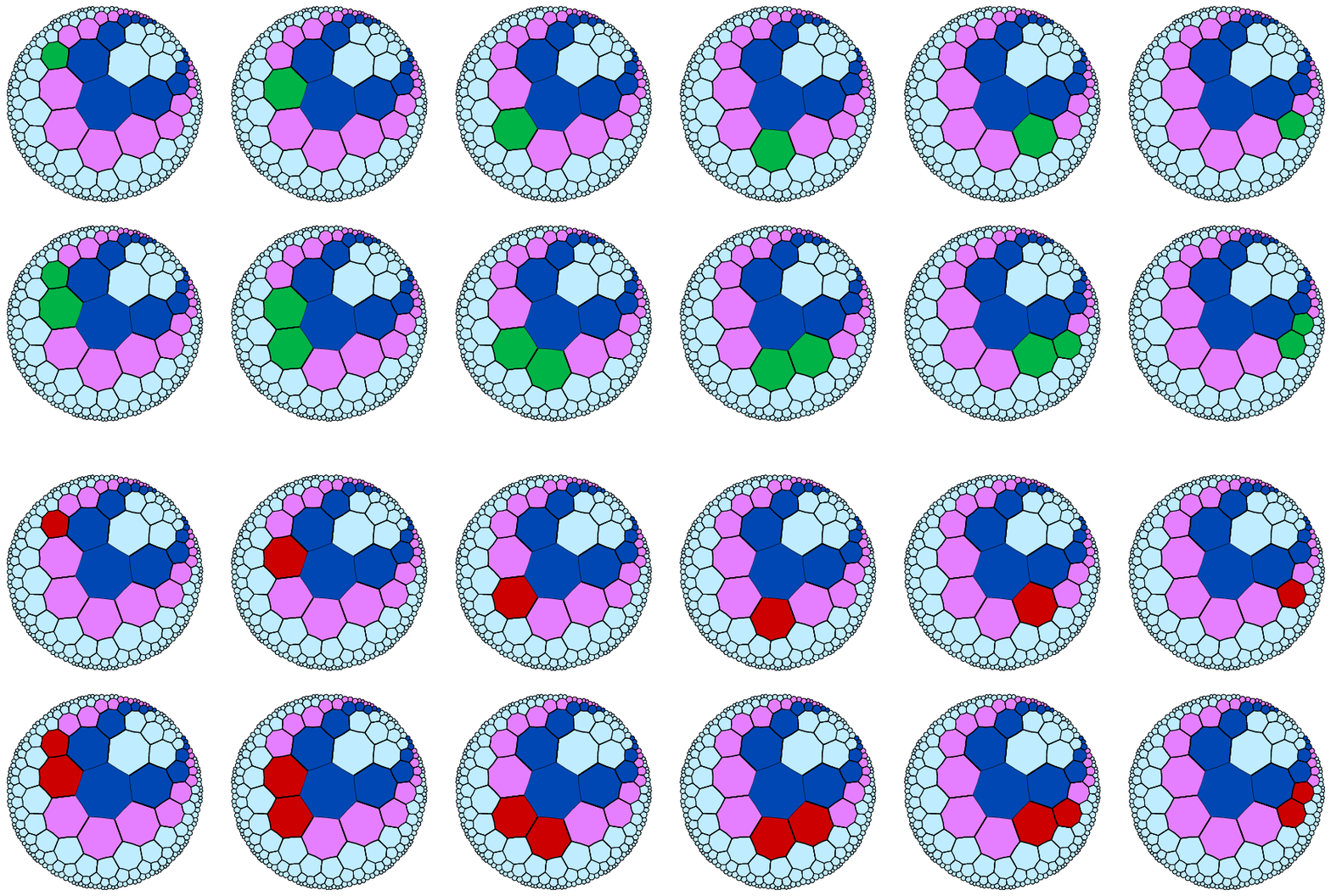}
\hfill}
\vspace{-20pt}
\begin{fig}\label{fBpath}
\leurre
Locomotives on a $B$-path. From top to bottom: simple green one, double green one,
simple red one, double red one. On those pictures, the radius of the arc on which the
tracks lie is~$4$. 
\end{fig}
%   total partiel 89
}

As mentioned in the legend of the figures, the radius of the arc on which the mauve 
cells, those of the tracks, lie is 4. As the cells of the tracks have at least three
white neighbours and at most two neighbours on their support, the identification of
which rule is applied is easy.

The figures help us to see that, in Table~\ref{rtracks}, the conservative rules
14 to~17 concern the mauve cells of tracks, while the rules~18 to~21 concern the white
cells which are inside the circle supporting the support of the tracks. The blue or
orange cells of the support are applied one of the witness rules from~22 to~25.
It is easy to check that the 45 meta-rules of Tables~\ref{rblank} and~\ref{rtracks}
represent 71 actual rules of the automaton.

\vtop{
\ligne{\hfill
\includegraphics[scale=0.5]{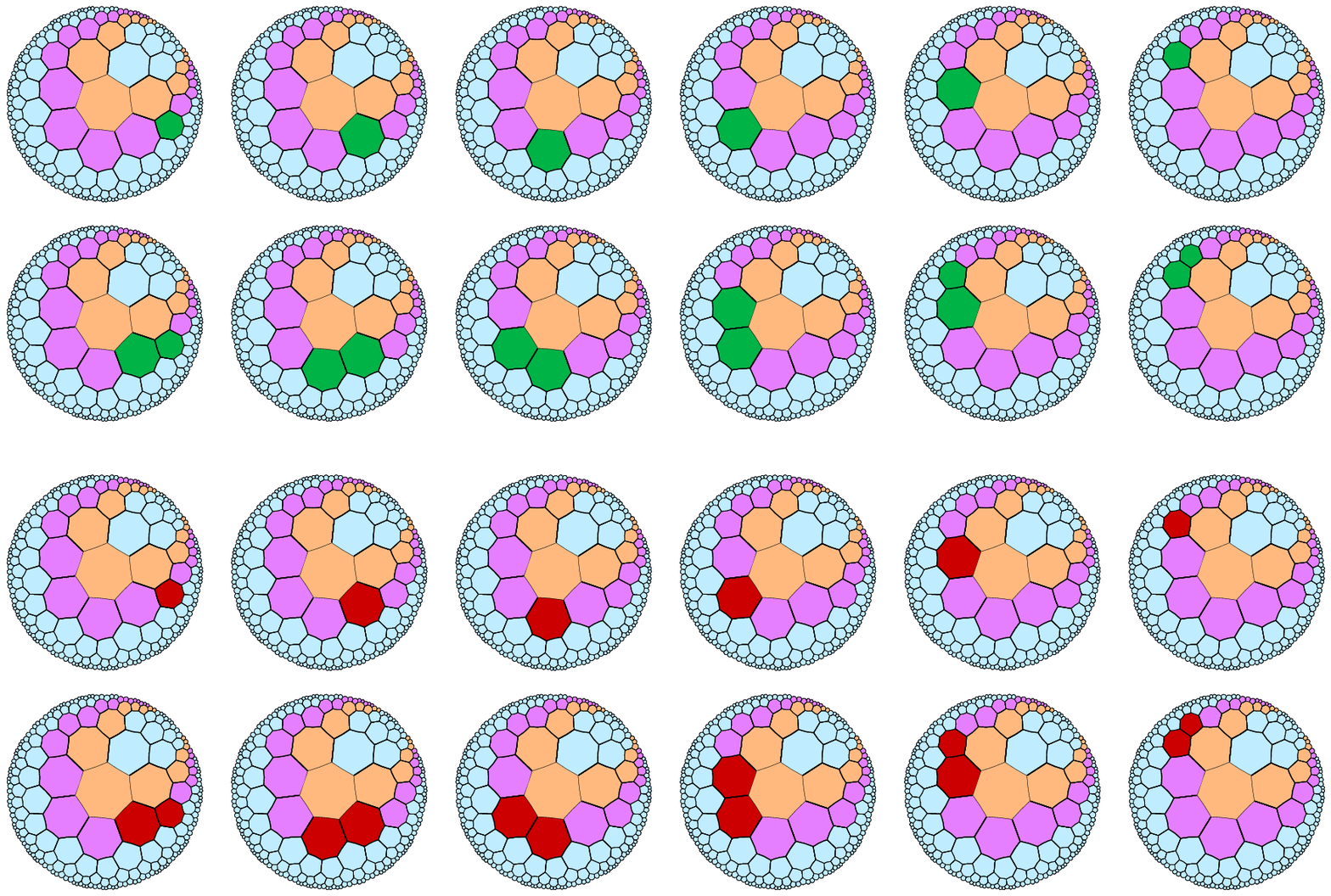}
\hfill}
\vspace{-20pt}
\begin{fig}\label{fOpath}
\leurre
Locomotives on a $O$-path. From top to bottom: simple green one, double green one,
simple red one, double red one. On those pictures too, the radius of the arc on which the
tracks lie is~$4$. 
\end{fig}
}

\vtop{
\begin{tab}\label{rlink}
\leurre
Rules for the links between $B$- and $O$-paths.
\end{tab}
\vspace{-7pt}
\ligne{\hfill $B$ $\rightarrow$ $O$  \hfill $O$ $\rightarrow$ $B$\hfill 
$B$ $\rightarrow$ $B$\hfill $O$ $\rightarrow$ $O$\hfill}
\vskip 5pt
\ligne{\hfill conservative rules and witness rules\hfill}
%   pour mémoire : total partiel 89
\ligne{
\vtop{\leftskip 0pt\parindent 0pt\hsize=70pt
\affRr {46} {M} {WBMWOOM} {M}            %V 70, 156, 200
\affn {B} {WB:MMMMM:} {B}              %         +10
\affn {O} {WWWWWOM} {O}                %V 70
\affn {W} {WWWBB:MM:} {W}              %         +4
\affn {W} {WWWWO:MM:} {W}              %V 159    +4
\affn {W} {WWOOOOO} {W}                %
%                                           +21
}
\hfill
\vtop{\leftskip 0pt\parindent 0pt\hsize=70pt
\affn {M} {WWMBBMO} {M}                %V 75
\affn {W} {WWBBBBB} {W}
\affn {W} {WWWWWMO} {W}                 
\affn {B} {BB:MMMMM:} {B}              %         +10
\affn {B} {WWW:MM:BB} {B}              %         +4
\affn {B} {WBB:MMM:B} {B}              %         +6
\affRr {58} {O} {WW:MMMM:O} {O}              %         +8
%                                           +31
}
\hfill
\vtop{\leftskip 0pt\parindent 0pt\hsize=70pt
\affn {M} {WWBMWMB} {M}                %V 80               
\affRr {60} {W} {WWWW:MMM:} {W}          %V*65, 95, 125*     +6
\affn {W} {WWWWWMB} {W}
\affRr {62} {W} {WWWWWBM} {W}            %V*124
\affn {W} {WWWWWBB} {W}
\affRr {64} {B} {WWB:MMMM:} {B}          %V*130              +8
\affRr {65} {B} {WW:MMMM:B} {B}              %                   +8
%                                           +25
}
\hfill
\vtop{\leftskip 0pt\parindent 0pt\hsize=70pt
\affn {M} {WMOWMOM} {M}                %V 85
\affRr {60} {W} {WWWW:MMM:} {W}          %V 59               +6
\affRr {68} {W} {WWWW:MM:O} {W}          %V 178              +4
\affn {W} {WWWWOOM} {W}
\affn {M} {WWWWWOM} {M}                %V 48               +2
%                                           +8
%  soit : 21+31+25+8 = 85
}
\hfill}
\vskip 5pt
\ligne{\hfill motion rules \hfill}
\ligne{\hfill
\vtop{\leftskip 0pt\parindent 0pt\hsize=61pt
\affRr {71} {M} {WBLWOOM} {L}
\affRr {72} {L} {WBMWOOM} {M}                %V 46
\affRr {73} {M} {WBMWOOL} {M}
\affRr {74} {L} {WBLWOOM} {L}
\affRr {75} {L} {WBMWOOL} {M}
%                                                     +10
}
\hfill
\vtop{\leftskip 0pt\parindent 0pt\hsize=61pt
\affn {M} {WWMBBLO} {L}
\affn {L} {WWMBBMO} {M}                %V 51
\affn {M} {WWLBBMO} {M}
\affn {L} {WWMBBLO} {L}
\affn {L} {WWLBBMO} {M}
%                                                     +10
}
\hfill
\vtop{\leftskip 0pt\parindent 0pt\hsize=61pt
\affn {M} {WWBLWMB} {L}
\affn {L} {WWBMWMB} {M}                %V 58
\affn {M} {WWBMWLB} {M}
\affn {L} {WWBLWMB} {L}
\affn {L} {WWBMWLB} {M}
%                                                     +10
}
\hfill
\vtop{\leftskip 0pt\parindent 0pt\hsize=61pt
\affn {M} {WLOWMOM} {L}
\affn {L} {WMOWMOM} {M}
\affn {M} {WMOWMOL} {M}
\affn {L} {WLOWMOM} {L}
\affn {L} {WMOWMOL} {M}
%                                                     +10
%   partiel ici : 85+40
}
\hfill}
%
%   total partiel 89+85+40 = 214
}
\vskip 10pt
Now, as already noticed, the tracks are composed of arcs belonging to different circles
with possibly different radiuses. The link between an arc to another one is an important
point which should not be forgotten. Table~\ref{rlink} provides us with the corresponding
rules, conservative, witness and motion ones.

Figures~\ref{fB2Opath}, \ref{fO2Bpath}, \ref{fB2Bpath} and~\ref{fO2Opath} provide us with
an illustration of how rules are applied to the cells of the different configurations.

\vtop{
\ligne{\hfill
\includegraphics[scale=0.5]{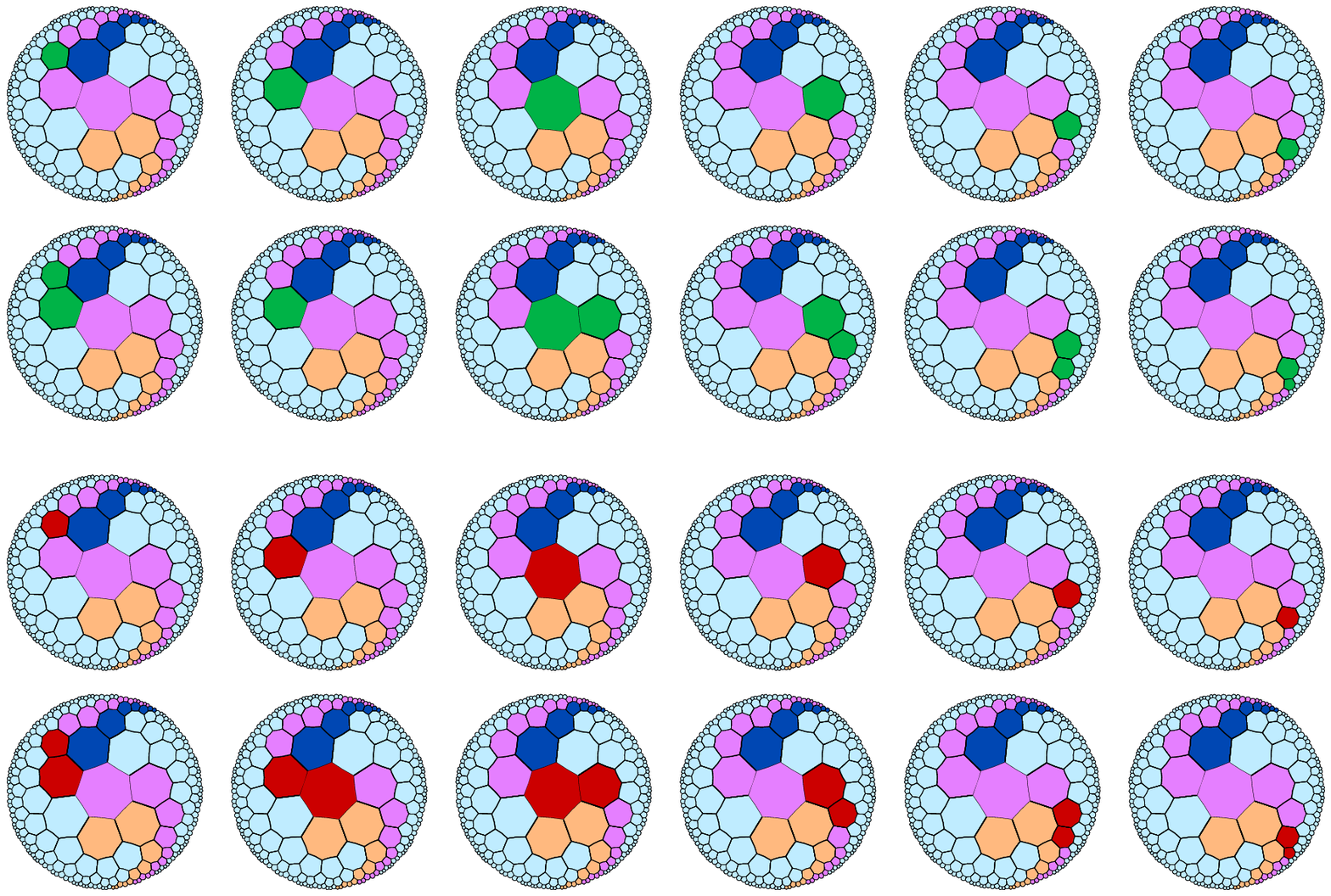}
\hfill}
\vspace{-20pt}
\begin{fig}\label{fB2Opath}
\leurre
Locomotives going from a $B$-path to an $O$-one. From top to bottom: simple green one, 
double green one, simple red one, double red one. On those pictures, the radius of the 
blue arc on which the tracks lie is~$4$ while that of the orange path is~$3$. 
\end{fig}
}

\vtop{
\ligne{\hfill
\includegraphics[scale=0.5]{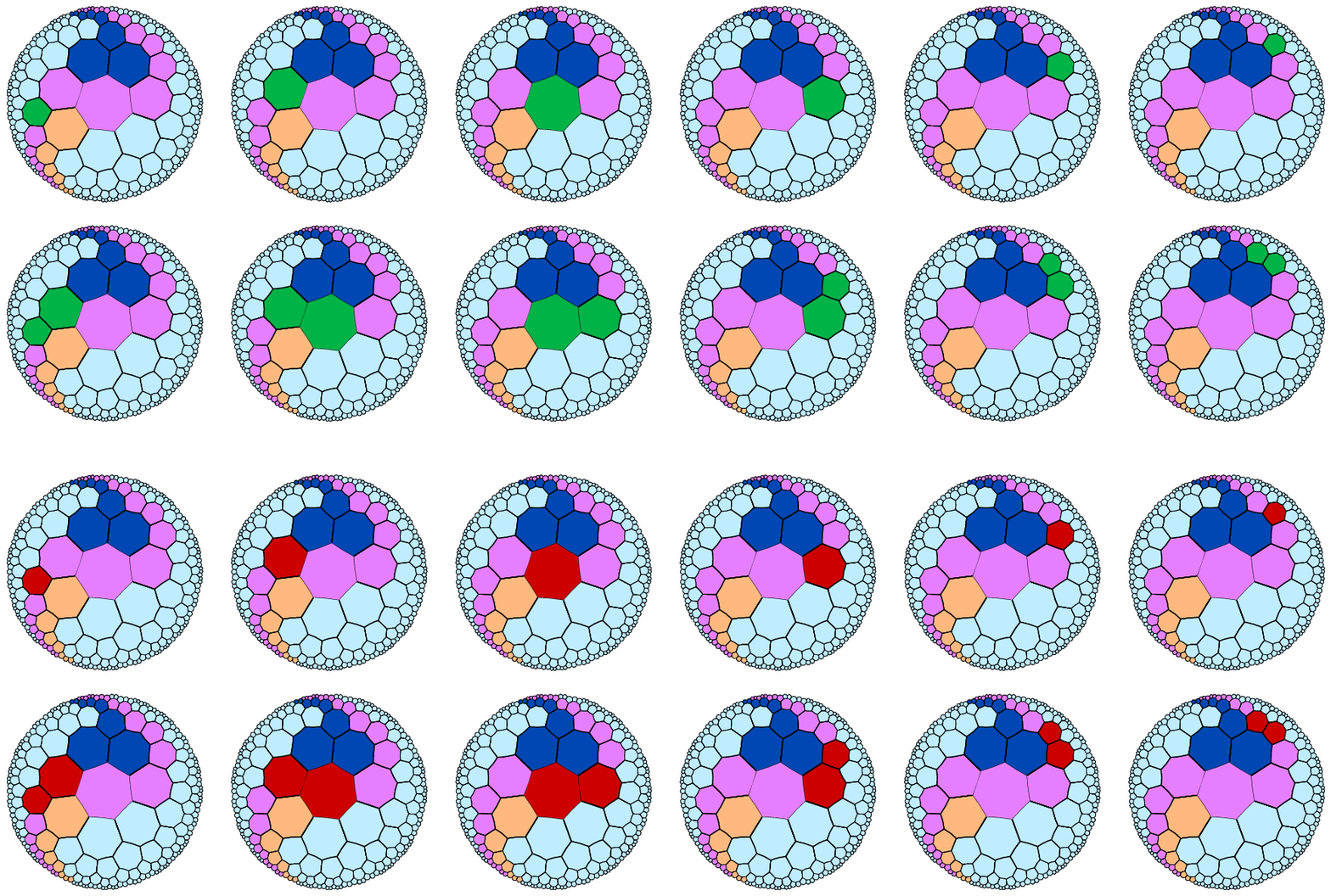}
\hfill}
\vspace{-20pt}
\begin{fig}\label{fO2Bpath}
\leurre
Locomotives going from an $O$-path to a $B$-one. From top to bottom: simple green one, 
double green one, simple red one, double red one. On those pictures, the radius of the 
blue arc on which the tracks lie is~$4$ while that of the orange path is~$3$. 
\end{fig}
}

As mentioned in the legends of the figures, the radiuses of the circle supporting the arcs
may be different. Note that the link requires some adaptation between the arcs. The 
flexibility on the radiuses allow us to give the configurations the same picture
around the mauve cell of the link as in the figures. Note that Figure~\ref{fO2Opath} 
is different from Figure~\ref{fB2Bpath} not only by the change of colour but also
by the occurrence of a mauve witness only present in Figure~\ref{fO2Opath}. That 
conditions is needed for compatibility of the rules.

\vtop{
\ligne{\hfill
\includegraphics[scale=0.5]{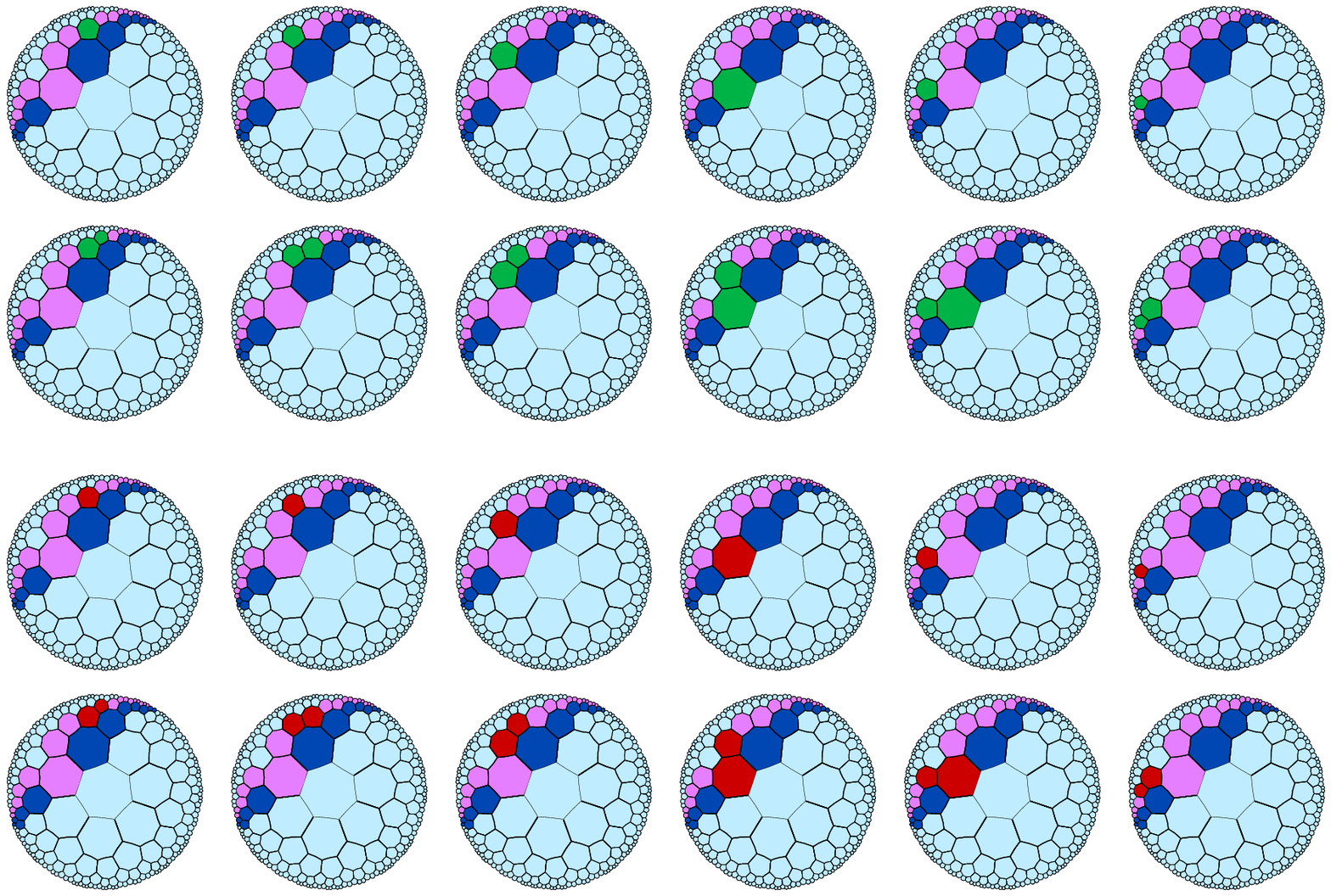}
\hfill}
\vspace{-20pt}
\begin{fig}\label{fB2Bpath}
\leurre
Locomotives going from a $B$-path to another one. From top to bottom: simple green one, 
double green one, simple red one, double red one. 
\end{fig}
}

\vtop{
\ligne{\hfill
\includegraphics[scale=0.5]{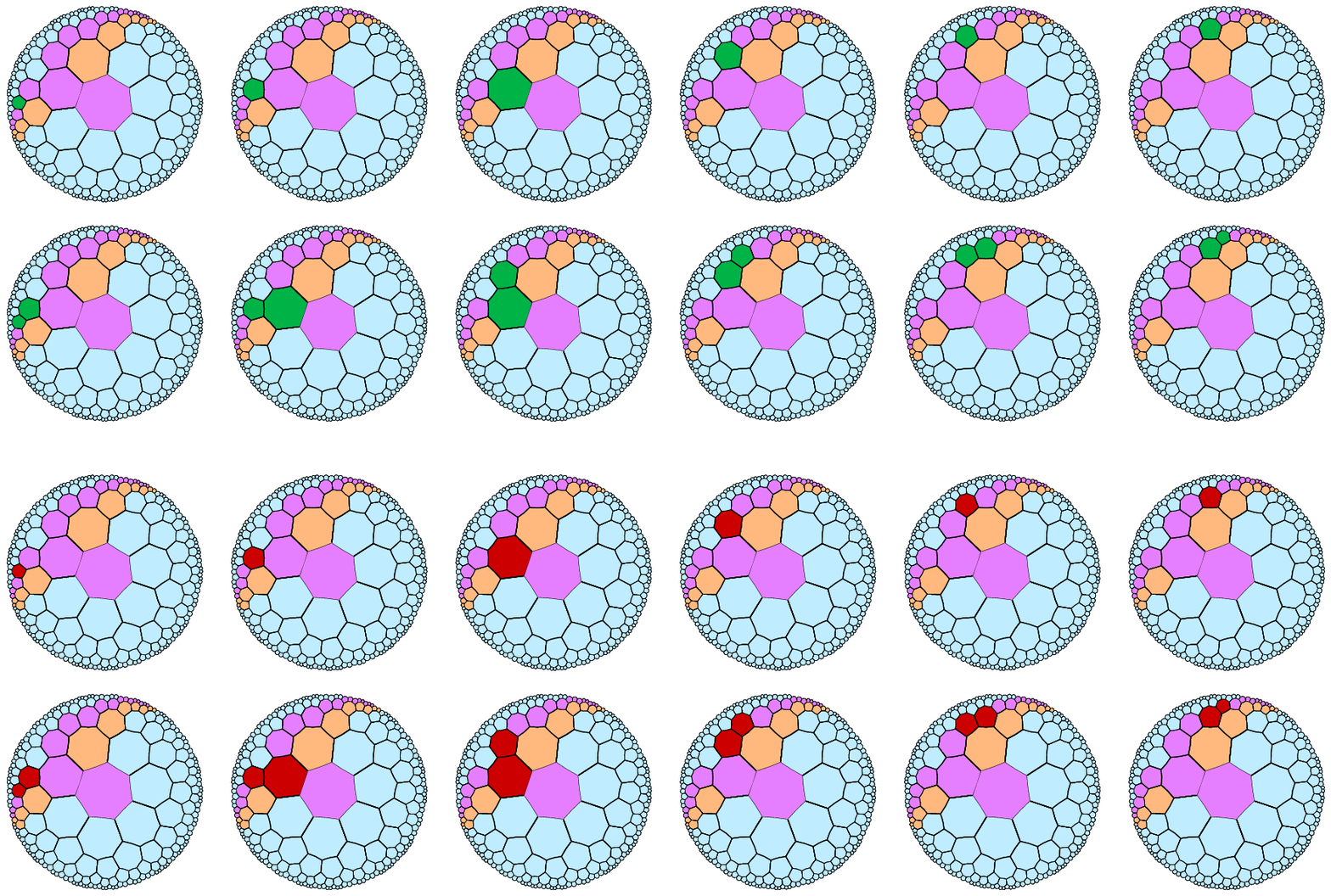}
\hfill}
\vspace{-20pt}
\begin{fig}\label{fO2Opath}
\leurre
Locomotives going from a $O$-path to another one. From top to bottom: simple green one, 
double green one, simple red one, double red one. 
\end{fig}
}

We remain with the rules devoted to the zig-zag lines which we defined in
Section~\ref{scenario}. They are given by Table~\ref{rzigzag}.

Note that the rules for the $O$-path can be deduced from those for the $B$-paths.
Also note that the connection between $B$, $O$-paths in the figures are a bit different
from those of Figures~\ref{fzigzag}. Indeed, the conservative rules
\affH {94} {M} {WWMBWBM} {M} {} and \affH {59} {M} {WWBMWMB} {M} {} together with the
corresponding motion rules give all the possible connections between two $B$-paths
as far as the situation \affH {\hskip -50pt} {M} {WWWBMMB} {M} is excluded. A similar 
remark holds for $O$-paths.

\vtop{
\begin{tab}\label{rzigzag}
\leurre
Rules for the zigzag-lines. 
\end{tab}
\vskip-7pt
\ligne{\hfill conservative and witness rules\hfill\hfill
motion rules\hfill}
\ligne{\hfill
\vtop{\leftskip 0pt\parindent 0pt\hsize=70pt
\affn {W} {WWWWBMB} {W}
\affRr {65} {B} {WW:MMMM:B} {B}
\affRr {64} {B} {WWB:MMMM:} {B}
\affn {M} {WWMBWBM} {M}
}
\hfill
\vtop{\leftskip 0pt\parindent 0pt\hsize=70pt
\affn {W} {WWWWOMO} {W}
\affRr {58} {O} {WW:MMMM:O} {O}
\affn {O} {WWO:MMMM:} {O}
\affn {M} {WWMOWOM} {M}
}
\hfill\hfill
\vtop{\leftskip 0pt\parindent 0pt\hsize=61pt
\affn {M} {WWMBWBL} {L}
\affn {L} {WWMBWBM} {M}
\affn {M} {WWLBWBM} {M}
\affn {L} {WWMBWBL} {L}
\affn {L} {WWLBWBM} {M}
}
\hfill
\vtop{\leftskip 0pt\parindent 0pt\hsize=61pt
\affn {M} {WWMOWOL} {L}
\affn {L} {WWMOWOM} {M}
\affn {M} {WWLOWOM} {M}
\affn {L} {WWMOWOL} {L}
\affn {L} {WWLOWOM} {M}
}
\hfill}
}
\vskip 10pt
\vtop{
\ligne{\hfill
\includegraphics[scale=0.5]{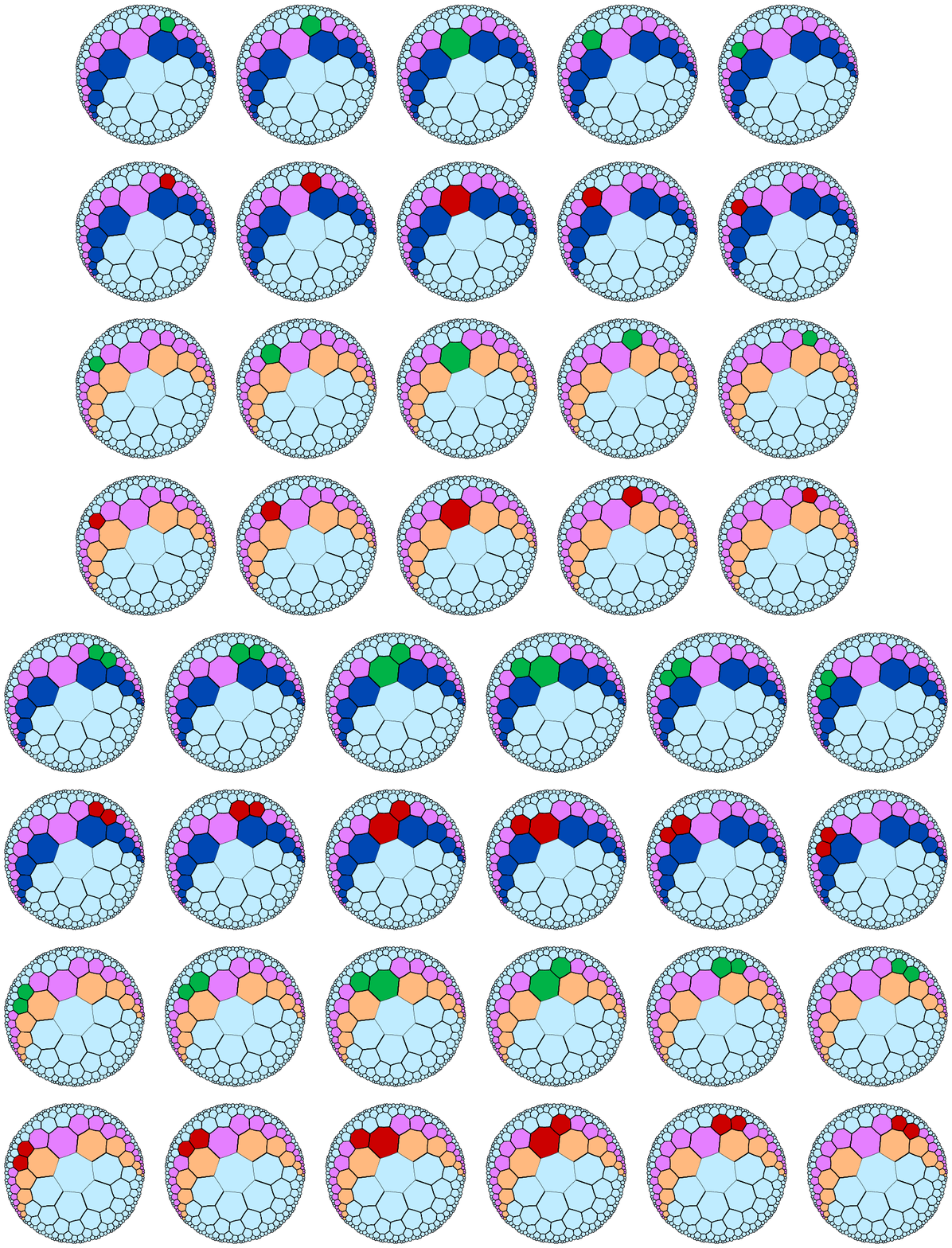}
\hfill}
\vspace{-10pt}
\begin{fig}\label{fzigzag}
\leurre
Locomotives going along the epicycles of a zig-zag line.
\end{fig}
}

\subsection{Passive fixed switch}\label{sbsfx}

As noticed in Section~\ref{scenar}, there is no active switch in our simulation as
return paths are strongly separated from the direct ones. We need to separately check 
the crossing from both branches of the switch. 

Table~\ref{rpfx} provides us with the rules whose application is illustrated by 
Figures~\ref{fFxgpath} and~\ref{fFxdpath}.

\vtop{
\begin{tab}\label{rpfx}
\leurre
Rules for the passive fixed switch.
\end{tab}
\vskip -7pt
\ligne{\hfill conservative and witness rules\hfill}
%   pour mémoire, total partiel 214
\ligne{\hfill
\vtop{\leftskip 0pt\parindent 0pt\hsize=61pt
\affn {M} {WMOYMOM} {M}                %V 100, 119
\affn {W} {WWWBBYO} {W}
\affn {M} {WWWMYBM} {M}                %V 105
\affn {M} {WWWWMYM} {M}                %V 110
\affn {M} {WWWOMYM} {M}                %V 115
%                                                     +5
}
\vtop{\leftskip 0pt\parindent 0pt\hsize=70pt
\affn {W} {WWOOOOM} {W}
\affRr {60} {W} {WWWW:MMM:} {W}          %V 59     +6
\affn {Y} {WB:MMMM:O} {Y}              %V 135    +8
\affn {B} {WB:MMMM:Y} {B}              %         +8
\affRr {118} {O} {WWY:MMM:O} {O}              %         +6
%                                           +23
}
\hfill}
\vskip 5pt
\ligne{\hfill motion rules : \hfill}
\ligne{\hfill from the left:\hfill\hfill\hskip 150pt from the right:\hfill}
\ligne{\hfill
\vtop{\leftskip 0pt\parindent 0pt\hsize=61pt
\affn {M} {WMOYLOM} {L}
\affRr {120} {L} {WMOYMOM} {M}           %V 89, 119
\affRr {121} {M} {WMOYMOL} {M}
\affn {L} {WMOYLOM} {L}
\affRr {123} {L} {WMOYMOL} {M}
%                                                     +10
}
\vtop{\leftskip 0pt\parindent 0pt\hsize=61pt
\affn {M} {WWWMYBL} {L}
\affn {L} {WWWMYBM} {M}                %V 91
\affn {M} {WWWLYBM} {M}
\affn {L} {WWWMYBL} {L}
\affn {L} {WWWLYBM} {M}
%                                                     +10
}
\vtop{\leftskip 0pt\parindent 0pt\hsize=61pt
\affn {M} {WWWWMYL} {L}
\affn {L} {WWWWMYM} {M}                %V 92
\affn {M} {WWWWLYM} {M}
\affn {L} {WWWWMYL} {L}
\affn {L} {WWWWLYM} {M}
%                                                     +10
}
\vtop{\leftskip 0pt\parindent 0pt\hsize=61pt
\affn {M} {WWWOMYL} {L}
\affn {L} {WWWOMYM} {M}                %V 93
\affn {M} {WWWOLYM} {M}
\affn {L} {WWWOMYL} {L}
\affn {L} {WWWOLYM} {M}
%                                                     +10
%  somme partielle +40
}
\hfill\hfill
\vtop{\leftskip 0pt\parindent 0pt\hsize=61pt
\affn {M} {WLOYMOM} {L}
\affRr {120} {L} {WMOYMOM} {M}           %V 89, 100
\affRr {121} {M} {WMOYMOL} {M}
\affn {L} {WLOYMOM} {L}
\affRr {123} {L} {WMOYMOL} {M}
%                                                     +4
}
\hfill}
%  ici, somme locale totale : 23+40+4
%  total partiel 214+67 = 281                           %  revérifier
}
\vskip 10pt
\vtop{
\ligne{\hfill
\includegraphics[scale=0.5]{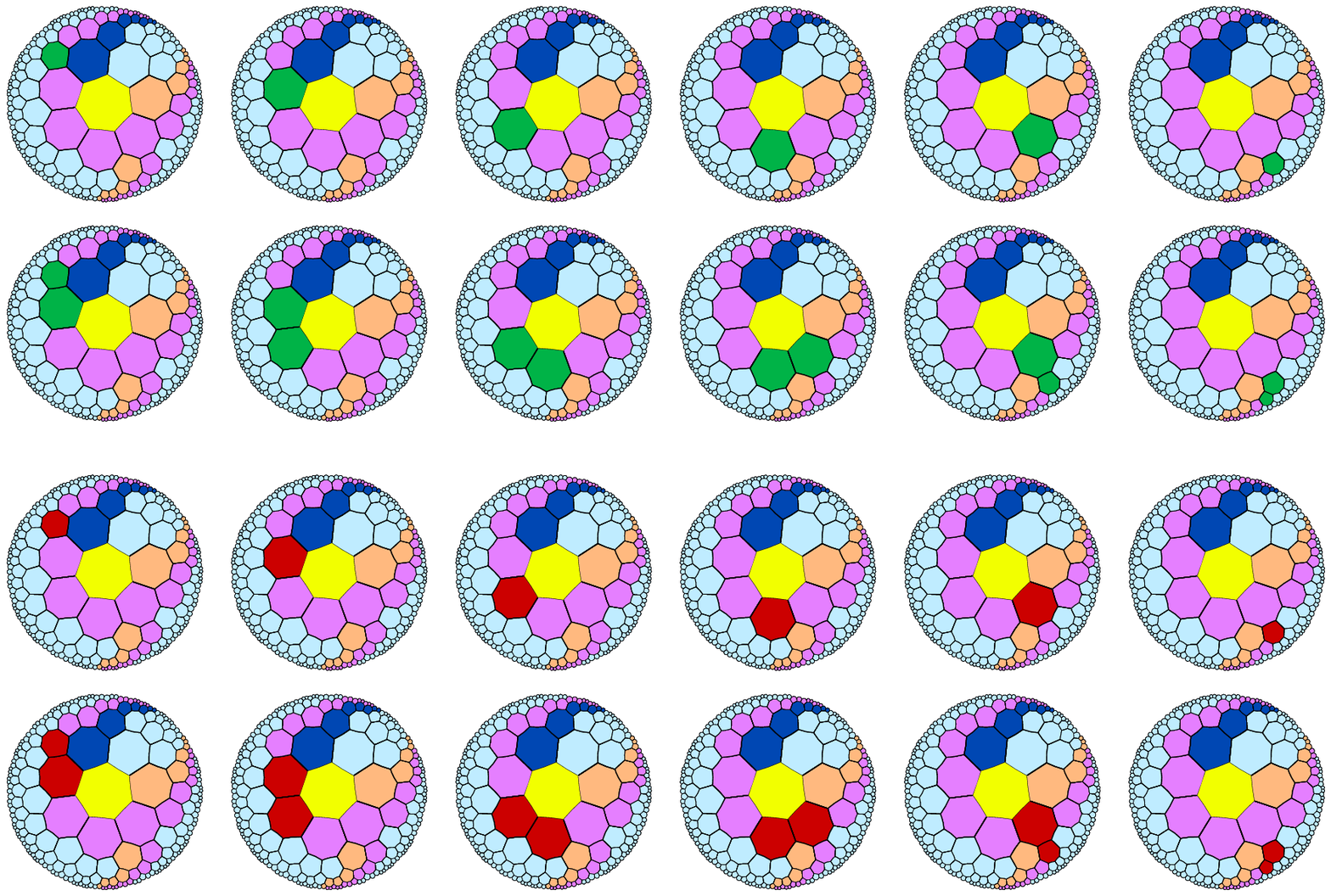}
\hfill}
\vspace{-20pt}
\begin{fig}\label{fFxgpath}
\leurre
Locomotives crossing a passive fixed switch through its left-hand side track.
From top to bottom: simple green one, double green one, simple red one, double red one. 
Note the yellow cell which separates the $B$- and the $O$-paths.
\end{fig}
}
\vskip 10pt
Note that the rules allow both a simple and a double locomotive to cross the switch.
From Figure~\ref{froundabout} we know that from~$A$ a double locomotive passively
crosses the fixed switch through its left-hand side branch while if the locomotive comes
to the round-about from~$B$, a double locomotive does the same through the right-hand side
branch of the switch. Also, those locomotives may be either green or red.

The passive fixed switch occurs in the passive memory switch, in the one-bit memory 
and, consequently, in the discriminating structures which allow a returning locomotive
from a register to go back to the right next instruction.
\vskip 10pt
\vtop{
\ligne{\hfill
\includegraphics[scale=0.5]{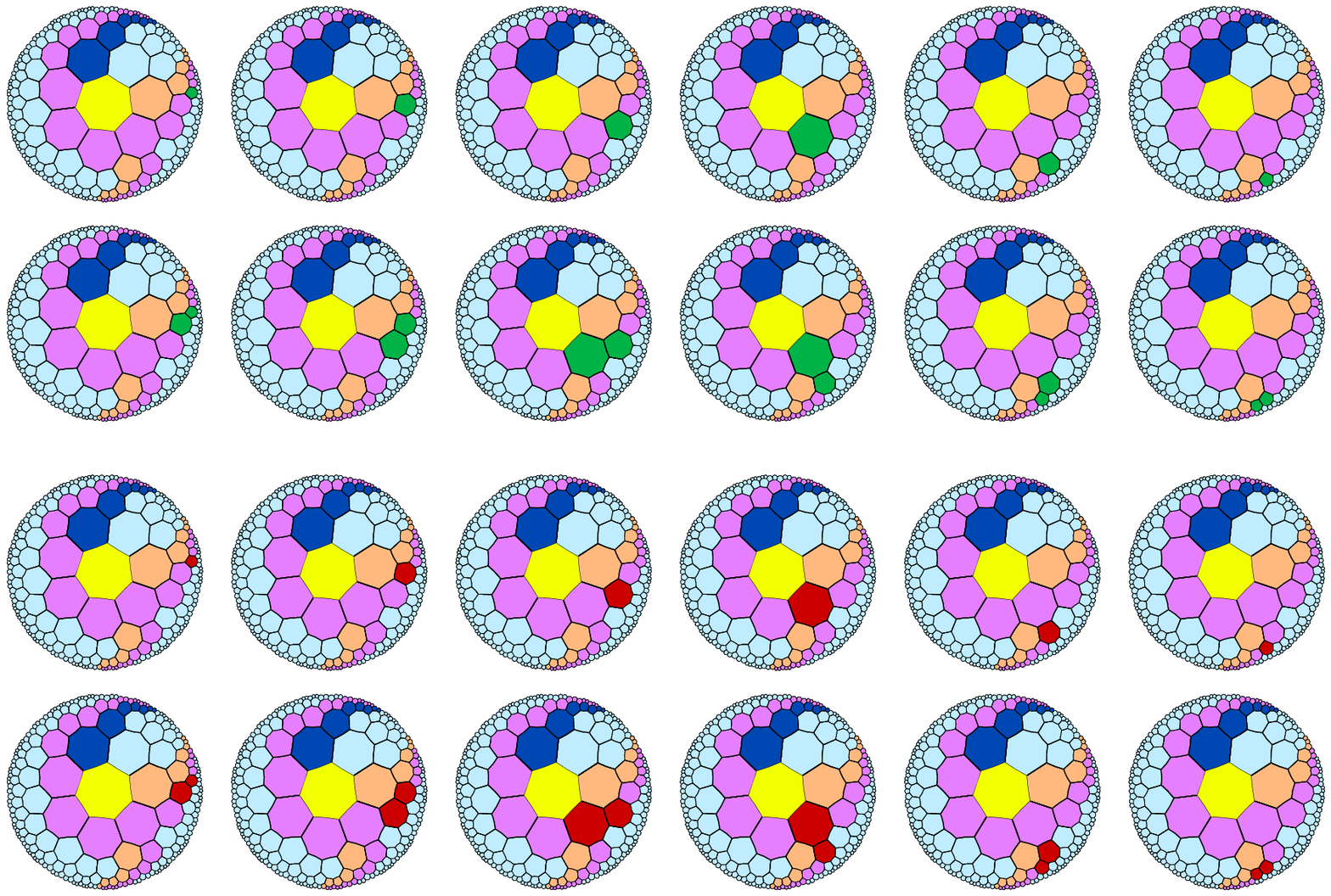}
\hfill}
\vspace{-20pt}
\begin{fig}\label{fFxdpath}
\leurre
Locomotives crossing a passive fixed switch through its right-hand side track.
From top to bottom: simple green one, double green one, simple red one, double red one. 
Note the yellow cell which separates the $B$- and the $O$-paths.
\end{fig}
}
\subsection{Fork and doubler}\label{sbsfrkdbl}

Presently, we deal with the rules managing the fork and the doubler. The rules are given
by Table~\ref{rfrkdbl}. Figure~\ref{fFrkpath} illustrates the application of the rules
managing the fork while Figure~\ref{fFdblpath} illustrates the application of those
managing the doubler.

\vtop{
\begin{tab}\label{rfrkdbl}
\leurre
Rules for the fork and for the doubler.
\end{tab}
\vskip-7pt
\ligne{\hfill fork\hfill\hfill doubler\hfill}
\vskip 5pt
\ligne{\hfill conservative and witness rules\hfill}
%  pour mémoire, total partiel 281
\ligne{\hfill
\vtop{\leftskip 0pt\parindent 0pt\hsize=61pt
\affn {M} {WMBOMBM} {M}                %V 142
\affn {M} {WWWBMOM} {M}                %V 147
\affn {W} {WWWOBBB} {W}              
\affRr {21} {W} {WWWWOOO} {W}           %V 21, 244
\affRr {62} {W} {WWWWWBM} {W}            %V 61
%                                           +4
}
\vtop{\leftskip 0pt\parindent 0pt\hsize=70pt
\affRr {60} {W} {WWWW:MMM:} {W}          %V 59
\affRr {64} {B} {WWB:MMMM:} {B}          %V 62
\affn {B} {WO:MMMM:B} {B}              %         +8
\affn {O} {WWO:MMM:B} {O}              %         +6
%                                           +14
}
\hfill\hfill
\vtop{\leftskip 0pt\parindent 0pt\hsize=61pt
\affn {M} {WMBMOBM} {M}
\affn {M} {WWWMOMB} {M}                %V 143-145
\affRr {19} {W} {WWWWBBB} {W}                %V 19
\affn {W} {WWWBBOO} {W}
\affn {W} {WWWMBBB} {W}
%                                           +5        
}
\vtop{\leftskip 0pt\parindent 0pt\hsize=70pt
\affn {B} {W:MMMMM:B} {B}              %         +10
\affn {B} {WB:MMMM:O} {B}              %V 96     +8
\affn {O} {WB:MMMM:O} {O}              %V 96     +8
%                                           +26
%     somme locale : 4+14+5+26 = 49
}
\hfill}
\vskip 5pt
\ligne{\hfill motion rules\hfill}
\ligne{\hfill
\vtop{\leftskip 0pt\parindent 0pt\hsize=61pt   
%\affn {} {M} {W$\ddot {\ftt M}$B$\ddot {\ftt M}$OB$\dot {\ftt M}$} {M}
\affn {M} {WMBOMBL} {L}
\affn {L} {WMBOMBM} {M}                %V 124
\affn {M} {WLBOLBM} {M}
\affn {L} {WMBOMBL} {L}
\affn {L} {WLBOLBM} {M}
%                                                     +10
}
\vtop{\leftskip 0pt\parindent 0pt\hsize=61pt
\affn {M} {WWWBLOM} {L}
\affn {L} {WWWBMOM} {M}                %V 125
\affn {M} {WWWBMOL} {M}
\affn {L} {WWWBLOM} {L}
\affn {L} {WWWBMOL} {M}
%                                                     +10
}
\hfill\hfill
\vtop{\leftskip 0pt\parindent 0pt\hsize=61pt
\affn {M} {WMBMOBL} {L}
\affn {L} {WMBLOBM} {L}
\affn {L} {WLBMOBM} {M}
\affn {M} {WLBMOBM} {M}
%                                                     +8
}
\vtop{\leftskip 0pt\parindent 0pt\hsize=61pt
\affn {M} {WWWLOMB} {L}                
\affn {L} {WWWMOLB} {M}                %
\affn {M} {WWWMOLB} {M}                %
%                                                     +6
}
%                                           +34
\hfill}
%  total partiel 281+49+34 = 364
}
\vskip 10pt
Note that the fork can be crossed by a double locomotive: that happens when a double 
locomotive arrives to the selector in a round-about. Contrarily to that circumstance,
the doubler receives only simple locomotives, however two of them at the same time.
The synchronization explained in Sub-subsection~\ref{double} is illustrated on 
Figure~\ref{fFdblpath}. by the fact that the green locomotive is, on each side,
at the same distance from the cell where the double locomotive is formed.
\vskip 10pt
\vtop{
\ligne{\hfill
\includegraphics[scale=0.5]{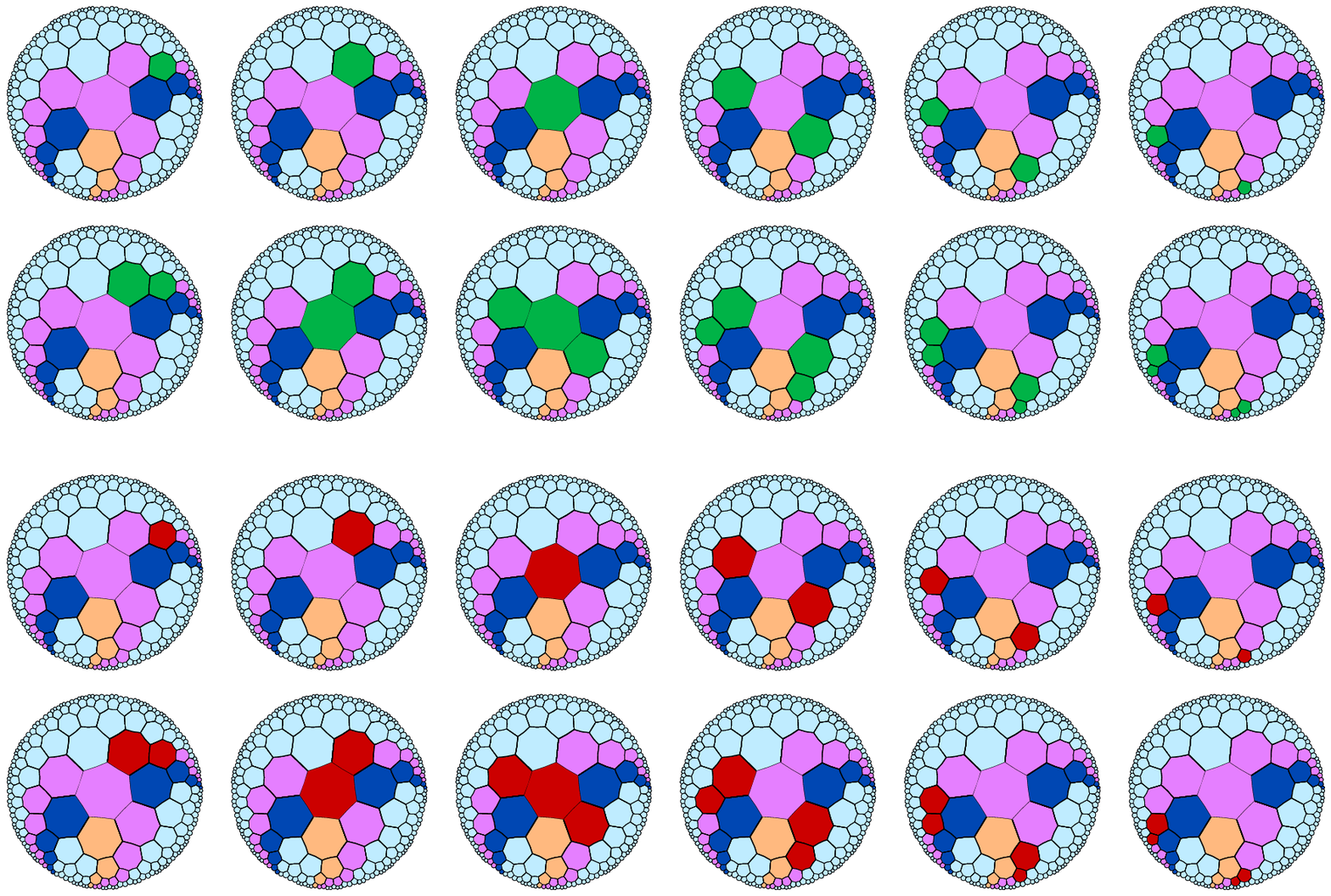}
\hfill}
\vspace{-20pt}
\begin{fig}\label{fFrkpath}
\leurre
Locomotives crossing a fork.
From top to bottom: simple green one, double green one, simple red one, double red one. 
\end{fig}
}

\vtop{
\ligne{\hfill
\includegraphics[scale=0.5]{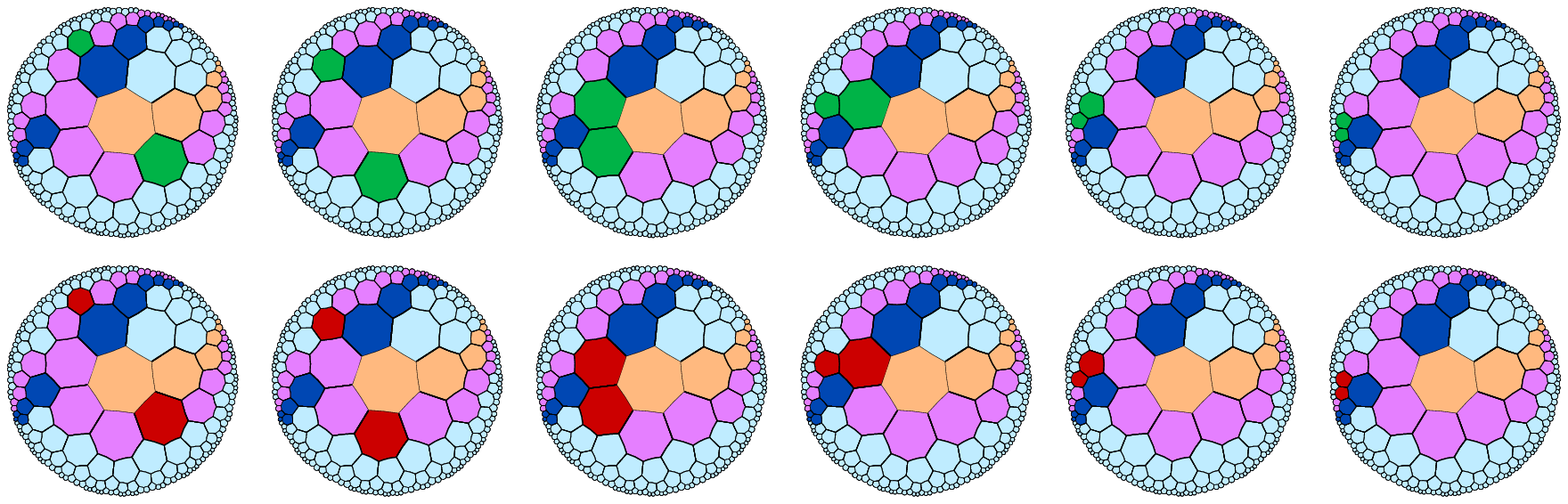}
\hfill}
\vspace{-10pt}
\begin{fig}\label{fFdblpath}
\leurre
Locomotives crossing a doubler.
Top row: a simple green locomotive becomes a double one. Bottom row: a simple red
locomotive becomes a double one.
\end{fig}
}

\subsection{Selector}\label{sbsrsel}

Here, we examine the rules managing the crossing of the selectors. As explained in
Sub-subsection~\ref{select}, a fork distributes the arriving locomotive into
two copies. The latter ones are sent to controllers which work on opposite ways. Both of 
them are fixed. One selector let a simple locomotive go and kills a double locomotive. 
The other selector performs the opposite action: it let a double locomotive go and it 
kills a simple locomotive. Accordingly, a locomotive arriving to the fork leading to the 
controllers eventually goes on its way on a single track, as required in our scenario.

\vtop{
\begin{tab}\label{rsel}
\leurre
Rules managing the controllers of the selector. We remind the reader that a meta-rule
indicating a window replaces the rules with a locomotive at the different places it 
occupies, whether it is green, red, simple or double.
\end{tab}
\vskip-7pt
\ligne{\hfill\hskip 30pt blue controller\hfill both controllers\hfill mauve controller
\hfill}
\vskip 5pt
\ligne{\hfill conservative and witness rules\hfill}
%  pour mémoire, total partiel 364
\ligne{\hfill
\vtop{\leftskip 0pt\parindent 0pt\hsize=70pt
\affn {W} {WWYBB:MM:} {W}              %         +4
\affn {B} {WYYYYBB} {B}
\affn {B} {BBBYYYY} {B}
\affn {Y} {WWWWWYB} {Y}
\affn {Y} {WWWWYBY} {Y}                %V 239
\affRr {183} {Y} {WWWYBBY} {Y}           %V 219, 250, 271
\affn {Y} {WWWWBBY} {Y}
\affn {B} {WBBB:MMM:} {B}              %V 169    +6
\affn {B} {W:MMM:BBY} {B}         %V 168    +6
%                                           +22
}
\hfill
\vtop{\leftskip 0pt\parindent 0pt\hsize=70pt
\affn {M} {WMBWMOO} {M}                %V 175
\affRr {46} {M} {WBMWOOM} {M}
\affn {W} {WWW:MM:BY} {W}              %         +4
\affRr {68} {W} {WWWW:MM:O} {W}          %         +4
\affn {O} {WWWWO:MM:} {O}              %V 50     +4
%                                           +9
}
\hfill
\vtop{\leftskip 0pt\parindent 0pt\hsize=70pt
\affn {M} {WYYYYMB} {M}
\affn {M} {BBMYYYY} {M}
\affRr {194} {Y} {WWWWWYM} {Y}
\affn {Y} {WWWWYMY} {Y}
\affn {Y} {WWWYMMY} {Y}
\affn {W} {WWYMB:MM:} {W}              %         +4
\affRr {198} {B} {W:MMM:BMY} {B}              %V 190    +6
\affRr {199} {B} {WMMB:MMM:} {B}              %V 191    +6
%                                           +21
%  somme locale : +22+9+21 = +52
}
\hfill}
\vskip 5pt
\ligne{\hfill motion rules\hfill}
\ligne{\hfill
\vtop{\leftskip 0pt\parindent 0pt\hsize=70pt
\affn {B} {WLLMBBY} {W}                 
\affn {W} {WMLLBBY} {B}
\affn {B} {WBBBMLM} {W}
\affn {W} {WBBBMMM} {B}
%                                                     +7
%
}
\hfill
\vtop{\leftskip 0pt\parindent 0pt\hsize=61pt
\affn {M} {WLOOWMB} {L}                % order changed as WL < WM 
\affn {L} {WMBWMOO} {M}                %V 155
\affn {M} {WLBWMOO} {M}
\affn {L} {WLOOWMB} {L}
\affn {L} {WLBWMOO} {M}
%                                                     +10
}
\vtop{\leftskip 0pt\parindent 0pt\hsize=61pt   %% renuméroter à partir d'ici
\affRr {71} {M} {WBLWOOM} {L}
\affRr {72} {L} {WBMWOOM} {M}                %V 46, 70
\affRr {73} {M} {WBMWOOL} {M}
\affRr {74} {L} {WBLWOOM} {L}
\affRr {75} {L} {WBMWOOL} {M}
\affn {M} {WWLWOOM} {M}                %** effacement de la locomotive
\affn {L} {WWWWMWM} {M}                %***effacement de la locomotive 
%                                                     +14
}
\hfill
\vtop{\leftskip 0pt\parindent 0pt\hsize=70pt
\affn {B} {WMLMBMY} {W}                % the blocking configuration
\affn {W} {WMMLBMY} {B}                % for a single locomotive
\affn {B} {WLMMWMM} {W}                %
\affn {W} {WMMBMLM} {B}                % 
\affRr {198} {B} {W:MMM:BMY} {B}
\affRr {199} {B} {WMMB:MMM:} {B}
%                                                     +8
%  somme locale : 7+10+14+8 = 39
}
\hfill}
%  total partiel : 364+52+39 = 455
}
\vskip 10pt
\vtop{
\ligne{\hfill
\includegraphics[scale=0.5]{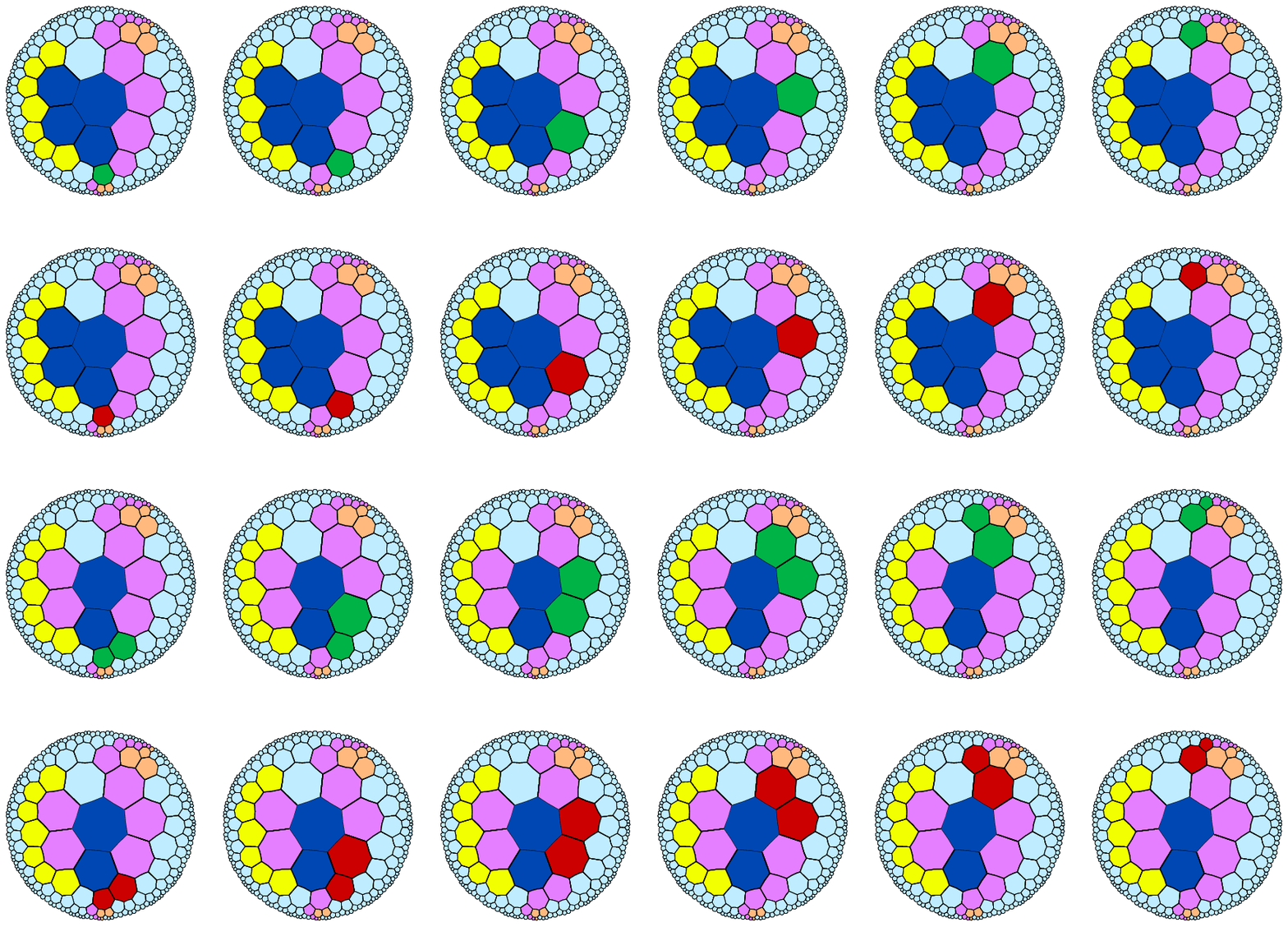}
\hfill}
\vspace{-10pt}
\begin{fig}\label{fSlokpath}
\leurre
Locomotives crossing the control of a selector when the control is open.
From top to bottom: simple green one, simple red one, double green one, double red one. 
Note that the control which let a double locomotive go has not the same configuration
than the one which let a simple locomotive go.
\end{fig}
}
\vskip 10pt
Note that in the table, the meta-rules are replaced by rules when the witness cell is
changed by the passage of the locomotive, see rules 182 up to 185 for the blue controller
and rules 198 up to 201 for the mauve controller. On the table, rule 186 shows us
the importance of the order taken to choose the representing instruction among its
rotated forms. It is different from the order we can see on rule 187. The reason is
that \hbox{\ftt WL $<$ WM} whatever the value of~{\ftt L}.

Figures~\ref{fSlokpath} and~\ref{fSlblpath} illustrate the application of the rules
given by Table~\ref{rsel}. We can check the key roles of the just mentioned rules
together with rules 196 and 197. Figure~\ref{fSlokpath} gathers the configurations when
both controllers let the locomotive go: this is why on that figure, the top two rows
show us the blue controller letting a simple locomotive go, whether it is green or red
while the bottom two rows show us the mauve controller letting a double locomotive go.

\vskip 10pt
\vtop{
\ligne{\hfill
\includegraphics[scale=0.5]{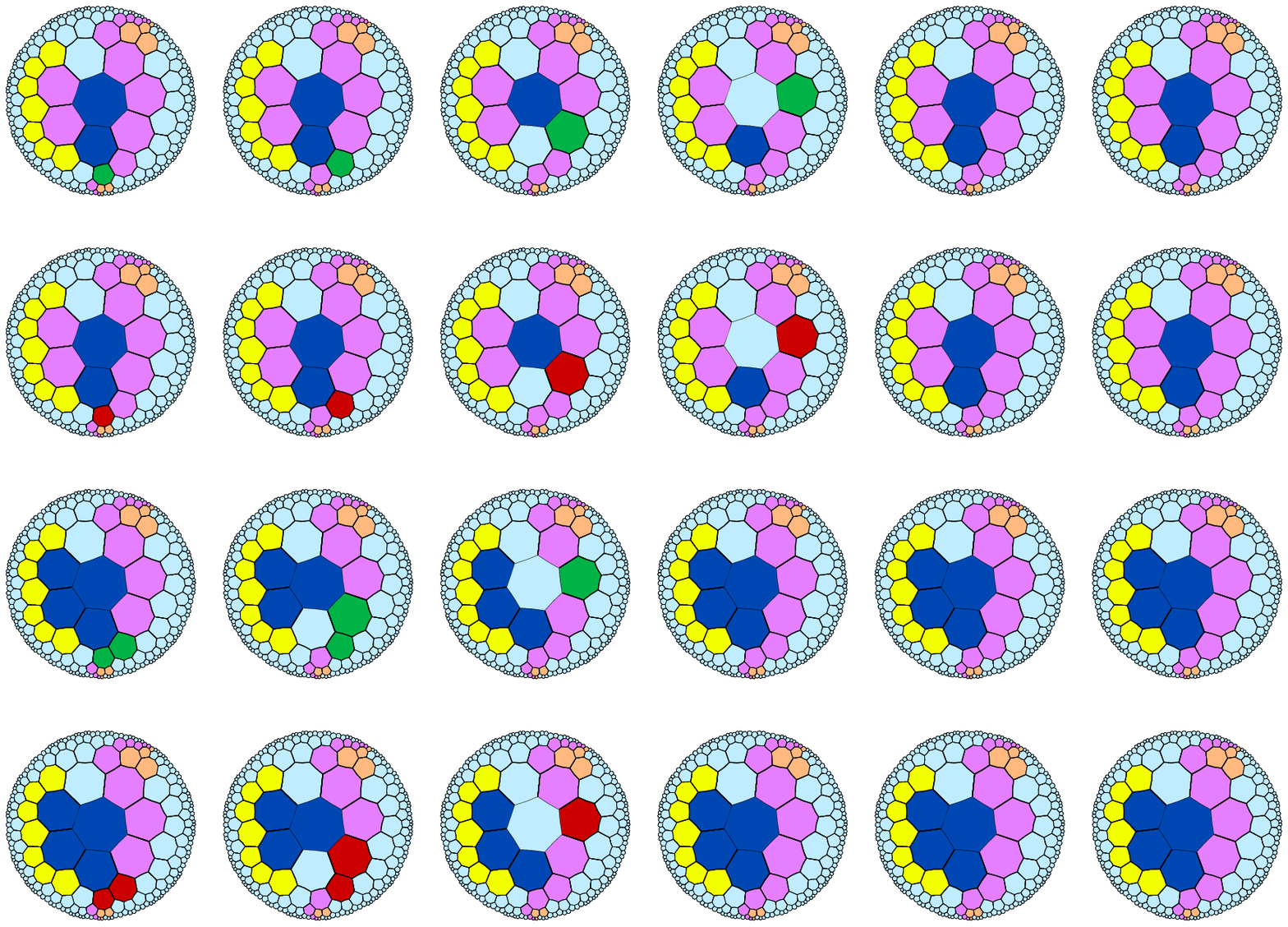}
\hfill}
\vspace{-10pt}
\begin{fig}\label{fSlblpath}
\leurre
Locomotives crossing the control of a selector when the control is closed.
From top to bottom: simple green one, simple red one, double green one, double red one. 
Note that the control which kills a double locomotive has not the same configuration
than the one which kills a simple locomotive. Comparing with 
Figure~{\rm{\ref{fSlokpath}}},
we can see that the controller which kills a double locomotive is the same as the one
which let a single locomotive go and that the controller which let a double locomotive go
kills a simple locomotive.
\end{fig}
}
\vskip 10pt
Figure~\ref{fSlblpath} shows the opposite working of the controllers, when they kill
the locomotive: a double one is killed by a blue controller, see the bottom two rows
of the figure while a simple locomotive is killed a mauve controller, see the top two
rows of the figure.
\vskip 10pt

We conclude the present subsection with the rules allowing a double locomotive to be
changed into a simple one. We remind the reader that Figure~\ref{selctrl} indicates
the need of such a structure as far as, according to what we just reminded the reader
about the controllers of a selector, the controller which let a double locomotive 
go on its way does not change it. Table~\ref{rchds} gives the few rules which manage
that structure together with the motion of the double locomotive when it crosses it.
Note that rule~213 is the key one which transforms the double locomotive into a simple
one. Figure~\ref{fchds} shows us the application of the rules to the configuration
which we have seen on the rightmost picture of Figure~\ref{selctrl}.

\vskip 10pt
\vtop{
\ligne{\hfill
\includegraphics[scale=0.5]{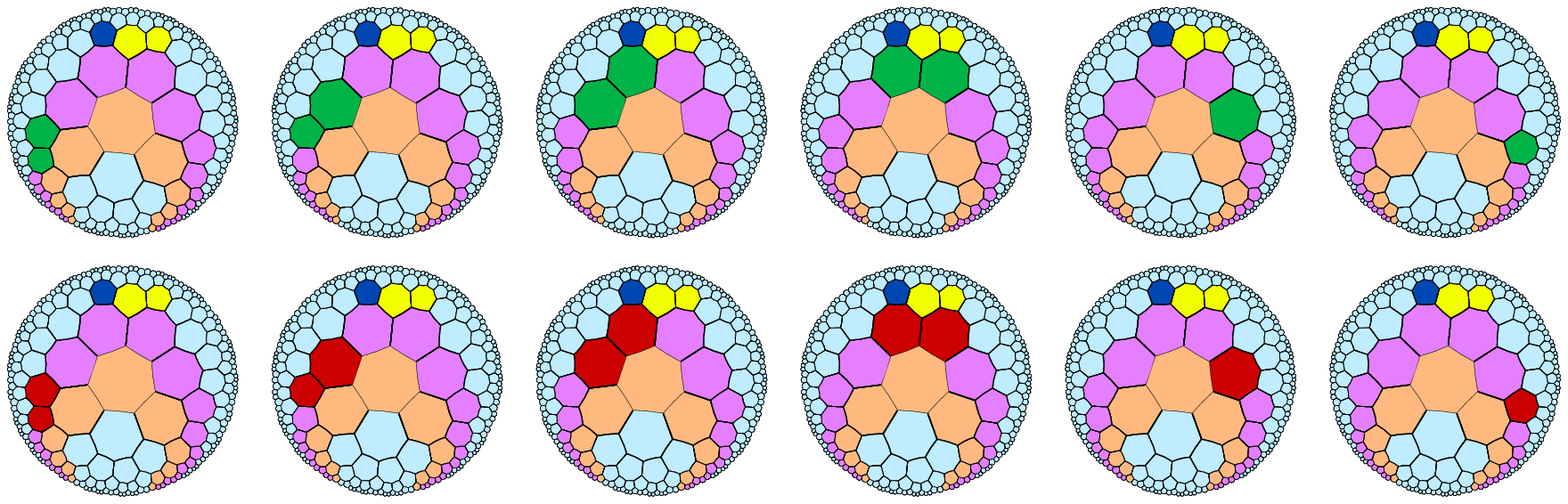}
\hfill}
\vspace{-10pt}
\begin{fig}\label{fchds}
\leurre
Top row: a green double locomotive crosses the structure; bottom row: a red double
locomotive does the same. In both cases, a simple locomotive with the same colour 
leaves the structure.
\end{fig}
}

\vtop{
\begin{tab}\label{rchds}
\leurre
Rules managing the conversion of a double locomotive into a simple one. 
\end{tab}
\vskip-7pt
\ligne{\hfill conservative and witness rules\hfill\hfill motion rules\hfill}
\ligne{\hfill
%  pour mémoire, somme totale : 455
\vtop{\leftskip 0pt\parindent 0pt\hsize=61pt  
\affn {B} {WWWWWMY} {B}                %
\affn {Y} {WWWB:MM:Y} {Y}              %         +4
\affRr {194} {Y} {WWWWWYM} {Y}
\affn {M} {WWMOMYB} {M}
\affn {M} {WWYYMOM} {M}
%                                                     +7
}
\hfill
\vtop{\leftskip 0pt\parindent 0pt\hsize=61pt  
\affn {M} {WWLOMYB} {L}
\affn {L} {WWLOMYB} {L}
\affn {L} {WWMOLYB} {M}
\affn {M} {WWYYLOM} {L}
\affn {L} {WWYYLOM} {M}                % stops the second cell of the double loco
%                                                     +10
}
\hfill}
%  somme totale : 455+11+10 = 476
}

\subsection{Controllers in the flip-flop and the memory switches}\label{sbsflflmemo}

In the present sub-section, we consider the rules managing the controllers of the
flip-flop switch and of the memory switch. Both switches use the same controllers but
their place in the circuit with respect to the tracks they control is opposite.
While the controllers of a selector are fixed, the controllers of those switches are
in some sense programmable. The blue controller may become mauve and the mauve one
may become blue. Also, those controllers can see a simple locomotive only. When
a controller is blue, it let the locomotive go, when it is mauve, it kills the locomotive.
On Figure~\ref{fCtrlf}, we can see that the controllers under study are rather close
to those we have studied for the selector. It is the reason why in Table~\ref{rctrl}
we do not reproduce a few conservative rules for yellow cells which have the same
neighbourhood here and in that former case as, for instance rules~163 and~164.

In Table~\ref{rctrl}, the key rules are rules~229 and~230 on one side, changing the blue
controller into a mauve one, and rules~233 and~234 on another side, changing the mauve
controller into a blue one. 

\vtop{
\begin{tab}\label{rctrl}
\leurre
Rules which manage the change of colour in the controllers of the flip-flop switch and 
of the memory one.
\end{tab}
\vskip -7pt
\ligne{\hfill blue to mauve\hfill mauve to blue\hfill}
\vskip 5pt
\ligne{\hfill conservative and witness rules\hfill}
\ligne{\hfill
%  pour mémoire, somme totale : 476
\vtop{\leftskip 0pt\parindent 0pt\hsize=61pt
\affn {B} {WYYYMBB} {B}
\affn {B} {BBBMYYY} {B}
\affn {M} {WMOYBBY} {M}                %V 217
%                                                     +3
}
\hfill
\vtop{\leftskip 0pt\parindent 0pt\hsize=70pt
\affn {Y} {WWWWMBY} {Y}
\affn {Y} {WWWYBMO} {Y}
\affn {W} {WWWW:MM:Y} {W}         %V 213, 215    +4
\affRr {238} {W} {WWWWWYO} {W}           %V 216, 2..
\affRr {118} {O} {WWY:MMM:O} {O}          %         
%                                                     +7
}
\hfill\hfill
\vtop{\leftskip 0pt\parindent 0pt\hsize=61pt
\affn {M} {WYYYMMB} {M}
\affn {M} {BBMMYYY} {M}
\affn {M} {WMOYMMY} {M}
%                                                     +3
}
\hfill
\vtop{\leftskip 0pt\parindent 0pt\hsize=61pt
\affn {Y} {WWWWMMY} {Y}
\affn {Y} {WWWYMMO} {Y}
%                                                     +2
}
%                                           +15
\hfill}
\vskip 5pt
\ligne{\hfill motion rules\hfill}
\ligne{\hfill
\vtop{\leftskip 0pt\parindent 0pt\hsize=61pt
\affn {M} {WLOYBBY} {L}
\affn {L} {WMOYBBY} {M}
\affn {B} {WYYYLBB} {M}
\affn {B} {BBBLYYY} {M}
%                                                     +8
}
\hfill\hfill
\vtop{\leftskip 0pt\parindent 0pt\hsize=61pt
\affn {M} {WLOYMMY} {L}
\affn {L} {WMOYMMY} {M}
\affn {M} {WYYYLMB} {B}
\affn {M} {BBMLYYY} {B}
%                                                     +8
}
\hfill}
%                                           +16
}
%  somme totale : 476+15+16 = 507
\vskip 10pt
\vtop{
\ligne{\hfill
\includegraphics[scale=0.5]{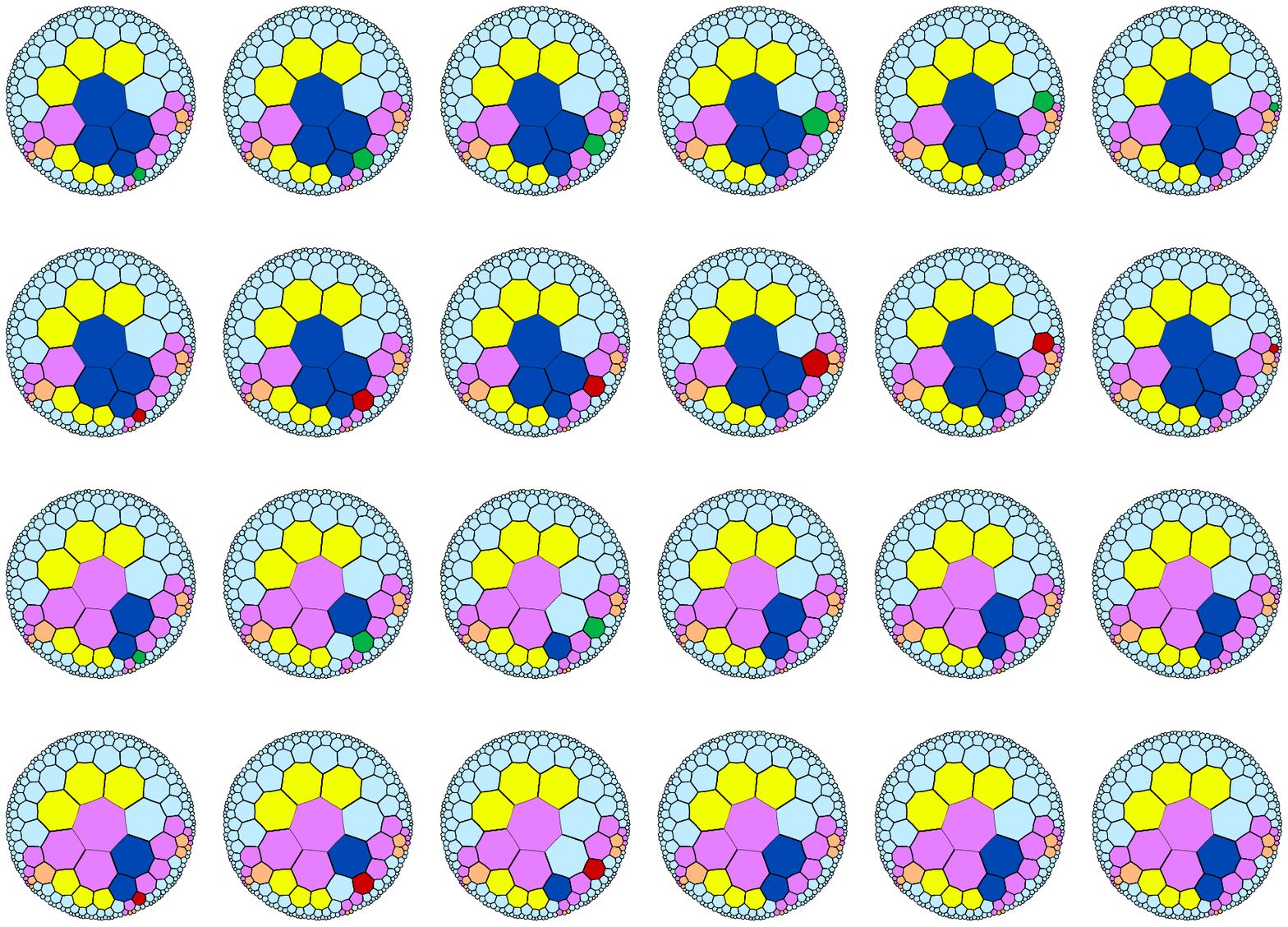}
\hfill}
\vspace{-10pt}
\begin{fig}\label{fCtrlf}
\leurre
Locomotives crossing the control of a passive memory switch when it let the 
locomotives go.
From top to bottom: simple green one, simple red one, double green one, double red one. 
Note that the control works as that of the selector. However, its decorations are
not the same and a path is getting out from the controller. See the use of that path
on Figure~{\rm\ref{fCtrlch}}.
\end{fig}
}

\vskip 10pt
As the working of those controllers are the same as those of a selector, Table~\ref{rctrl}
does not remind us the rules associated with the motion of the locomotive on the ordinary
tracks. The table gives the rules connected with the change of the colour only.
First the conservative and witness rules which are specific to those controllers.
Then, it gives the rules for the motion of the signal-locomotive and also the rules
managing the change of colour of the controllers.

Figure~\ref{fCtrlf} illustrates the working of the controllers. Figure~\ref{fCtrlch}
illustrates how the change of colour is performed.
\vskip 10pt
\vtop{
\ligne{\hfill
\includegraphics[scale=0.5]{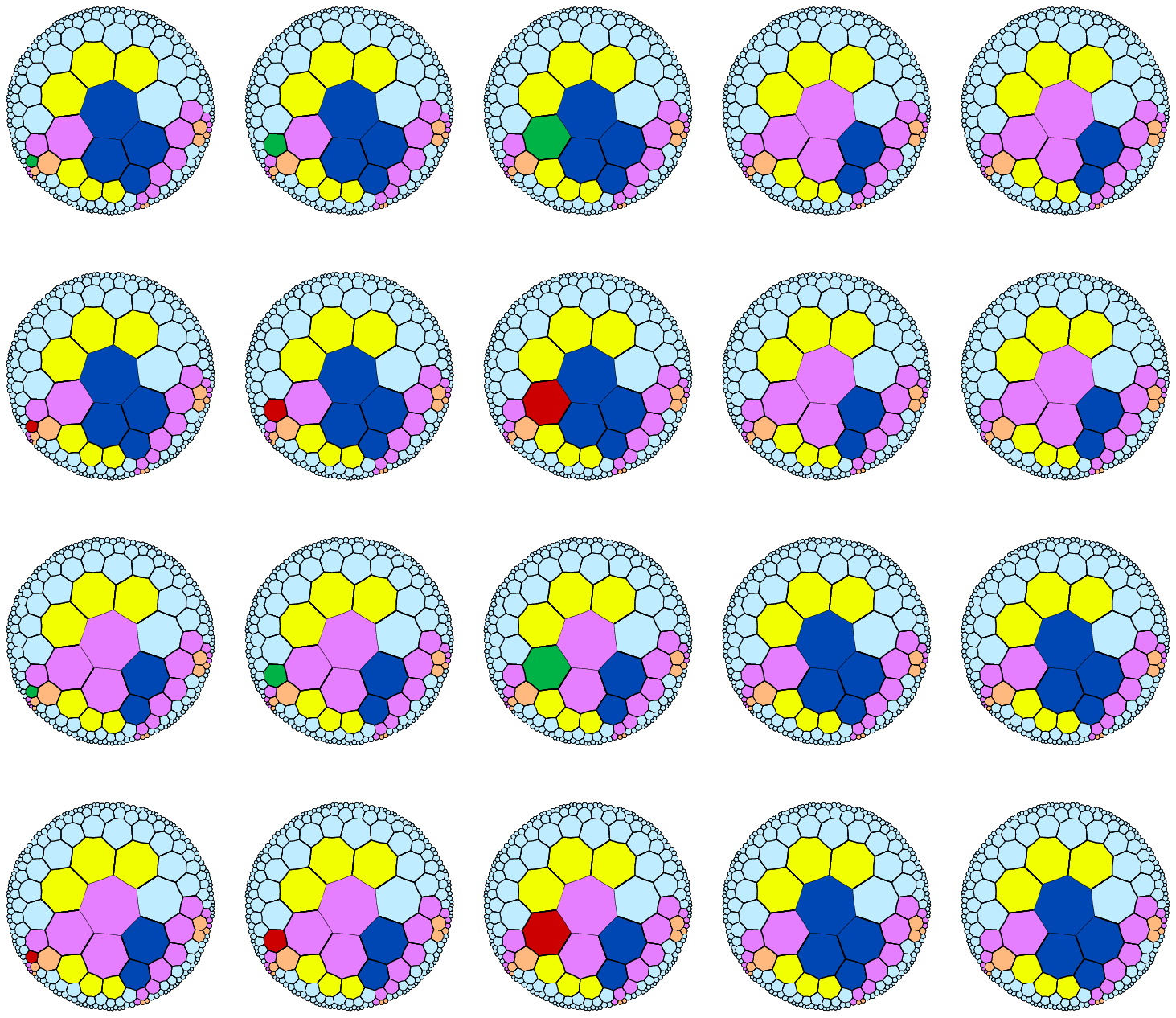}
\hfill}
\vspace{-10pt}
\begin{fig}\label{fCtrlch}
\leurre
Signal-locomotives arriving at the control of a passive memory switch.
From top to bottom: green signal to the blue control, red one to the same control, 
green signal to the mauve control, red one to that control.  
Compare with the working of the control on Figure~{\rm\ref{fCtrlf}}.
\end{fig}
}

\subsection{Registers}\label{sbsreg}

We give the rules for the registers in the present sub-section. We distinguish three
main cases: when a locomotive arrives and performs an instruction, the case when 
the arriving locomotive finds out an empty register, the case when the register contains
value~1 and the locomotive arrives to perform an instruction. As we have many rules,
we split the table into two ones: a small table, Table~\ref{rregcw}, for the 
conservative and for the witness rules, a much bigger one, Table~\ref{rregmr}, for the 
motion rules.

\vtop{
\begin{tab}\label{rregcw}
\leurre
Conservative and witness rules for a register.
\end{tab}
\vskip-7pt
%\ligne{\hfill action\hfill finding~0\hfill finding~1\hfill}
%\vskip 5pt
%\ligne{\hfill conservative and witness rules\hfill}
\ligne{\hfill
\vtop{\leftskip 0pt\parindent 0pt\hsize=61pt
% pour mémoire, somme totale : 507
\affn {W} {WWWWWYY} {W}                %V 272
\affn {W} {WWWWWOY} {W} 
\affn {W} {WWWWWOO} {W} 
\affn {B} {BYYBOOO} {B} 
\affn {B} {BYYOOOO} {B} 
\affRr {183} {Y} {WWWYBBY} {Y} 
\affRr {259} {Y} {WWWWOBY} {Y}           %V 250, 256, 263, 276
\affn {O} {WWWWOBO} {O}                %V 252, 295
\affn {O} {WWWOBBO} {O}
%                                                 +8
}
\hfill
\vtop{\leftskip 0pt\parindent 0pt\hsize=70pt % registre à 0
\affn {W} {WWWWBOO} {W}
\affRr {263} {W} {WWWWWMY} {W}
\affn {W} {WWWWYYB} {W}
\affn {W} {WWWWYOO} {W}
\affn {W} {WWWMYBY} {W}
\affn {W} {WWWWOYO} {W}
\affn {B} {WWWWWYO} {B}                %V 228
\affn {B} {WWY:MMM:B} {B}              %         +6
\affn {B} {WY:MM:YOY} {B}              %         +4
\affn {Y} {WWWWMOB} {Y}
\affn {Y} {WYMMOMB} {Y}                
%                                                +19
}
\vtop{\leftskip 0pt\parindent 0pt\hsize=70pt
\affn {Y} {WWWW:MM:Y} {Y}              %         +4
\affn {Y} {WWY:MMM:O} {Y}              %         +6
\affn {Y} {WWWOBMY} {Y}
\affn {Y} {WOMBWMO} {Y}                %V 302, 323
\affn {Y} {WWWWWBO} {Y}
\affn {O} {WWWWOMY} {O}                %V
\affn {O} {WWWWOMO} {O}                %V 332
\affn {O} {WBYMYMO} {O}                %V 397
\affn {O} {WY:MMMM:O} {O}              %         +4  R seulement
\affn {O} {WWWWYBY} {O}                %V 150
\affn {O} {WWWY:MM:O} {O}              %         +4
%                                                +25
}
\hfill
\vtop{\leftskip 0pt\parindent 0pt\hsize=70pt
\affRr {21} {W} {WWWWOOO} {W}
\affRr {238} {W} {WWWWWYO} {W}
\affn {B} {:YY:OOOOM} {B}              %         +3 car une règle change B
\affn {Y} {WMOYBMB} {Y}                %V 349, 370
\affn {Y} {WWWOBYO} {Y}                %V 309, 355, 375
\affn {O} {WWYYMMO} {O}
\affn {O} {WWWWOBY} {O}
\affn {O} {WWWOMBO} {O}
%                                                +8
}
\hfill}
% somme totale : 507+8+19+25+8              +567
}

\vtop{
\begin{tab}\label{rregmr}
\leurre
Motion rules for a register. First, to increment or decrement the register.
Then, the particular cases: when the register contains~$0$ and then when it contains~$1$.
\end{tab}
\vskip -7pt
\ligne{\hfill incrementing\hfill\hfill decrementing\hfill}
% pour mémoire, somme totale : 567
\ligne{\hfill
\vtop{\leftskip 0pt\parindent 0pt\hsize=61pt  % incrémentation
\affn {Y} {WWWYBBG} {G}
\affn {G} {WWWYBBY} {Y}                %V 151, 229
\affn {Y} {WWWGBBY} {Y}
\affRr {313} {O} {WWWWOBR} {R}           %V 282
\affn {R} {WWWWOBO} {O}                %V 221, 295
\affn {O} {WWWWRBO} {O}
\affn {O} {WWWOBBR} {R}
\affRr {299} {R} {WWWOBBO} {O}
\affn {O} {WWWRBBO} {O}
\affn {Y} {WWWWOBG} {G}                %V 257
\affn {G} {WWWWOBY} {G}                %V 268
\affn {W} {1WWWWOG} {R}
\affRr {304} {W} {WWWWWOB} {O}
%                                                +13
}
\hfill
\vtop{\leftskip 0pt\parindent 0pt\hsize=61pt  % incrémentation
\affRr {305} {O} {WWWWOBG} {B}           %V 356
\affRr {306} {B} {WWWOBGR} {R}           %V 359
\affn {G} {WWWRBBY} {Y}
\affRr {308} {R} {WWWWWBG} {Y}                %V 360
\affn {Y} {WWWYRBY} {Y}
\affn {Y} {WWWWWRY} {Y}               %V 272**
\affRr {311} {W} {WWWWWRY} {O}                %V 271, 362
\affRr {312} {R} {WWOOBYY} {B}
\affRr {313} {O} {WWWOBRO} {R}                %V 371
\affn {B} {WORBYYO} {B}
\affRr {259} {Y} {WWWWOBY} {Y}
%                                                +10
%                                           +23
}
\hfill\hfill
\vtop{\leftskip 0pt\parindent 0pt\hsize=61pt  % décrémentation
\affn {Y} {WWWYBBR} {R}
\affn {R} {WWWYBBY} {Y}                %V 151, 219, 250
\affn {Y} {WWRBBYO} {Y}                %V 361  303 a créé O au temps précédent
\affn {O} {WWWWWYY} {W}                %V 224
\affn {B} {BYROOOO} {G}
\affn {R} {WWWWOBY} {W}
\affRr {313} {O} {WWWWOBR} {R}
\affn {B} {BYYGOOO} {B}
\affRr {306} {R} {WWWWWOG} {O}
\affn {G} {WROOOBY} {G}
\affRr {308} {O} {WWWWOGR} {R}
\affn {O} {WWWWWWY} {W}                %V 10
%                                                +11
}
\hfill
\vtop{\leftskip 0pt\parindent 0pt\hsize=61pt  
\affn {O} {WWWWOGO} {O}                %V 384
\affn {O} {WWWOBGO} {O}                %V 350
\affn {Y} {WWWWGBY} {Y}
\affn {G} {WOROOBY} {G}
\affRr {332} {O} {WWWWWRG} {W}               
\affRr {333} {W} {WWWWWOG} {R}
\affn {G} {WWOROBY} {G}
\affn {O} {WWWOBGR} {R}
\affRr {336} {R} {WWWWWWG} {W}           %V 11
\affn {G} {WRWORBY} {O}                % fin de la décrémentation
%                                                +8
%                                           +19
}
\hfill}
%  somme partielle 13+23+19                 +55
\vskip 5pt
\ligne{\hskip 60pt 0 $\rightarrow$ 1\hfill\hfill testing 0\hskip 80pt}
\ligne{\hfill
\vtop{\leftskip 0pt\parindent 0pt\hsize=61pt  
\affn {M} {WWWWGOY} {G}
\affn {Y} {WGOWOMB} {B}
\affn {G} {WWWWMOY} {M}
\affn {B} {WBMMYOY} {B}
\affn {B} {WOMBWMO} {Y}                %V 236, 323
\affRr {343} {W} {WWWWOBO} {O}
\affn {M} {BBOOOYM} {B}
\affn {W} {WWWWWBO} {O}
\affRr {304} {W} {WWWWWOB} {O}
%                                           +8
}
\hfill
\vtop{\leftskip 0pt\parindent 0pt\hsize=61pt  
\affn {B} {WYBMYOY} {B}
\affn {B} {BYBOOYM} {B}
\affn {B} {WWWOBYO} {R}               %V 248, 355, 375
\affn {O} {WBYMYBO} {R}               % 'non contagion'
\affn {O} {WWWWBYO} {Y}               % construction d'un Y additionnel
\affRr {352} {W} {WWWWORO} {W}
\affn {B} {BYRORYM} {M}
\affn {R} {WOOBYYO} {B}                % création du 1
\affn {Y} {WMOYRBB} {Y}
\affn {Y} {WWWORYO} {Y}
\affRr {343} {W} {WWWWOBO} {O}               
\affn {B} {WOOMYYO} {B}
%                                           +11
}
\hfill\hfill
\vtop{\leftskip 0pt\parindent 0pt\hsize=61pt  
\affn {M} {WWWWROY} {R}
\affn {R} {WWWWMOY} {M}
\affn {M} {WWWWMOR} {M}
\affn {Y} {WROWOMB} {R}
\affn {R} {WOMBWMO} {Y}                %V 236, 302
\affRr {352} {W} {WWWWORO} {W}
\affRr {365} {B} {WRMMYOY} {B}
\affn {O} {WWWWOMR} {R}
\affn {M} {BROOOYM} {M}
\affn {W} {WWWWRYO} {W}
\affn {Y} {WRMBWMO} {Y}
%                                           +10
}
\hfill
\vtop{\leftskip 0pt\parindent 0pt\hsize=61pt  
\affn {R} {WWWWOMO} {O}                %V 239
\affn {O} {WWWWRMY} {O}
\affRr {372} {O} {WBYMYMR} {R}
\affn {M} {BYOROYM} {R}                % signal pour la locomotive
\affn {R} {WBYMYRO} {O}                % départ comme d'habitude
\affn {R} {BYOORYM} {M}
\affRr {376} {M} {WWMBYRY} {R}                % signal standard
\affn {M} {BRYYMYY} {R}                % départ du signal de zéro
%                                           +8
}
\hfill
\vtop{\leftskip 0pt\parindent 0pt\hsize=61pt  
\affn {R} {BMYYMYY} {M}
\affn {M} {BYOOOYR} {M}
\affn {M} {WWWMYRY} {R}
\affn {M} {WWWMOYR} {R}
\affn {R} {WWWMOYM} {M}
\affn {M} {WWRYMYO} {M}                %  272 a crée un O qui sera gommé après
\affn {O} {WWWWWMY} {W}                %  effacement du O superflu
%                                           +7
}
\hfill}
%   8+11+10+8+7 =                           +44
\vskip 5pt
\ligne{\hfill 1 $\rightarrow$ 2\hfill \hfill\hskip 30pt 1 $\rightarrow$ 0\hfill\hskip 40pt}
\ligne{\hfill
\vtop{\leftskip 0pt\parindent 0pt\hsize=61pt  
\affn {Y} {WGOYBMB} {G}
\affn {B} {YYOOOOM} {B} 
\affn {B} {WGMMYOY} {B} 
\affn {B} {YOOOOMG} {B}                %déjà comptée, notez l'ordre
\affn {G} {WMOYBMB} {Y}                %V 247, 370
\affn {Y} {WWWOBGO} {G}                %V 289
\affn {M} {BGBOOYM} {M}
\affRr {333} {W} {WWWWWOG} {R}           % utile quand même pour loco rouge
\affn {B} {WYMMYOY} {B}
\affn {B} {YGOOOOM} {B} 
\affn {G} {WWWOBYO} {G}                %V 248, 309, 375
\affRr {305} {O} {WWWWOBG} {B}         
%                                           +11
}
\hfill
\vtop{\leftskip 0pt\parindent 0pt\hsize=61pt  
\affRr {304} {W} {WWWWWOB} {O}
\affn {B} {BOOOMYG} {B}
\affRr {305} {B} {WWWOBGR} {R}         
\affRr {308} {R} {WWWWWBG} {Y}            
\affn {G} {WWRBBYO} {Y}                %V 281
\affRr {311} {W} {WWWWWRY} {O}           %V 271
\affn {B} {ROOOMYY} {B}
\affRr {312} {R} {WWOOBYY} {B}
\affRr {313} {O} {WWWOBRO} {R}
\affRr {299} {R} {WWWOBBO} {O}
\affRr {343} {W} {WWWWOBO} {O}           % configuration achevée
%                                           +4
}
\hfill\hfill
\vtop{\leftskip 0pt\parindent 0pt\hsize=61pt  
\affn {Y} {WROYBMB} {R}
\affn {R} {WMOYBMB} {Y}                %V 247, 349
\affRr {365} {B} {WRMMYOY} {B}
\affn {B} {RYOOOOM} {B}
\affn {Y} {WWWOBRO} {R}                %V 274,365
\affn {M} {BRBOOYM} {M}
\affn {O} {WWYRMMO} {O}
\affn {B} {ROOOOMY} {G}                % effacement provisoire du 1
\affn {R} {WWWOBYO} {W}                % effacement des O superflus
\affn {Y} {WMORBMB} {Y}                %V 274,365
\affRr {313} {O} {WWWWOBR} {R}           %V 
%                                           +10
}
\hfill
\vtop{\leftskip 0pt\parindent 0pt\hsize=61pt  
\affn {W} {WWWRGYO} {W}
\affn {G} {WROOOMY} {G}
\affRr {306} {R} {WWWWWOG} {O}
\affn {Y} {WGMBWMO} {Y}
\affRr {308} {O} {WWWWOGR} {R}
\affn {W} {WWWOGYO} {W}
\affn {R} {WWWWOGO} {O}                %V 288
\affn {G} {WOROOMY} {G}
\affRr {332} {O} {WWWWWRG} {W}           %  effacement d'un O
\affRr {308} {O} {WWWWOGR} {R}           %  la loco rouge se déplace
%                                           +6
}
\hfill
\vtop{\leftskip 0pt\parindent 0pt\hsize=61pt  
\affn {W} {WWWWGYO} {W}
\affRr {333} {W} {WWWWWOG} {R}           % ce R sera effacé ensuite
\affn {G} {WWOROMY} {G}
\affRr {332} {O} {WWWWWRG} {W}           % O encore effacé
\affn {O} {WWWOMGR} {R}
\affn {R} {WWWOMGO} {O}                % le O libéré sera effacé par 296
\affRr {336} {R} {WWWWWWG} {W}           %V  11
\affn {G} {WRWORMY} {O}                % le 1 complètement effacé
\affRr {372} {O} {WBYMYMR} {R}
\affn {R} {WBYMYMO} {O}                %V 240
\affn {O} {WBYRYMO} {O}
\affRr {376} {M} {WWMBYRY} {R}
%                                           +9
}
%      11+4+10+6+9 =                   +40
\hfill}
%  somme totale : 567+55+44+40 = 706
}
\vskip 10pt
In Table~\ref{rregmr}, we can see several rules whose number is red according to the 
convention we already mentioned.
%and is not the number
%correesponding of the order of the place of those rules in the table. The reason is 
%that the number in red reminds us that the rule was already used, either in the same table
%but at a previous step or in a previous table. 
As an example, rule~315 transforms a white
cell into a red one if that latter has two contiguous neighbours {\ftt O} and {\ftt G} 
in that order, the other neighbours being white. That rule applies three times and when
the created this way red cell is superfluous, it is returned to white by rule~318.
A few rules play the key role: rule~336 creates the~1 when an empty registered is
visited by a green locomotive. Rule~359 triggers the locomotive which will stop the
returning red locomotive as the former one signals the zero-state of the register and
arrives to the memory of decrementing the register by the way devoted to the zero-test.
At last, Rule~397 starts the process of returning the content of the register to~0
when its content is~1 and at that time, a red locomotive arrives at the register.
\vskip 5pt
\vtop{
\ligne{\hfill
\includegraphics[scale=0.5]{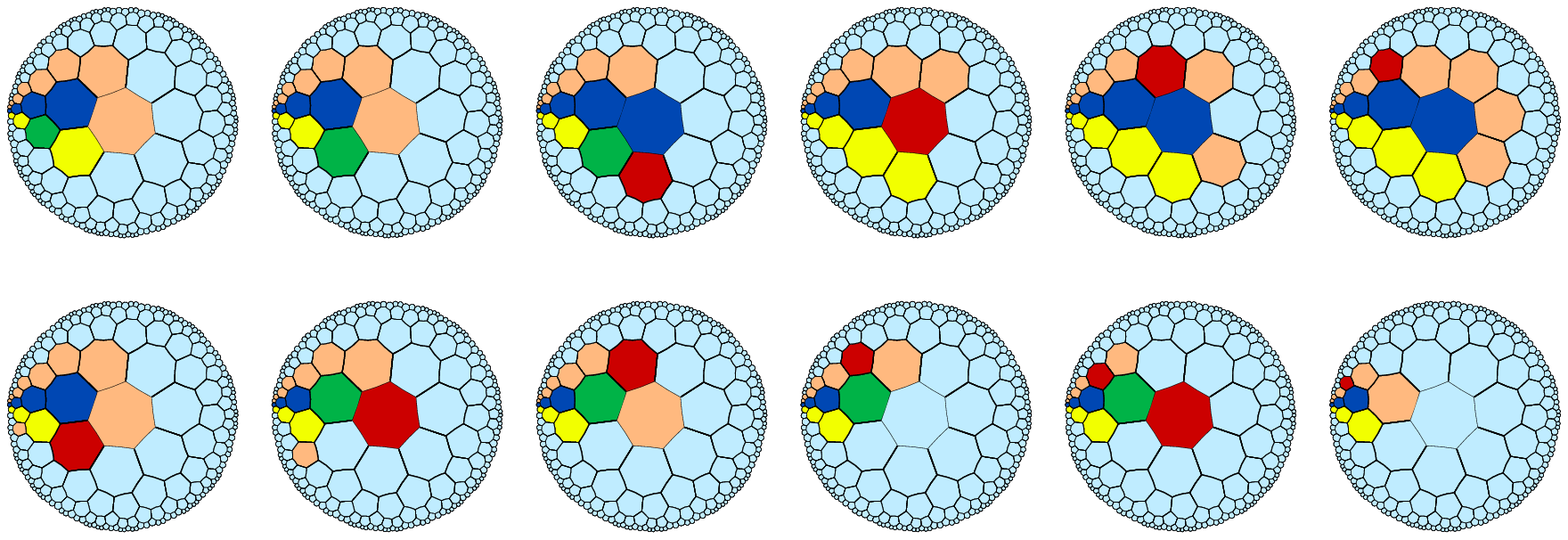}
\hfill}
\vspace{-0pt}
\begin{fig}\label{fRegfn}
\leurre
Working on a register. Top row, a green locomotive arrives: it increments the register.
Bottom row, a red locomotive arrives: it decrements the register. As decrementing a 
register needs one step more than incrementing it, the illustration of decrementing 
starts one step further after the arrival of the red locomotive to the register.
\end{fig}
}

\vtop{
\ligne{\hfill
\includegraphics[scale=0.5]{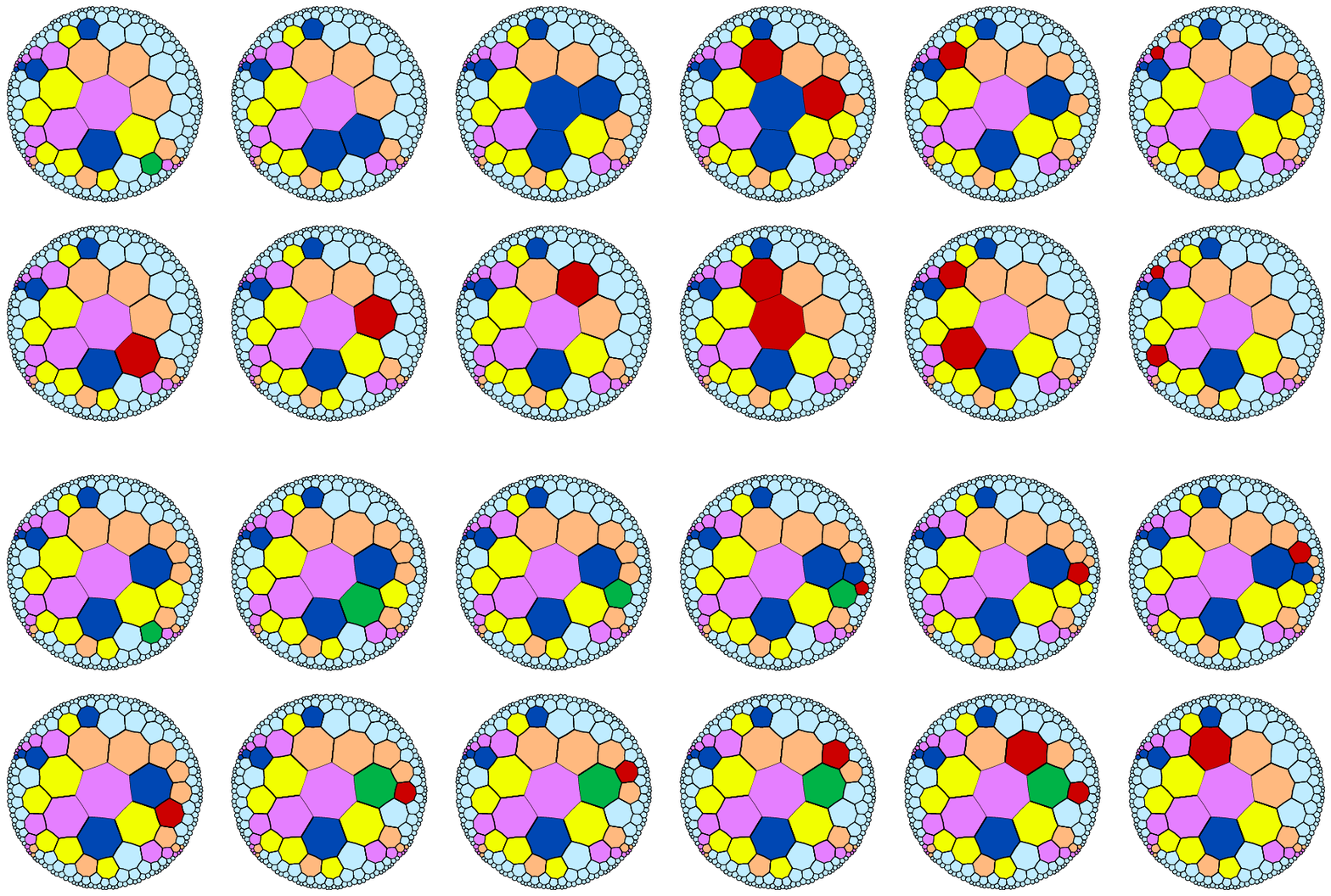}
\hfill}
\vspace{-20pt}
\begin{fig}\label{fRegspe}
\leurre
Working on a register, special cases. First two rows the register is empty, last two rows
it contains value~$1$. Topmost row: a green locomotive arrives; the content of the
register grows from~$0$ to~$1$. Next row: a red locomotive arrives; it can see that the
register is empty, so that it triggers the locomotive exits through a different way.
Third row from the top, a green locomotive arrives: the content of the register grows
from~$1$ to~$2$. Last row, a red locomotive arrives: the content of the register goes
from~$1$ to~$0$. Here too, there is a delay in the progression of the red locomotive
when compared with the arrival of the green one.
\end{fig}
}

\newdimen\largesept\largesept=21.5pt
\def\execneuf #1 #2 #3 #4 #5 #6 #7 #8 #9 {
\hbox{
\hbox to 25pt{\hfill#1\hskip 4pt}
\hbox to \largesept{\hfill#2\hskip 4pt}
\hbox to \largesept{\hfill#3\hskip 4pt}
\hbox to \largesept{\hfill#4\hskip 4pt}
\hbox to \largesept{\hfill#5\hskip 4pt}
\hbox to \largesept{\hfill#6\hskip 4pt}
\hbox to \largesept{\hfill#7\hskip 4pt}
\hbox to \largesept{\hfill#8\hskip 4pt}
\hbox to \largesept{\hfill#9\hskip 4pt}
}
}

\def\exechuit #1 #2 #3 #4 #5 #6 #7 #8 {
\hbox{
\hbox to 25pt{\hfill#1\hskip 4pt}
\hbox to \largesept{\hfill#2\hskip 4pt}
\hbox to \largesept{\hfill#3\hskip 4pt}
\hbox to \largesept{\hfill#4\hskip 4pt}
\hbox to \largesept{\hfill#5\hskip 4pt}
\hbox to \largesept{\hfill#6\hskip 4pt}
\hbox to \largesept{\hfill#7\hskip 4pt}
\hbox to \largesept{\hfill#8\hskip 4pt}
}
}

\def\execsept #1 #2 #3 #4 #5 #6 #7 {
\hbox{
\hbox to 25pt{\hfill#1\hskip 4pt}
\hbox to \largesept{\hfill#2\hskip 4pt}
\hbox to \largesept{\hfill#3\hskip 4pt}
\hbox to \largesept{\hfill#4\hskip 4pt}
\hbox to \largesept{\hfill#5\hskip 4pt}
\hbox to \largesept{\hfill#6\hskip 4pt}
\hbox to \largesept{\hfill#7\hskip 4pt}
}
}

\def\execsix #1 #2 #3 #4 #5 #6 {
\hbox{
\hbox to \largesept{\hfill#1\hskip 4pt}
\hbox to \largesept{\hfill#2\hskip 4pt}
\hbox to \largesept{\hfill#3\hskip 4pt}
\hbox to \largesept{\hfill#4\hskip 4pt}
\hbox to \largesept{\hfill#5\hskip 4pt}
\hbox to \largesept{\hfill#6\hskip 4pt}
}
}

\def\execcinq #1 #2 #3 #4 #5 {
\hbox{
\hbox to \largesept{\hfill#1\hskip 4pt}
\hbox to \largesept{\hfill#2\hskip 4pt}
\hbox to \largesept{\hfill#3\hskip 4pt}
\hbox to \largesept{\hfill#4\hskip 4pt}
\hbox to \largesept{\hfill#5\hskip 4pt}
}
}

\def\execquat #1 #2 #3 #4 {
\hbox{
\hbox to \largesept{\hfill#1\hskip 4pt}
\hbox to \largesept{\hfill#2\hskip 4pt}
\hbox to \largesept{\hfill#3\hskip 4pt}
\hbox to \largesept{\hfill#4\hskip 4pt}
}
}

\def\exectroi #1 #2 #3 {
\hbox{
\hbox to \largesept{\hfill#1\hskip 4pt}
\hbox to \largesept{\hfill#2\hskip 4pt}
\hbox to \largesept{\hfill#3\hskip 4pt}
}
}

\subsection{Changing the colour of the locomotive}

   We finish that section with the rules managing the change of colours for a locomotive.
Such a transformation may be necessary when the locomotive goes back to the area of
the simulation where the instructions are placed. On the way back from a register, 
the locomotive is always red, even in the case when it was not possible to decrement
the register. If the next instruction increments a register, the locomotive must become 
green. So we need to convert a red locomotive to a green one.
We have already mentioned that the occurrence of a double locomotive happens within
a round-about only. Accordingly, the change of colour concerns simple locomotives only.
That needs a few instructions gathered in Table~\ref{rchrg}. Their application
is illustrated by Figure~\ref{fchrg}.

\vtop{
\begin{tab}\label{rchrg}
\leurre
Rules for changing a red simple locomotive into a green one.
\end{tab}
\vskip-7pt
\ligne{\hfill conservative and witness rules\hfill\hfill motion rules\hfill}
\ligne{\hfill
% pour mémoire, somme totale  : 706
\vtop{\leftskip 0pt\parindent 0pt\hsize=61pt
\affRr {263} {W} {WWWWWMY} {W}
\affn {W} {WWWWWYM} {W}
\affn {Y} {WWWWWMY} {Y}
\affn {Y} {WWWWYMM} {Y}
\affn {M} {WWMBMYY} {M}
\affn {M} {WWYMBBM} {M}
%                                           +5
}
\hfill\hfill
\vtop{\leftskip 0pt\parindent 0pt\hsize=61pt
\affn {M} {WWYMBBR} {R}
\affn {R} {WWYMBBM} {M}
\affn {M} {WWYGBBM} {M}                %  constate que la couleur a changé
\affn {M} {WWMBRYY} {G}                %  changement de couleur
\affn {G} {WWMBMYY} {M}
\affn {M} {WWGBMYY} {M}
%                                           +6
}
\hfill}
%  somme totale : 706+11 = 717+15(zigzag) = 732
}

\vskip 10pt
\vtop{
\ligne{\hfill
\includegraphics[scale=0.5]{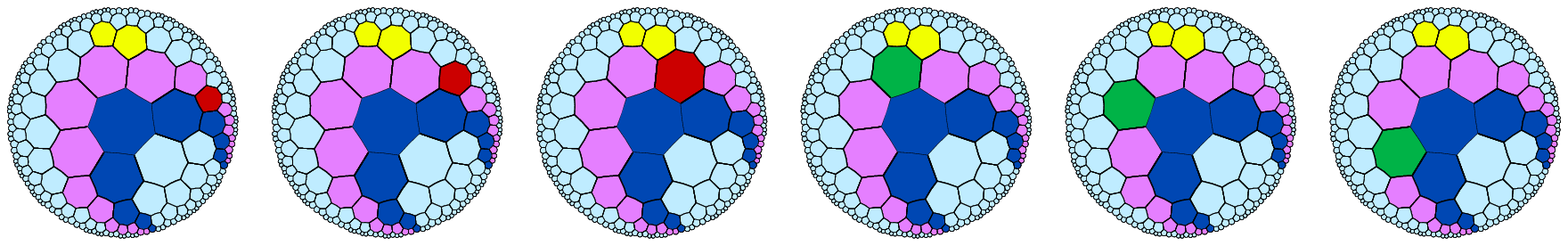}
\hfill}
\vspace{-10pt}
\begin{fig}\label{fchrg}
\leurre
Changing a red simple locomotive into a green one.
\end{fig}
}
\vskip 10pt
The last rule has number~452. However, 42~rules are repeated at least once, 32 of them 
exactly once. Accordingly, the total number of rules as indicated in the tables is~420. 
However, as already mentioned, many rules of the tables are schemes of rules, so that 
the actual number of rules is 732. Note that the total 
number of all possible different rules is \hbox{ 7$^9$ = 40,353,607} which is 
approximately reduced to \hbox{7$^8$ = 5,764,801} if we consider 
rules which are pairwise different under rotation invariance. It is interesting to note
that many rules can be gathered according to their neighbourhood. The neighbourhood
of a rule \hskip -20pt\affH {} {x$_0$} {x$_1$..x$_7$} {x$_8$} is the word 
\hbox{\ftt x$_1$..x$_7$}. As an example, the neighbourhood {\ftt WWWWWOG} occurs
in rules~306 and~333, each of them being repeated in the tables. Rule~306 
changes~{\ftt R} into~{\ftt O} while rule~333 changes~{\ftt W} into~{\ftt R}.
In the same line, the neighbourhood \hbox{\ftt WWWWWRY} occurs in rules~310 and~311.
In rule~311 it changes~{\ftt W} into~{\ftt O} while in rule~310 it remains {\ftt Y}
unchanged. Note that rule~311 entails parasitic occurrences of~{\ftt O} which are erased
by \affH {327} {O} {WWWWWWY} {W} {}. That latter feature reminds me a similarity
with the correction of errors in the replication of DNA in natural processes.

\section*{Conclusion}

There are several questions raised by this result. Is it possible to reduce the number
of states in this context? The huge number of arbitrary rules seems to say that it
could be possible. However, that would be at the cost at reducing several facilities we
had here although the number of freedom is reduced, compared with what was previously
achieved. Perhaps another model should be used for that purpose. What could be done
in the pentagrid is also an open question.

\end{document}